\pdfoutput=1                       
\documentclass[12pt]{article}

\usepackage[letterpaper,left=0.984in,right=0.984in,top=0.984in,bottom=0.7875in]{geometry}

\usepackage[T1]{fontenc}
\usepackage[utf8]{inputenc}
\usepackage{newtxtext}
\DeclareUnicodeCharacter{03A6}{\ensuremath{\Phi}}

\usepackage[table,xcdraw]{xcolor}
\usepackage[normalem]{ulem}
\usepackage{amsmath}
\usepackage{amssymb}
\usepackage{amsbsy}
\usepackage{upgreek}
\usepackage{bm}
\usepackage{physics}
\usepackage{graphicx}
\usepackage{microtype}
\usepackage{float}
\usepackage{setspace}
\usepackage{enumitem}
\usepackage[labelfont=bf]{caption}
\usepackage{tocloft}
\usepackage{hyperref} 
\hypersetup{hidelinks, colorlinks=true, linkcolor=black, citecolor=blue,
            bookmarks=true, bookmarksnumbered=true, bookmarksopen=true, bookmarksopenlevel=1,
            bookmarksdepth=2}

\def\equationautorefname~#1\null{Eq. (#1)\null}

\usepackage[backend=biber, style=nature, sorting=none, maxnames=22, minnames=1,
            url=false, doi=false, isbn=false, refsection=none]{biblatex}
\AtEveryBibitem{%
  \clearfield{month}%
  \clearfield{day}%
  \clearfield{note}%
  \clearfield{eprint}%
  \clearfield{url}%
  \clearfield{urldate}%
  \clearfield{issn}%
  \clearfield{doi}%
  \clearlist{language}%
}
\DeclareFieldFormat{labelnumber}{#1}
\DeclareFieldFormat{labelnumberwidth}{\mkbibbrackets{#1}}

\definecolor{darkgreen}{rgb}{0,0.7,0}

\makeatletter
\let\ORIG@seccntformat\@seccntformat         

\AtBeginDocument{\let\ORIG@contentsline\contentsline}
\newcommand{\SkipTocEntries}{\renewcommand{\contentsline}[4]{}}
\newcommand{\ResumeTocEntries}{\let\contentsline\ORIG@contentsline}

\newcommand{\MAIN@section}{\@startsection{section}{1}{\parindent}%
  {0.1\baselineskip}{0.1\baselineskip}{\normalfont\bfseries}}
\newcommand{\MAIN@subsection}{\@startsection{subsection}{2}{\parindent}%
  {0.1\baselineskip}{0.1\baselineskip}{\normalfont\bfseries}}
\newcommand{\MAIN@seccntformat}[1]{%
  \ifcsname the#1\endcsname\csname the#1\endcsname.\ \fi}

\newcommand{\SI@sectionbody}[1]{%
  \refstepcounter{section}%
  \addcontentsline{toc}{section}{\protect\numberline{\thesection}#1}%
  \vspace{1.0\baselineskip}%
  \noindent\hspace*{\parindent}\textbf{\uline{\thesection .-\ #1}}\par
  \vspace{0.2\baselineskip}%
}
\newcommand{\SI@sectionstar}[1]{%
  \vspace{1.0\baselineskip}%
  \noindent\hspace*{\parindent}\textbf{\uline{#1}}\par
  \vspace{0.2\baselineskip}%
}
\newcommand{\SI@section}{\@ifstar{\SI@sectionstar}{\SI@sectionbody}}

\newcommand{\SI@subsectionbody}[1]{%
  \refstepcounter{subsection}%
  \vspace{0.5\baselineskip}%
  \noindent\hspace*{\parindent}\textbf{\thesubsection .-\ #1}\par
  \vspace{0.1\baselineskip}%
}
\newcommand{\SI@subsectionstar}[1]{%
  \vspace{0.5\baselineskip}%
  \noindent\hspace*{\parindent}\textbf{#1}\par
  \vspace{0.1\baselineskip}%
}
\newcommand{\SI@subsection}{\@ifstar{\SI@subsectionstar}{\SI@subsectionbody}}

\cftpagenumbersoff{section}
\newcommand{\ActivateMainStyle}{%
  \let\section\MAIN@section
  \let\subsection\MAIN@subsection
  \let\@seccntformat\MAIN@seccntformat
  \setcounter{secnumdepth}{2}%
  \setcounter{tocdepth}{2}
  \renewcommand{\thesection}{\Roman{section}}%
  \renewcommand{\thesubsection}{\Alph{subsection}}%
  \renewcommand{\thefigure}{\arabic{figure}}%
  \renewcommand{\thetable}{\arabic{table}}%
  \captionsetup[figure]{labelfont=bf,labelsep=colon}%
  \captionsetup[table]{labelfont=bf,labelsep=colon}%
}

\newcommand{\ActivateSIStyle}{%
  \let\section\SI@section
  \let\subsection\SI@subsection
  \let\@seccntformat\ORIG@seccntformat
  \setcounter{secnumdepth}{3}
  \setcounter{tocdepth}{1}%
  \renewcommand{\thesection}{\Alph{section}}%
  \renewcommand{\thesubsection}{\thesection.\arabic{subsection}}%
  \renewcommand{\thefigure}{S\arabic{figure}}%
  \renewcommand{\thetable}{S\arabic{table}}%
  \captionsetup[figure]{labelfont=bf,labelsep=period}%
  \captionsetup[table]{labelfont=bf,labelsep=period}%
  \setcounter{section}{0}\setcounter{subsection}{0}%
  \setcounter{subsubsection}{0}%
  \setcounter{figure}{0}\setcounter{table}{0}\setcounter{equation}{0}%
  \renewcommand{\theHsection}{SI.\arabic{section}}%
  \renewcommand{\theHsubsection}{SI.\arabic{section}.\arabic{subsection}}%
  \renewcommand{\theHsubsubsection}{SI.\arabic{section}.\arabic{subsection}.\arabic{subsubsection}}%
  \renewcommand{\theHfigure}{SI.\arabic{figure}}%
  \renewcommand{\theHtable}{SI.\arabic{table}}%
  \renewcommand{\theHequation}{SI.\arabic{equation}}%
}
\makeatother

\usepackage{docmute}

\begin{document}

\addtocontents{toc}{\protect\SkipTocEntries}
\ActivateMainStyle
\begin{refsection}

%
%

{\noindent\fontsize{16}{19} \flushleft \bf 
High-field Josephson effect enabled by a moiré Hofstadter spectrum}
\vspace{0.8\baselineskip}

\vspace{-0.5\baselineskip}
{\noindent A. Díez-Carlón\textsuperscript{1,2,3*}, 
M. Cárdenes Wuttig\textsuperscript{4,5}, 
N. Wei\textsuperscript{4}, 
D. Ivanov\textsuperscript{1,2}, 
P. Altpeter\textsuperscript{1,2}, 
P. Hakonen\textsuperscript{6}, 
K. Watanabe\textsuperscript{7}, 
T. Taniguchi\textsuperscript{8}, 
L. I. Glazman\textsuperscript{4} 
and D. K. Efetov\textsuperscript {1,2*}\par}

\begin{enumerate}[label=\arabic*., align=left, nosep]
\item Fakultät für Physik, Ludwig-Maximilians-Universität, Schellingstrasse 4, 80799 München, Germany
\item Munich Center for Quantum Science and Technology (MCQST), München, Germany
\item Nanomaterials and Nanotechnology Research Center (CINN-CSIC), Universidad de Oviedo (UO), Principado de Asturias, 33940 El Entrego, Spain
\item Department of Physics and Yale Quantum Institute, Yale University, New Haven, Connecticut 06520, USA
\item Department of Applied Physics, Yale University,  New Haven, Connecticut 06520, USA
\item Department of Applied Physics, Aalto University, Espoo, 02150, Finland
\item Research Center for Functional Materials, National Institute for Materials Science, 1-1 Namiki, Tsukuba 305-0044, Japan
\item International Center for Materials Nanoarchitectonics, National Institute for Materials Science, 1-1 Namiki, Tsukuba 305-0044, Japan
\end{enumerate}

{\noindent *Correspondence to: andres.diez@cinn.es and dmitri.efetov@lmu.de\par}
\vspace{1.0\baselineskip}

\setstretch{1.1} 


\section*{ABSTRACT}

\textbf{Magnetic fields generally suppress phase-coherent Josephson transport, limiting superconducting interferometry to relatively low fields. Here we show that moiré-engineered graphene Josephson junctions can overcome this constraint. Using ballistic graphene/hBN junctions, we establish phase-coherent Andreev transport through Fabry-Pérot oscillations and Fraunhofer interference that persist across both the primary Dirac cone and the reconstructed moiré minibands. We then demonstrate phase-coherent Josephson interference up to $\sim$6 T in the fractal Hofstadter-butterfly regime, well beyond the range expected for conventional ballistic graphene junctions. Comparison with Hofstadter-spectrum calculations reveals that superconductivity survives where the moiré potential transforms Landau levels with quenched group velocity into dispersive magnetic Bloch bands with finite quasiparticle group velocity, enabling extended electron–hole Andreev trajectories across the junction. Our results show that Hofstadter minibands can stabilize phase-coherent superconductivity deep into the parameter domain conventionally associated with the quantum Hall regime, establishing a new platform for high-field superconducting interferometry.}
\vspace{1.0\baselineskip}

%
%
%
%
%
\section{INTRODUCTION}


Josephson junctions (JJs) are among the most sensitive phase-coherent quantum devices, forming the basis of superconducting interferometry and enabling technologies ranging from SQUID magnetometers to quantum processors \cite{clarke_squid_2004, kjaergaard_superconducting_2020}, while providing a direct probe of macroscopic quantum coherence \cite{tinkham_introduction_1996}. In conventional JJs, however, magnetic fields rapidly suppress coherent supercurrents through destructive interference and electron trajectory bending, restricting interferometric operation to relatively low fields. Yet many of the most exotic quantum phases of matter—including quantum Hall states, topological superconductors, and strongly correlated electron systems—emerge only in strong magnetic fields, where few experimental probes can directly access the phase structure of quantum many-body states \cite{tsui_two-dimensional_1982, levy_magnetic_2005, pierce_unconventional_2021, tsui_direct_2024, nuckolls_spectroscopy_2025}. A Josephson interferometer that remains phase coherent at multi-tesla $B$-fields could therefore provide a valuable platform for phase-sensitive studies of quantum matter in regimes that have long remained inaccessible to superconducting quantum devices. 

Hybrid superconductor--graphene JJs provide a uniquely versatile platform for exploring superconductivity in high magnetic fields. Their gate-tunable carrier density, ballistic transport, and intrinsically two-dimensional nature enable direct access to phase-coherent Andreev bound states (ABS) across a broad range of magnetic-field regimes, from electron-optics phenomena such as Fabry-Pérot and Fraunhofer interference \cite{calado_ballistic_2015, allen_spatially_2016, bretheau_tunnelling_2017, jung_tunneling_2025}, to Landau quantization and the quantum Hall effect \cite{amet_supercurrent_2016, lee_inducing_2017, wei_chiral_2019, seredinski_quantum_2019, zhao_interference_2020, vignaud_evidence_2023, barrier_one-dimensional_2024}. The survival of the proximity effect at elevated $B$ is highly sensitive to the trajectories of electron--hole Andreev pairs between the superconducting leads. In ballistic monolayer graphene JJs, bulk ABS have been shown to sustain supercurrent up to $\sim2.5$~T, with the limiting field determined primarily by mesoscopic device dimensions \cite{ben_shalom_quantum_2016, amet_supercurrent_2016, villani_quasi-0_2025}. Beyond this regime, magnetic field confines the motion of charge carriers to cyclotron orbits and progressively suppresses coherent bulk transport, ultimately driving a crossover from conventional Josephson coupling to the Andreev pair propagation along defects running across the junction~\cite{barrier_one-dimensional_2024}, or transport along chiral edge states in the regime of Landau quantization~\cite{amet_supercurrent_2016, vignaud_evidence_2023}.

While Landau quantization of simple electronic bands, such as graphene's Dirac cone, generally suppresses bulk superconducting transport, van der Waals materials offer a unique opportunity to explore to test this constraint by engineering the electronic spectrum through moiré superlattices. In graphene aligned to hBN, the moiré potential reconstructs the Dirac dispersion into minibands hosting satellite Dirac points, van Hove singularities, and, at high magnetic fields, a fractal Hofstadter spectrum with emergent subbands at experimentally accessible densities \cite{yankowitz_emergence_2012, ponomarenko_cloning_2013, hunt_massive_2013, dean_hofstadters_2013, yu_hierarchy_2014, wang_evidence_2015, wallbank_generic_2013, moon_electronic_2014}. By reshaping the electronic structure on the same energy scale as magnetic quantization, moiré minibands provide a natural platform to investigate whether Hofstadter subbands can stabilize and even enhance phase-coherent superconducting transport beyond the conventional quantum Hall crossover.

Here we address this question by engineering ballistic JJs whose weak link is a graphene/hBN moiré superlattice. We first establish phase-coherent Andreev transport through measurements of Fabry-Pérot oscillations in the critical current that persist across both the primary Dirac cone and the reconstructed minibands. We then map the Josephson effect at high magnetic fields, revealing supercurrents that survive up to $\sim6$~T, far beyond the range expected for the conventional semiclassical mechanism in ballistic graphene JJs \cite{ben_shalom_quantum_2016}. Comparison with Hofstadter-spectrum calculations reveals that the enhanced high-field robustness of superconductivity arises where the moiré potential transforms Landau levels with quenched group velocity into dispersive magnetic Bloch bands. By interrupting Landau quantization and restoring finite quasiparticle group velocity, these Hofstadter minibands enable extended electron–hole Andreev trajectories across the moiré weak link, thereby stabilizing phase-coherent Josephson transport deep into the quantum Hall regime.

%
%
\vspace{1.0\baselineskip}
\section{RESULTS}
%
%
\subsection{Phase-coherent transport in a ballistic graphene/hBN moiré JJ}

\begin{figure}[H]
  \centering
  \includegraphics[width=\linewidth]{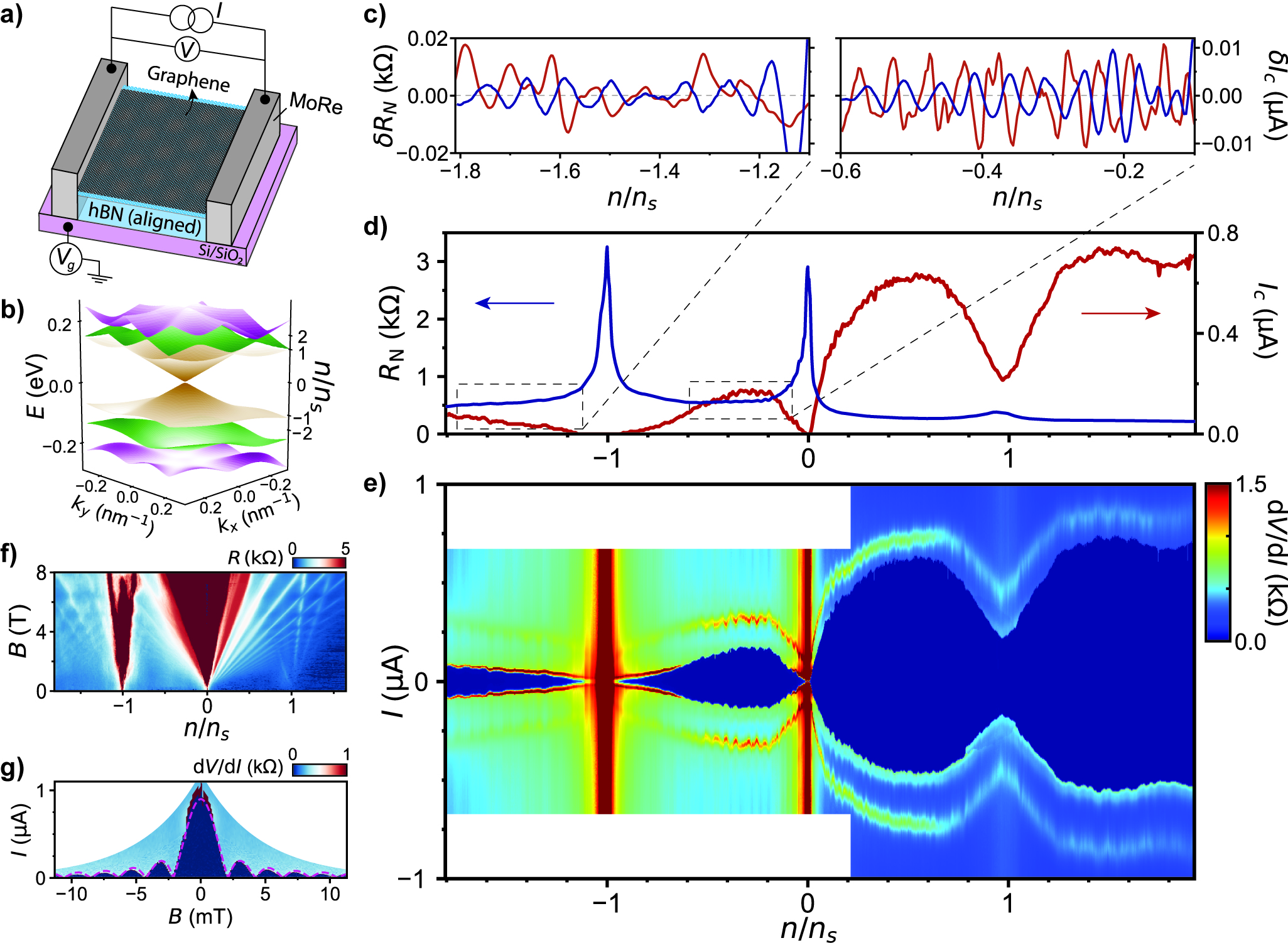}
  \caption{
  \textbf{Ballistic superconducting proximity effect in a moiré graphene/hBN Josephson junction.}
  \textbf{a)} Device schematic of a graphene/hBN moiré superlattice acting as the weak link of a JJ. The voltage $V$ across the junction is recorded as a current bias $I$ is applied through the superconducting electrodes in a two-probe measurement. The carrier density $n$ is tuned by a gate voltage $V_g$ applied to the doped Si.
  \textbf{b)} Electronic band structure of a graphene/hBN moiré superlattice with twist angle $\theta=0.21$°, with the main Dirac cone in brown and the moiré minibands in green and violet.
  \textbf{c)} Fabry-Pérot oscillations in the moiré minibands (left) and the main Dirac cone (right) shown as a function of carrier density in units of the superlattice density $n/n_s$. These are obtained from \textbf{d} by subtracting a smooth fit of the normal-state resistance $R_N$ (blue) and critical current $I_c$ (red) (see SI for more details).
  \textbf{d)} $R_N$ at 10~K in blue (left axis), and $I_c$ at 2~K in red (right axis), both as a function of $n/n_s$.
  \textbf{e)} Differential resistance $\mathrm{d}V/\mathrm{d}I$ map at 2~K, where dark blue regions are superconducting. Their contour along positive d.c. current bias $I$ yields the $I_c$ in \textbf{d}. 
  \textbf{f)} Hofstadter-butterfly Landau fan at 100~mK, obtained from the two-probe resistance $R$ vs $n/n_s$ and vs an applied perpendicular magnetic field $B$ to the graphene sheet.
  \textbf{g)} Interference pattern at $n/n_s\sim1.6$, where $\mathrm{d}V/\mathrm{d}I$ is measured vs $B$ and vs $I$ at 100~mK. The $I_c$ follows well the dashed Fraunhofer curve expected for a uniform junction of the device area.
  }
\label{fig1:device_MoRe}
\end{figure}

Devices investigated in this work consist of Josephson junctions defined in hBN-encapsulated monolayer graphene crystals aligned to one hBN layer, such that the weak link experiences a moiré superlattice potential. Superconducting MoRe leads, with critical temperature $T_c\sim9$~K and critical field $\sim8$~T, are coupled to the graphene/hBN stack via one-dimensional edge contacts \cite{calado_ballistic_2015}, enabling measurements of the Josephson effect over a wide range of carrier density $n$ and perpendicular magnetic field $B$. A schematic of the device architecture and measurement configuration is shown in \autoref{fig1:device_MoRe}a. The principal observations reported here were reproduced across multiple additional devices with different twist angles and geometries (see Supplementary Information), however in the following we focus on device GH1, which has a twist angle $\theta\sim0.21\pm 0.01^\circ$, junction length $L\sim200$ nm, and width $W\sim1.5$~µm.

Above the superconducting transition temperature of MoRe, at $10$~K, the normal-state resistance of the device, $R_N$, exhibits satellite Dirac peaks at carrier densities corresponding to full filling of the moiré unit cell, $n/n_s=\pm1$ (\autoref{fig1:device_MoRe}d) \cite{hunt_massive_2013, ponomarenko_cloning_2013}. At high magnetic fields, Landau fans emerge from these satellite peaks, confirming the superlattice reconstruction of the Dirac band (see \autoref{fig1:device_MoRe}f) \cite{dean_hofstadters_2013, yu_hierarchy_2014, wang_evidence_2015}. Upon cooling into the superconducting state, current-biased $\mathrm{d}V/\mathrm{d}I$ maps measured at $2$~K (\autoref{fig1:device_MoRe}e) reveal a robust zero-resistance state, demonstrating the Josephson effect across the entire accessible density range, spanning from the main Dirac band into both the electron ($n/n_s>1$) and hole ($n/n_s<-1$) moir\'e minibands (see also \autoref{fig1:device_MoRe}b).
The critical current $I_c$, shown in \autoref{fig1:device_MoRe}d, is extracted from \autoref{fig1:device_MoRe}e as the d.c. bias current $I$ at which $\mathrm{d}V/\mathrm{d}I$ reaches 50\% of $R_N$. Furthermore, \autoref{fig1:device_MoRe}g shows the characteristic Fraunhofer interference pattern at $n/n_s\sim1.6$, where $I_c$ oscillates with a period of $\sim2.2\pm0.3$~mT, corresponding to one superconducting flux quantum $\phi_0=h/2e$ threading the weak-link area when flux-focusing effects are taken into account (see SI for details). The agreement with the standard Fraunhofer pattern confirms a uniform supercurrent distribution across the device.


A defining hallmark of ballistic transport in our devices is the presence of Fabry-Pérot interference: phase-coherent oscillations arising from multiple reflections of electron waves between the device leads, so long as the electronic mean free path exceeds the junction length. In our JJs, this electronic resonant cavity is formed by the n-doping of graphene from the MoRe leads, which creates a n-p-n junction when the graphene channel is hole-doped. Such resonances modulate not only $R_N$ but also the critical current $I_c$, which oscillates together with the electron transmission coefficient \cite{calado_ballistic_2015, tinkham_introduction_1996}.
These Fabry-Pérot oscillations are observed in \autoref{fig1:device_MoRe}c as a function of carrier density, appearing near the charge neutrality point and extending across almost the entire main Dirac band (see right panel of \autoref{fig1:device_MoRe}c and SI), consistent with previous studies of ballistic graphene JJs \cite{calado_ballistic_2015, ben_shalom_quantum_2016, borzenets_ballistic_2016}.

Crucially, we also observe a second set of oscillations in the moiré minibands at $n/n_s<-1$ (left panel of \autoref{fig1:device_MoRe}c). Analysis of their periodicity (see SI) reveals a reduced miniband Fermi velocity, $v_F\sim0.5\times10^6$~m/s compared to $v_F\sim0.8\times10^6$~m/s in the main Dirac cone, together with a breakdown of the periodicity near the van Hove singularities, consistent with previous studies \cite{lee_ballistic_2016, handschin_fabry-perot_2017, kraft_anomalous_2020, mrenca-kolasinska_probing_2023, moon_electronic_2014}. We confirm that the origin of these oscillations lies on Fabry-Pérot interferometry by their persistence in the normal state at voltage bias above the MoRe superconducting gap, and that our JJs belong to the long-ballistic regime by studying the temperature dependence of $I_c$ (see SI) \cite{dubos_josephson_2001, borzenets_ballistic_2016, salim_revisiting_2023}. While Fabry-Pérot oscillations have previously been reported in graphene/hBN moiré superlattices under transverse electron focusing \cite{lee_ballistic_2016, kraft_anomalous_2020} and in p-n-p junctions \cite{handschin_fabry-perot_2017, mrenca-kolasinska_probing_2023}, here we observe Fabry-Pérot interference in the supercurrent itself. To our knowledge, this constitutes the first realization of a ballistic Josephson effect across the moiré minibands of graphene/hBN \cite{diez_carlon_study_2025, indolese_signatures_2018}. As we show below, the ballistic, phase coherent charge carrier propagation is a key prerequisite for stabilizing the Josephson effect in the high-magnetic-field limit \cite{ben_shalom_quantum_2016, amet_supercurrent_2016}.

%
%
\vspace{1.0\baselineskip}
\subsection{Interplay between Landau quantization and superconductivity in  primary Dirac cones}

\begin{figure}[H]
    \centering
    \includegraphics[width=\linewidth]{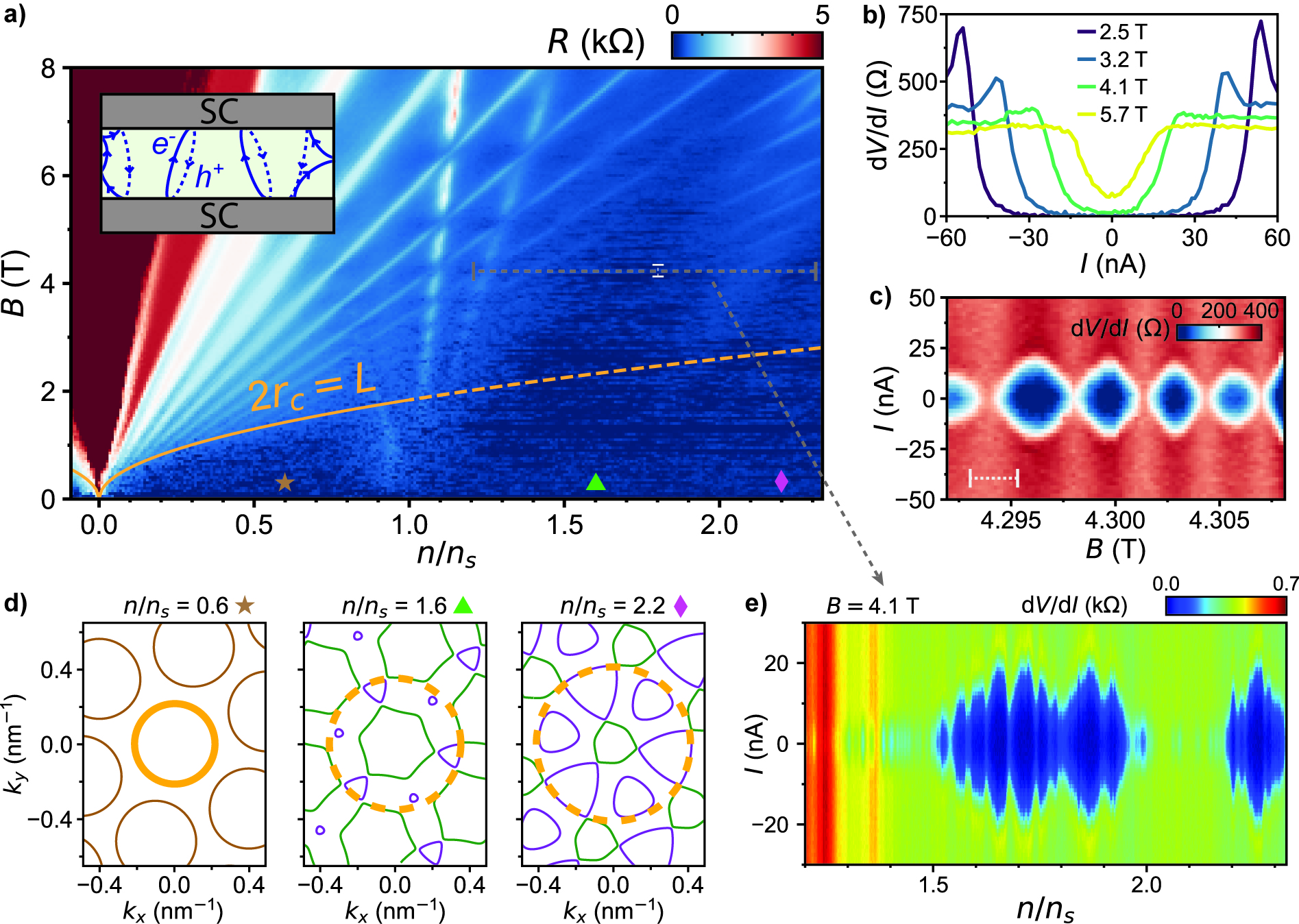}
    \caption{
    \textbf{High field superconductivity at the moiré bands.}
    \textbf{a)} Resistance $R$ vs perpendicular magnetic field $B$ and carrier density in units of the superlattice density $n/n_s$. The dark-blue superconducting pockets are well bounded by the orange solid line in the main Dirac cone band ($0 < n/n_s < 1$), whereas in the moiré minibands ($n/n_s > 1$) superconductivity persists up to $\sim6$ T, surpassing the threshold (orange dashed line) set by the Fermi-surface trajectories shown in \textbf{d}. Inset: sketch of semiclassical trajectories followed by the ABS at high fields.
    \textbf{b)} Differential resistance $\mathrm{d}V/\mathrm{d}I$ vs d.c. current $I$ at a fixed $n/n_s \sim 1.8$ for several values of magnetic field.
    \textbf{c)} Interference pattern measured at $n/n_s \sim 1.8$ and around $B = 4.3$~T. The location in the phase diagram \textbf{a} is indicated by a vertical dashed white line.
    \textbf{d)} Calculated equal-energy contours at $n/n_s=0.6$, $1.6$, and $2.2$; following the same color code as the bands in \autoref{fig1:device_MoRe}b. The solid orange line in the left panel represents the semiclassical reciprocal-space trajectory of quasiparticles in the main Dirac band. The dashed orange lines in the middle and right panels represent the trajectories at densities $n/n_s>1$, after the main Dirac band dynamics is restored by full magnetic breakdown. The field corresponding to these trajectories under the condition $2r_c=L$ is plotted in \textbf{a}. 
    \textbf{e)} $\mathrm{d}V/\mathrm{d}I$ map vs $I$ and vs $n/n_s$ at $B=4.1$~T, where superconducting pockets with oscillating $I_c$ are observed. The location in \textbf{a} is marked by a horizontal dashed gray line.
    All experimental data were obtained at 100 mK.
    }
    \label{fig2:highfield_data}
\end{figure}

\autoref{fig2:highfield_data}a shows the resistance $R$ as a function of $n/n_s$ and $B$, measured at a temperature of $100$~mK. The resulting Landau fan diagram clearly reveals the emergence of Landau levels originating from both the primary and satellite Dirac points induced by the hBN-aligned moiré superlattice. At sufficiently high magnetic fields, these Landau fans evolve into a Hofstadter butterfly spectrum, reflecting the interplay between the magnetic length scale and the moiré superlattice periodicity. 

Strikingly, across a large region of the $n$--$B$ phase space, which mostly does not coincide with clearly formed Landau levels, we also observe robust signatures of superconductivity, which remarkably persist up to fields as high as $B\sim6$~T. These superconducting features manifest as scattered pockets of vanishing resistance $R$, that are visible as dark-blue specks in \autoref{fig2:highfield_data}a, and which respectively show non-linear $\mathrm{d}V/\mathrm{d}I$ characteristics in \autoref{fig2:highfield_data}b. Even in this extreme high-field regime, these states show phase-coherent Josephson transport, as evidenced by the oscillating $I_c$ with both magnetic field (see \autoref{fig2:highfield_data}c), and carrier density (\autoref{fig2:highfield_data}e).

How can the high-field interplay between Landau quantization and superconductivity be understood? We first focus on the region $n/n_s<1$, where the electronic structure is described by a single Dirac cone and has been extensively studied in previous works with graphene JJs \cite{ben_shalom_quantum_2016, amet_supercurrent_2016, villani_quasi-0_2025}. In the presence of a magnetic field, charge carriers undergo cyclotron motion with radius $r_c = \hbar k_F/eB$ in real space, while in momentum space they traverse circular trajectories defined by the contour of the Fermi surface (left panel of \autoref{fig2:highfield_data}d). The size of these orbits is set by the Fermi wave vector, which scales as $k_F=\sqrt{\pi n}$.
In our JJs, for well-defined Landau levels to emerge, charge carriers must be able to complete a cyclotron orbit before encountering the device boundaries, which requires the cyclotron diameter to remain smaller than the junction length, i.e. $2r_c < L$. Conversely, when $2r_c > L$, quasiparticles reach the superconducting contacts before completing a full orbit, preventing the formation of closed phase-coherent trajectories and thereby suppressing Landau quantization.

To emphasize this crossover, we plot the corresponding $2r_c=L$ boundary in \autoref{fig2:highfield_data}a as an orange solid line within the unreconstructed Dirac cone, and extend it as a dashed line into the minibands. At $n/n_s<1$ we observe that this boundary remarkably matches the onset of Landau quantization and how it evolves to higher fields with $\sqrt{n}$, consistent with previous studies of ballistic graphene JJs \cite{ben_shalom_quantum_2016, amet_supercurrent_2016, villani_quasi-0_2025}.
At the same time, this same boundary $2r_c = L$ marks the loss of the observed superconducting pockets.
Under a semiclassical picture, this occurs because the electron-hole pairs forming the ABS also move along the cyclotron orbits, and thus their trajectories are able to connect the two superconducting leads only if $2r_c>L$.
As the cyclotron radius $r_c$ shrinks upon raising of the $B$-field, an increasingly intricate network of paths connecting the superconducting electrodes forms (see inset of \autoref{fig2:highfield_data}a), rendering the interference patterns irregular in amplitude and period \cite{ben_shalom_quantum_2016, amet_supercurrent_2016}. 

Consequently, for a \textit{B}-field above this limit, ABS can no longer complete the Andreev processes required to establish a supercurrent between the electrodes, leading to the suppression of the bulk proximity effect and signaling the transition to transport dominated by Landau quantization. 
Although other works have shown that superconductivity can still be sustained for \textit{B}-fields above this limit within the quantum Hall regime by chiral Andreev states with $I_c\lesssim1$~nA \cite{amet_supercurrent_2016, seredinski_quantum_2019, vignaud_evidence_2023, zhao_interference_2020}, we do not observe this in our experiment, since their corresponding Josephson energy $E_J/k_B\sim30$~mK is washed out by thermal fluctuations at our 100 mK base temperature \cite{tinkham_introduction_1996, amet_supercurrent_2016, zhao_thermal_2025, zhao_loss_2023}.

\vspace{1.0\baselineskip}
\subsection{High-field superconductivity and Hofstadter butterfly in moiré minibands}

Strikingly, the picture described above breaks down once the electron density reaches $n/n_s>1$ and the Fermi level enters the moiré minibands. In this regime, the phase space supporting high-field superconductivity in \autoref{fig2:highfield_data}a expands dramatically, well beyond the limit set by $2r_c = L$, extending to magnetic fields as high as $\sim6$~T. We provide evidence in the following that even at these extremely high fields, the ABS transport mechanism remains dominated by mesoscopic electron-hole orbits. 

In these pockets and at this density range, the supercurrent reaches values as large as $I_c\sim48\pm2$~nA (see the $2.5$~T trace in \autoref{fig2:highfield_data}b), comparable to the quantum ballistic limit of a single ABS mode, $I_Q=e\Delta/h\sim53\pm1$~nA \cite{ben_shalom_quantum_2016, tinkham_introduction_1996}, where $\Delta\sim1.4$~meV is the BCS gap of our MoRe leads. 
We note that at $4.3$~T, the expected $I_c$ following the conventional Fraunhofer interference shown in \autoref{fig1:device_MoRe}g, would be $\sim0.01$~nA, far less than our observations in \autoref{fig2:highfield_data}c.
The irregular variation in the amplitude and periodicity of these oscillations, ranging from $\sim2.3\pm0.2$~mT to $\sim3.4\pm0.4$~mT (see also SI), can be associated with a chaotic billboard of Andreev-pair trajectories whose enclosed area varies randomly, modifying the field required to thread one $\phi_0$, as was observed in Ref.~\cite{ben_shalom_quantum_2016}. 
These observations contrast sharply with the quantum Hall regime of the Josephson effect, where supercurrents mediated by chiral edge states are typically far more fragile ($I_c\lesssim1$~nA) and their periodicity with field very regular \cite{amet_supercurrent_2016, vignaud_evidence_2023, zhao_interference_2020, kurilovich_disorder-enabled_2023, zhao_loss_2023}. This mechanism is further disfavored by measurements of the Landau fan diagram performed while quenching the superconducting pockets with a d.c. bias current of 200~nA (see SI), well above the maximum $I_c$ for $B>10$~mT. Under these conditions, the underlying normal state exhibits no signatures of Landau quantization. Thus, the high-field superconducting pockets in the moiré minibands emerge precisely where Landau quantization is absent.

This dramatic extension of the superconducting phase to high magnetic fields is nevertheless at odds with the semiclassical description employed for the Dirac band at $n/n_s<1$, where superconductivity is governed by electron cyclotron orbits. Upon entering the moiré minibands at $n/n_s>1$, the electronic structure evolves into a multi-band regime with strongly faceted Fermi surfaces, giving rise to intricate Fermi contours and irregular cyclotron trajectories (middle and right panels of \autoref{fig2:highfield_data}d) \cite{lee_ballistic_2016, handschin_fabry-perot_2017, kraft_anomalous_2020, de_vries_kagome_2024}. 
Because these Fermi pockets, and the corresponding real-space electron orbits, remain smaller than the orbits in the absence of moiré potential, one would thus naively expect the superconducting phase to be confined to even lower magnetic fields than for $n/n_s<1$. Furthermore, magnetic breakdown between neighboring contours \cite{alexandradinata_geometric_2017, bocarsly_haasvan_2024}, would result in trajectories still bound by the size of the main Dirac cone extrapolated to the same carrier density (orange dashed contours in \autoref{fig2:highfield_data}d).
However, the corresponding upper bound set by the condition $2r_c=L$, shown as the orange dashed line in \autoref{fig2:highfield_data}a, falls well below the observed extent of the superconducting phase.

%
%

\vspace{1.0\baselineskip}
\subsection{Charge propagation across dispersive Hofstadter bands}

To identify what sustains the proximity effect, the theoretical description calls for an analysis beyond considering the semiclassical motion controlled by the zero-field Fermi surfaces, and to account for the high-field Hofstadter spectrum \cite{wallbank_generic_2013, moon_electronic_2014}. 
For the parameters of our experiment, the ballistic Thouless energy $E_{\rm Th}=\hbar v_F/L$ is the controlling scale of the system, ranging from $E_{\rm Th}\sim 2.6$~meV at the main Dirac band, to $E_{\rm Th}\sim 1.6$~meV at the moiré minibands is $E_{\rm Th}\sim 2.6$~meV; both in the order of the superconducting gap $\Delta$ in the MoRe leads. That allows one to estimate the Josephson critical current by the ballistic version of the Ambegaokar-Baratoff relation, $I_c\sim Wn^{1/2}(e\Delta/\hbar)$, if $n<n_s$ and no magnetic field applied. 
In case $E_{\rm Th}\lesssim\Delta$, the Josephson current is limited by Thouless energy, $I_c\sim Wn^{1/2}(e E_{\rm Th}/\hbar)$ \cite{tinkham_introduction_1996}. At $E_{\rm Th}\ll\Delta$, this relation  can be derived by a full solution of the Schr\"odinger equation, or using the semiclassical real-space trajectories representation \cite{meier_edge_2016}.
This picture is consistent with the experimentally observed suppression $I_c$ at $2r_c<L$, where the wave packets are assembled from ``flat'' Landau levels having zero group velocity, leading to $E_{\rm Th}\sim 0$. 

The flatness of Landau levels is predicated on the irrelevance of the magnetic breakdown effects. This condition is violated once the Fermi lines associated with the primary Dirac cones approach the boundaries of the moiré Brillouin zone, which happens at $n\sim 0.9n_s$ (see left panel of \autoref{fig2:highfield_data}d). There, magnetic breakdown broadens Landau levels into bands of the Hofstadter spectrum, whose dispersion endows the carriers with
a finite group velocity $v_{\rm H}$ \cite{alexandradinata_geometric_2017}.
Remarkably, the wave packets picture of the real-space electron motion is applicable to the Hofstadter bands, as was shown in Ref. \cite{chang_berry_1996}. A finite $v_{\rm H}$ enables electron propagation between the leads, restores a finite $E_{\rm Th}$, and in turn removes the exponential suppression of the bulk contributions to the junction normal-state conductance $1/R$ and of the $I_c$ in the superconducting state.
We conjecture that the short-junction limit is preserved if the Thouless energy for a Hofstadter band, $E_{\rm Th}^{\rm H}\sim \hbar v_{\rm H}/L$ is large enough, $E_{\rm Th}^{\rm H}\gtrsim\Delta$. For the conditions of our experiment, the crossover regime $E_{\rm Th}^{\rm H}\sim\Delta$ is reached at $v_{\rm H}\sim 0.42 v_F$.

\begin{figure}[H]
  \centering
  \includegraphics[width=\linewidth]{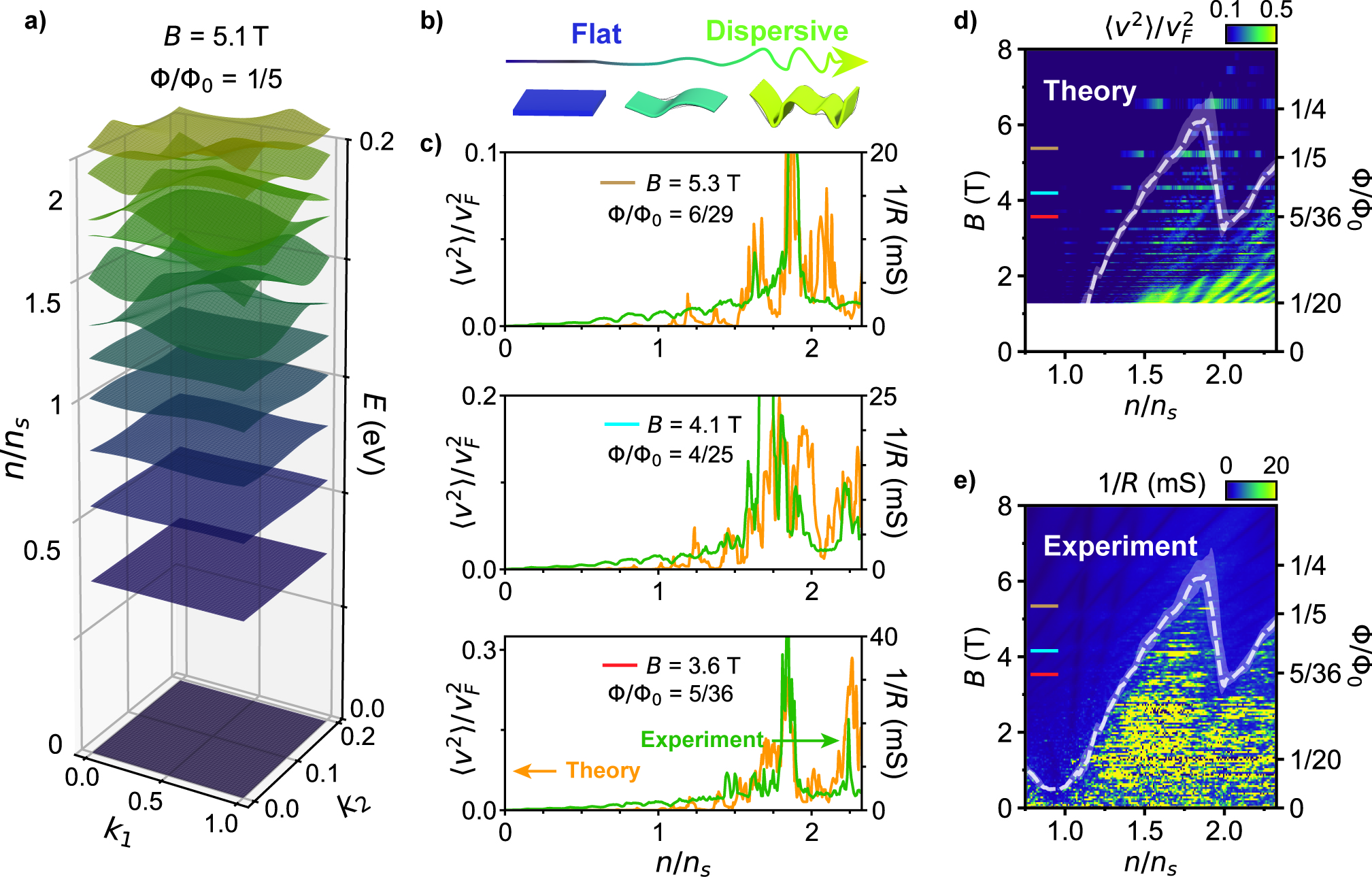}
  \caption{
  \textbf{Absence of Landau levels from dispersive bands in the superconducting regions.}
  \textbf{a)} Hofstadter electronic bands calculated at a flux per moiré unit cell $\Phi/\Phi_0=1/5$ corresponding to a magnetic field $B=5.1$~T. The quasimomentum ${\bf k}= k_1 {\bf G}_1+k_2{\bf G}_2$, where ${\bf G}_{1,2}$ are two primitive reciprocal lattice vectors of the graphene/hBN mini Brillouin zone.
  \textbf{b)} Sketch illustrating the increasing dispersion upon doping into the moiré minibands at high fields, according to the calculations in \textbf{a} and to their comparison with the experimental data in \textbf{c} and \textbf{d}-\textbf{e}.
  \textbf{c)} The three panels show the calculated average squared band velocity $\langle v^2 \rangle$ normalized by the square graphene Fermi velocity $v_F^2$ in orange (left) as a function of filling $n/n_s$, at a fixed magnetic flux $\Phi/\Phi_0 = 6/29$, $4/25$ and $5/36$, respectively. Their overlap with the experimental inverse resistance $1/R$ in green (right axis) at the corresponding magnetic fields $B = 5.3$~T, $4.1$~T and $3.6$~T, indicates that the observed superconductivity (high conductance $1/R$) and enhanced group velocity occur in the same parameter domains.
  \textbf{d)} Map of calculated $\langle v^2 \rangle / v_F^2 $ vs $n/n_s$ in the ($\xi = +$)-valley over a wide range of flux $\Phi/\Phi_0$ (right axis) and its corresponding $B$ (left axis). The lower and upper colorbar limits correspond to the limits imposed by the Thouless energy. The white dashed line marks the experimental maximum field of the superconducting regions in \textbf{e} within the extracted standard deviation of the transition (transparent white background).
  \textbf{e)} Map of measured inverse resistance $1/R$ from \autoref{fig2:highfield_data}a, showing that the superconducting regions coincide with the dispersive regions from \textbf{d}.
  The linecuts in \textbf{c} are indicated in \textbf{d} and \textbf{e} by horizontal lines following the same color code.
  }
  \label{fig3:theory_vs_data}
\end{figure}

Once the electron Fermi level enters the dense pack of minibands, the separation between the Fermi pockets becomes small (see \autoref{fig1:device_MoRe}b and \autoref{fig2:highfield_data}d) and easily breached by the magnetic breakdown. In the limit of strong field, breakdown restores the electron dynamics associated with the primary Dirac spectrum, which restores well-separated, nearly-flat Landau levels. However, in a broad range of intermediate fields the magnetic breakdown succeeds in creating a crowded spectrum of highly-dispersive Hofstadter bands capable of supporting large values of $v_{\rm H}$. 
An example of such Hofstadter band dispersion is shown in \autoref{fig3:theory_vs_data}a-b, which shows our numerical calculations at magnetic flux per moiré unit cell $\Phi/\Phi_0=1/5$, with $\Phi_0=h/e$. It can be seen that once doping reaches $n/n_s\sim 1.5$, the electron Fermi level enters the domain of wide, overlapping bands of the Hofstadter spectrum.

Since evaluating the dispersive spectrum at generic fields is impractical due to the fractionalization of the magnetic Brillouin zone, we instead follow Ref.~\cite{krishna_kumar_high-order_2018} and use the Fermi-level average $\langle v^2\rangle$ as a proxy for $v_{\rm H}^2$. 
\autoref{fig3:theory_vs_data}c-d shows our results, where $\langle v^2\rangle/v_F^2$ is represented as a function of $n/n_s$. 
According to our conjecture, we expect the survival of superconductivity beyond the geometrical limit $2r_c=L$ if $\langle v^2\rangle/v_F^2$ is not reduced to values substantially lower than $\sim0.17$, so that $E_{\rm Th}^{\rm H}$ is not much smaller than $\Delta$.
\autoref{fig3:theory_vs_data}c-d confirms that the range of densities $\delta n$ where $\langle v^2\rangle/v_F^2>0.1$ narrows as $B$ increases, dropping to $\delta n\sim 0.1 n_s$ at $B=5.3$~T; reflecting the overall trend towards flat Landau levels.

To relate the theoretical model to our experimental results, we use the base-temperature zero-bias conductance $1/R$ as a proxy for the onset of the superconducting phase, i.e. finite junction critical current. This measured $1/R$ map in \autoref{fig3:theory_vs_data}e clearly correlates with that of the numerical $\langle v^2\rangle/v_F^2$ in \autoref{fig3:theory_vs_data}d, where bright regions indicating superconductivity coincide with those of high group velocity, and hence high Thouless energy. 
This dispersive-band crowding therefore accounts for the experimentally observed survival of the Josephson effect up to $\sim6$~T, precisely in the field–density regions where Landau quantization is suppressed.

%
%
%
\vspace{1.0\baselineskip}

\section{CONCLUSIONS}


Our results thus establish ballistic moiré graphene/hBN JJs as a platform that extends the field limit of the semiclassical regime of the Josephson effect to $\sim6$~T when the Fermi level enters the moir\'e minibands, as compared to previous graphene JJ experiments that were as high as $\sim2.5$~T \cite{ben_shalom_quantum_2016, amet_supercurrent_2016, villani_quasi-0_2025}. We find that the onset of this phase cannot be explained within a semiclassical picture built on the zero-field Fermi surface, even when magnetic breakdown between neighboring Fermi contours is included. Hofstadter-spectrum calculations instead show that the observed superconducting regions coincide with those where multiple dispersive Hofstadter bands overlap in energy and Landau quantization is suppressed, such that open-orbit states remain available for ABS transport across the weak link. Our work therefore identifies a previously unexplored route for stabilizing supercurrents at high magnetic fields by using moiré-engineered magnetic Bloch bands to preserve finite-velocity quasiparticle propagation across the weak link.

Given that the critical currents remain comparatively large, and their amplitude and field periodicity indicate mesoscopic interference of high-field ABS trajectories, the regime studied here is distinct from the quantum Hall Josephson effect mediated by chiral Andreev edge states \cite{amet_supercurrent_2016, lee_inducing_2017, wei_chiral_2019, zhao_interference_2020, vignaud_evidence_2023}. Instead, the observed supercurrent is governed by ABS trajectories shaped by the Hofstadter miniband structure. Another manifestation of that structure are the Brown--Zak oscillations in the normal-state resistance, which are facilitated by the enhanced charge carrier velocity at commensurate fluxes \cite{krishna_kumar_high-order_2018, barrier_long-range_2020}.
A natural next step of our work would therefore be to extend the present approach to regimes where Hofstadter minibands, Brown--Zak transport, and quantum Hall edge physics coexist within the weak link, enabling controlled studies of Andreev processes in fractal spectra and of the interplay between miniband topology and superconducting correlations.

%
%
%
%
%
%
%
%
%
\pagebreak

%
%
\pagebreak

%
%
\pagebreak

%
%
%
%
%
\pagebreak
\section*{References}
\printbibliography[heading=none]

%
%
%
%
%
\pagebreak

\vspace{0.5cm}
\noindent{\textbf{Acknowledgements:}
D.K.E. acknowledges funding from the European Research Council (ERC) under the European Union's Horizon 2020 research and innovation program (grant agreement No. 852927), the German Research Foundation (DFG) under the priority program SPP2244 (project No. 535146365), the EU EIC Pathfinder Grant ``FLATS'' (grant agreement No. 101099139) and the Keele, Kavli, Tschira and Wells Foundations as part of the SuperC collaboration.
K.W. and T.T. acknowledge support from the Elemental Strategy Initiative conducted by the MEXT, Japan (grant number JPMXP0112101001) and JSPS KAKENHI (grant numbers 19H05790, 20H00354, and 21H05233).
M.C.W., N.W. and L.I.G. thanks Daniil Antonenko, Yongxin Zeng, Peter Ding, Tomohiro Soejima, and Kryštof Kolář for helpful discussions and the Yale Center for Research Computing for guidance and assistance in computations run on the Grace cluster. Work at Yale University was supported by NSF
Grant No. DMR-2410182 and by the Air Force Office of Scientific Research (AFOSR) under Award No.
FA95502510287.\par}

\vspace{0.5cm}
\noindent{\textbf{Author contributions:}
A.D.C. and D.K.E. conceived and designed the experiments; 
A.D.C. and D.I. fabricated the devices; 
A.D.C. performed the measurements and analyzed the data; 
M.C.W., N.W. and L.I.G. performed the theoretical calculations and analysis; 
P.A. helped with the fabrication process; 
T.T. and K.W. provided materials; 
D.K.E. supported the experiments; 
A.D.C., M.C.W., N.W., L.I.G. and D.K.E. wrote the paper.\par}

\vspace{0.5cm}
\noindent{\textbf{Competing interests:}
The authors declare no competing interests.\par}

\vspace{0.5cm}
\noindent{\textbf{Data availability:}
The data that support the findings of this study are available from the corresponding author upon reasonable request.\par}

\end{refsection}

\clearpage

\addtocontents{toc}{\protect\ResumeTocEntries}
\pdfbookmark[0]{Supplementary Information}{SIroot}   
\ActivateSIStyle
\begin{refsection}
  
%
%

{\noindent\fontsize{16}{19}\selectfont\textbf{Supplementary Information: High-field Josephson effect enabled by a moiré Hofstadter spectrum}\par}
\vspace{0.8\baselineskip}

\vspace{-0.5\baselineskip}
{\noindent A. Díez-Carlón\textsuperscript{1,2,3*}, 
M. Cárdenes Wuttig\textsuperscript{4,5}, 
N. Wei\textsuperscript{4}, 
D. Ivanov\textsuperscript{1,2}, 
P. Altpeter\textsuperscript{1,2}, 
P. Hakonen\textsuperscript{6}, 
K. Watanabe\textsuperscript{7}, 
T. Taniguchi\textsuperscript{8}, 
L. I. Glazman\textsuperscript{4} 
and D. K. Efetov\textsuperscript {1,2*}\par}

\begin{enumerate}[label=\arabic*., align=left, nosep]
\item Fakultät für Physik, Ludwig-Maximilians-Universität, Schellingstrasse 4, 80799 München, Germany
\item Munich Center for Quantum Science and Technology (MCQST), München, Germany
\item Nanomaterials and Nanotechnology Research Center (CINN-CSIC), Universidad de Oviedo (UO), Principado de Asturias, 33940 El Entrego, Spain
\item Department of Physics and Yale Quantum Institute, Yale University, New Haven, Connecticut 06520, USA
\item Department of Applied Physics, Yale University,  New Haven, Connecticut 06520, USA
\item Department of Applied Physics, Aalto University, Espoo, 02150, Finland
\item Research Center for Functional Materials, National Institute for Materials Science, 1-1 Namiki, Tsukuba 305-0044, Japan
\item International Center for Materials Nanoarchitectonics, National Institute for Materials Science, 1-1 Namiki, Tsukuba 305-0044, Japan
\end{enumerate}

{\noindent *Correspondence to: andres.diez@cinn.es and dmitri.efetov@lmu.de\par}
\vspace{2.0\baselineskip}

\setstretch{1.1}
\tableofcontents

\pagebreak

\section{Methods}\label{methods}

\vspace{-0.5\baselineskip}
\subsection{Device fabrication}\label{a.1.--device-fabrication}

The van der Waals heterostructures are fabricated using the ``dry transfer'' technique. Graphene and hBN flakes are first exfoliated on a Si\textsuperscript{++}/SiO\textsubscript{2} (285 nm) substrate and later picked up using a polycarbonate (PC)/polydimethylsiloxane (PDMS) stamp. The PC/PDMS stamp was used to pick-up subsequently the top hBN, the monolayer graphene, and finally the bottom hBN. All layers were picked up at a temperature of \textasciitilde100 °C. The finalized stack is dropped on a Si\textsuperscript{++}/SiO\textsubscript{2} substrate by melting the PC at 180 °C.

To fabricate the Josephson junctions, we spin coat a PMMA resist at 6000 rpm and bake at 150 °C for 2 min. After the e-beam lithography patterning, the resist development is carried out in an IPA:DI water (7:3) mixture at room temperature. The stack is shaped into a rectangular mesa by reactive ion etching using a CHF3/O2 mixture (40/4 sccm). The one-dimensional superconducting contacts are made by repeating the same process as with the mesa etch, followed by dc sputtering of MoRe (100 nm) and lift-off in acetone at 80 °C. A more detailed explanation of the whole fabrication process can be found in Ref. \cite{diez_carlon_study_2025}.

\begin{figure}[H]
  \centering
  \includegraphics[width=0.8\linewidth]{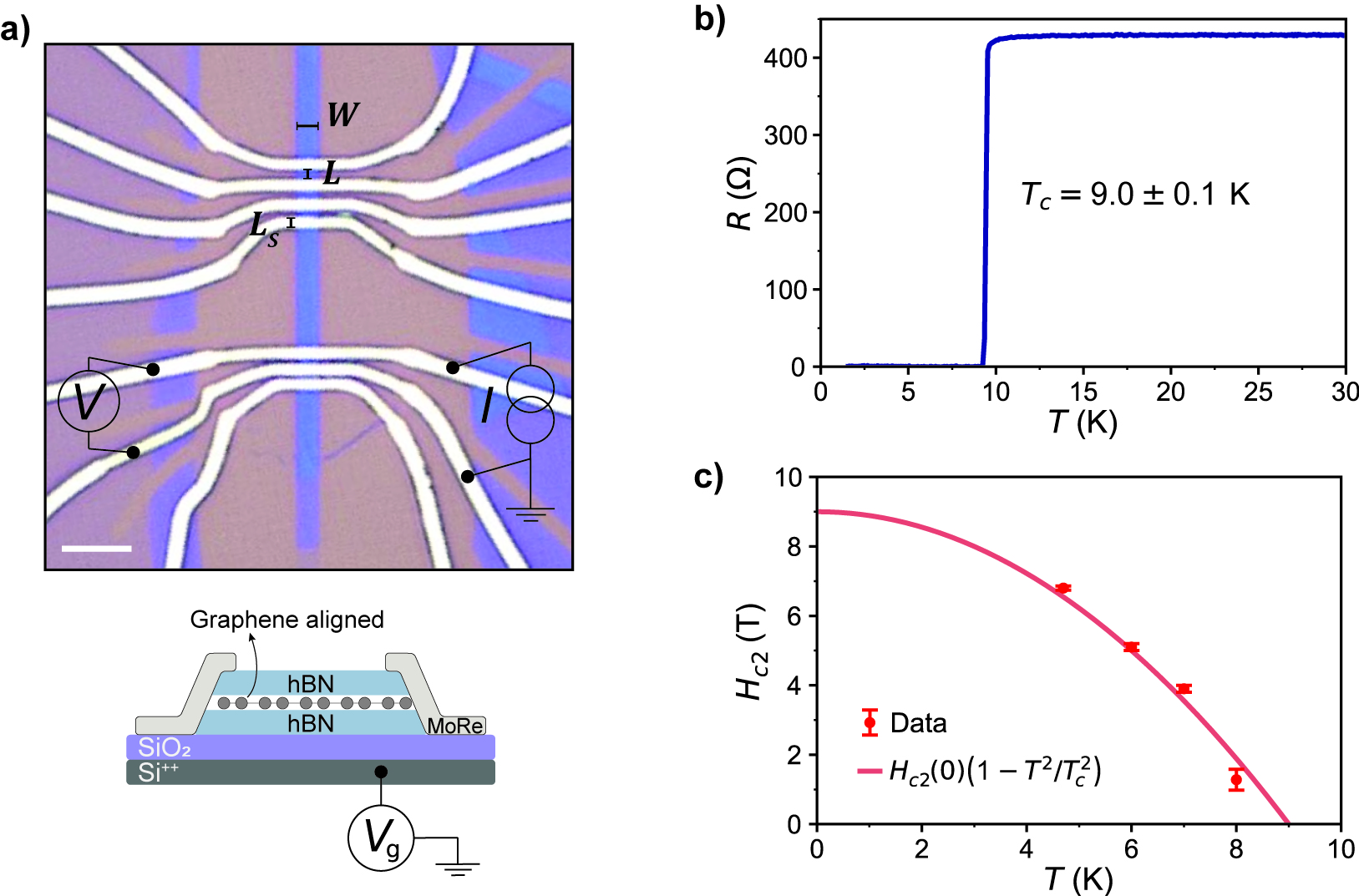}
  \caption{
  \textbf{Device example and superconducting MoRe.}
  \textbf{a)} Picture of a device, where multiple MoRe (white color leads) contacts are made to the van-der-Waals structure (blue color strips), which has already been etched in a rectangular mesa. The geometrical dimensions of a typical junction, with $L$ the length and $W$ the width are shown. The length of the superconducting electrodes across the junction is $L_s$. Scale bar is 5 µm.
  \textbf{b)} Resistance as a function of temperature of a MoRe film. The superconducting transition has a critical temperature $T_c = 9.0 \pm 0.1$ K.
  \textbf{c)} External critical magnetic field of MoRe as a function of temperature, where the fit is to a type-II superconductor \cite{tinkham_introduction_1996}. Below 4 K, the critical field exceeds the values that the superconducting magnet in our cryostat can provide ($>8$ T).}
  \label{fig:device_MoRe}
\end{figure}

An example of a device along with the experimental scheme is shown in a micrograph in \autoref{fig:device_MoRe}a, where the van-der-Waals heterostructure of hBN/graphene/hBN is contacted with multiple MoRe leads. The geometrical dimensions of a typical junction are also shown, where $L$ is the length and $W$ the width. The length of the superconducting electrodes across the junction is $L_s$. The superconducting gap of our MoRe leads $\mathrm{\Delta}\ \sim\ 1.4$ meV is deduced from its $T_c\ \sim\ 9.0\ $K by following the BCS formula $\mathrm{\Delta}\ \sim\ 1.764k_{B}T_c$ (see \autoref{fig:device_MoRe}b) \cite{tinkham_introduction_1996}. The observed superconductivity in our JJ devices at high magnetic fields can be achieved thanks to the high critical magnetic field of MoRe, which is above 8 T at zero temperature (see \autoref{fig:device_MoRe}c).

\vspace{0.2\baselineskip}
\subsection{Measurement techniques}\label{a.2.--measurement-techniques}

The measurement techniques and protocols of this work were previously described in Ref. \cite{diez-carlon_probing_2025}. The measurements were carried out in a dilution refrigerator (Bluefors SD250) with a base temperature of 100 mK. For the a.c. measurements we have used a standard low-frequency lock-in technique (Stanford Research SR860 amplifiers) with an excitation frequency $f=17.777$ Hz. The differential resistance $\mathrm{d}V/\mathrm{d}I$ measurement was performed by biasing an a.c. excitation current of 1 nA (unless stated otherwise), generated by the lock-in amplifier in combination with a 10 M$\mathrm{\Omega}$ resistor. A d.c. bias current was applied through another 10 M$\mathrm{\Omega}$ resistor before combining it with the a.c. excitation. The as-induced differential voltage and differential current were further measured at the same frequency with the standard lock-in technique. A Keithley 2400 source-meter was used to control the gate. All measured signals were filtered at the mixing chamber and still plate using commercially available low-pass RC and LC filters, respectively. After that, the signals were further filtered and amplified at room temperature by voltage-preamplifiers SR560 before entering the lock-in amplifiers or the multimeters. In measurements with a two-probe scheme, a finite known in-series resistance of 3447 $\mathrm{\Omega}$ was subtracted from the data.

The magnetic field was applied by providing current to an AMI superconducting magnet coil with the provided current source in the case of high fields ($>1$ T), and a Keithley 2400 in the case of low fields ($<1$ T). In both cases, and in order to minimize the amount of trapped flux in our superconducting magnet, we sweep the field at a rate of $\sim0.2$ mT/s. When measuring the interference patterns, first, the field is gradually ramped from zero to the desired maximum value; then the field is swept back and forth three times, at the same rate, between the maximum and minimum values at which the interference pattern will be measured. We empirically found that this procedure leads to more stable and reproducible interference patterns at very high fields.

\vspace{0.2\baselineskip}
\subsection{Twist-angle extraction}\label{a.3.--twist-angle-extraction}

The twist-angle is extracted from the high-field phase diagrams shown in \autoref{fig:Landaufans_fits}. From the slope of the Landau levels in magnetic field vs. gate-voltage $V_{g}$ and their filling factor we can extract the carrier density $n$. Then, we extract the carrier density corresponding to a fully filled superlattice unit cell $n_s$ by finding the position of the superlattice resistance peak, or by tracing the Landau levels emerging from that satellite Dirac cone back to zero field. By applying the relation $n_s = 8/\sqrt{3}\lambda_m^2$, where $\lambda_m = (1 + \delta)a/\sqrt{2(1 + \delta)\left( 1 - \cos\theta \right) + \delta^2}$ is the moiré wavelength, $a=0.246$ nm is the graphene lattice constant, and $\delta = 0.018$ is the lattice mismatch between graphene and hBN, we extract a twist-angle $\theta$ with an uncertainty of 0.01°.

\section{Transport characterization of all samples}\label{transport-characterization-of-all-samples}

Table S1 summarizes the main geometrical parameters of the four devices presented in this work, which all share the same width $W$, but have different twist angle $\theta$ and length $L$. To corroborate that the designed lengths of the junctions match with the real values, we measure the superconducting interference patterns (see \autoref{fig:FH_zeroField}) and compare their magnetic field periodicity $\mathrm{\Delta}B$ with the expected one $\mathrm{\Delta}B_{phys} = \phi_{0}/WL_{eff}$ by taking into account flux-focusing effects \cite{diez-carlon_probing_2025, amet_supercurrent_2016}. Here $\phi_{0} = h/2e$ is the superconducting flux quanta, $L_{eff} = L + L_s$, and $L_s \sim 0.45$ µm the length of the electrodes across the junction (see \autoref{fig:device_MoRe}a).

\begin{table}[H]
    \centering
    \small
    \setlength{\tabcolsep}{6pt}
    \begin{tabular}{llllll}
    \multicolumn{1}{c|}{\textbf{Device}} & \multicolumn{1}{c|}{$\boldsymbol{L}$ \textbf{(µm)}} & \multicolumn{1}{c|}{$\boldsymbol{W}$ \textbf{(µm)}} & \multicolumn{1}{c|}{$\boldsymbol{\theta}$ \textbf{(°)}} & \multicolumn{1}{c|}{$\boldsymbol{\Delta B}$ \textbf{(mT)}} & \multicolumn{1}{c|}{$\boldsymbol{\Delta B_{phys}}$ \textbf{(mT)}}    \\ \hline
    \multicolumn{1}{c|}{\textbf{GH1}}  & \multicolumn{1}{c|}{0.20}  & \multicolumn{1}{c|}{1.50}  & \multicolumn{1}{c|}{0.21} & \multicolumn{1}{c|}{$2.2\pm0.3$}  & \multicolumn{1}{c|}{$2.1\pm0.1$}   \\ \hline
    \multicolumn{1}{c|}{\cellcolor[HTML]{EFEFEF}\textbf{GH2}}  & \multicolumn{1}{c|}{\cellcolor[HTML]{EFEFEF}0.30}  & \multicolumn{1}{c|}{\cellcolor[HTML]{EFEFEF}1.50}  & \multicolumn{1}{c|}{\cellcolor[HTML]{EFEFEF}0.38}  & \multicolumn{1}{c|}{\cellcolor[HTML]{EFEFEF}$1.7\pm0.3$}  & \multicolumn{1}{c|}{\cellcolor[HTML]{EFEFEF}$1.8\pm0.1$}   \\ \hline      
    \multicolumn{1}{c|}{\textbf{GH3}}  & \multicolumn{1}{c|}{0.50}  & \multicolumn{1}{c|}{1.50}  & \multicolumn{1}{c|}{0.27} & \multicolumn{1}{c|}{--} & \multicolumn{1}{c|}{$1.45\pm0.05$}   \\ \hline
    \multicolumn{1}{c|}{\cellcolor[HTML]{EFEFEF}\textbf{GH4}}  & \multicolumn{1}{c|}{\cellcolor[HTML]{EFEFEF}0.60}  & \multicolumn{1}{c|}{\cellcolor[HTML]{EFEFEF}1.50}  & \multicolumn{1}{c|}{\cellcolor[HTML]{EFEFEF}0.22} & \multicolumn{1}{c|}{\cellcolor[HTML]{EFEFEF}--} & \multicolumn{1}{c|}{\cellcolor[HTML]{EFEFEF}$1.31\pm0.04$}   
    \end{tabular}
    \caption[Summary of all graphene/hBN moiré Josephson junction devices]
    {\textbf{Summary of all graphene/hBN moiré Josephson junction devices.} 
    Shown parameters include length $L$ ($\pm0.01$ µm), width $W$ ($\pm0.01$ µm), twist-angle $\theta$ ($\pm0.01$°), measured field periodicity in the interference patterns around zero field $\Delta B$, and their expected values from the size of the junction $\Delta B_{phys}$.}
    \label{tab:devices_GH1-4}
\end{table}

\begin{figure}[H]
  \centering
  \includegraphics[width=0.95\linewidth]{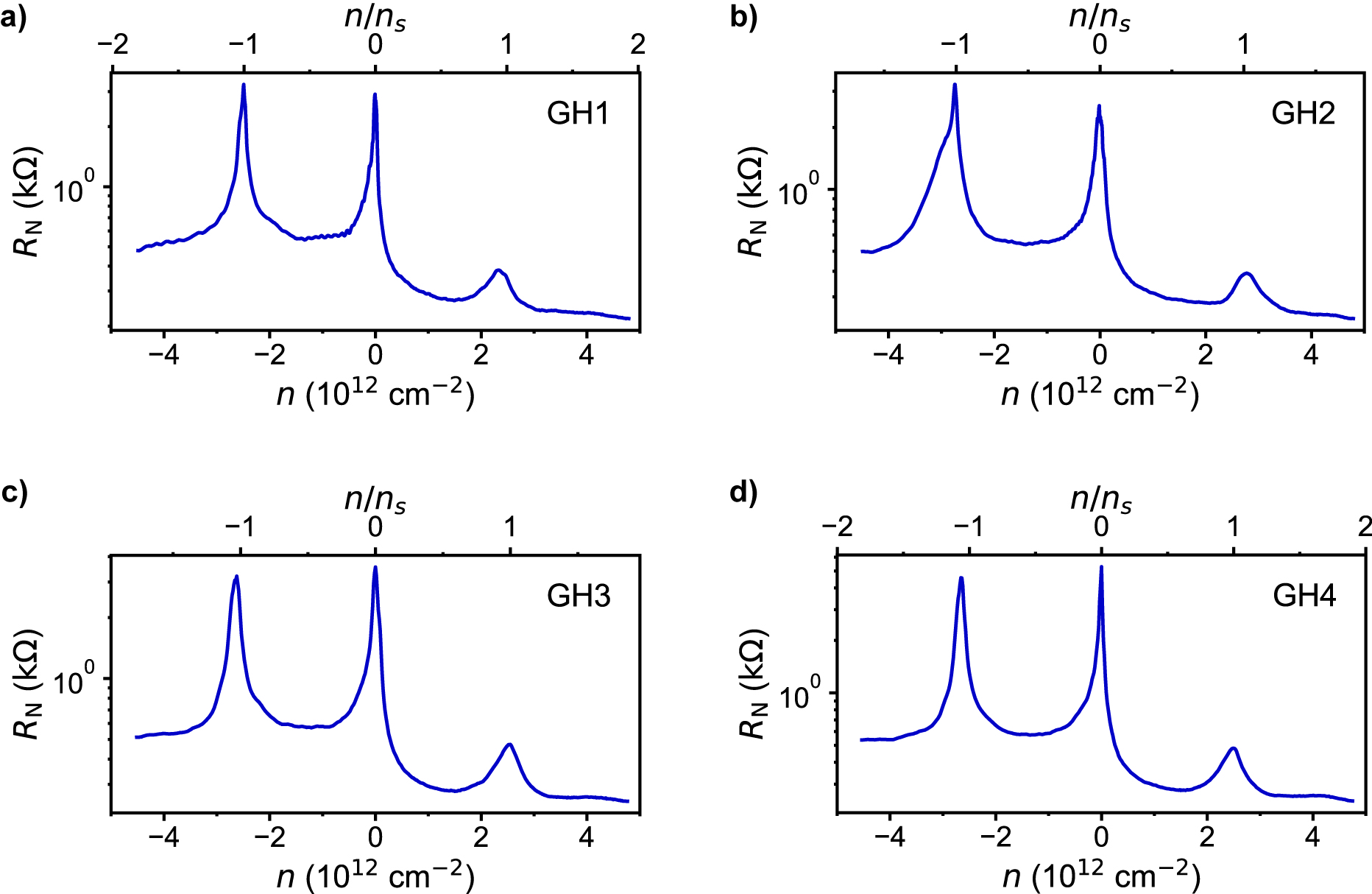}
  \caption{
  \textbf{Normal resistance of all junctions in this work.}
  \textbf{a-d)} Normal resistance $R_N$ measured at 10 K, vs. carrier density $n$ (bottom) and filling of the moiré superlattice bands $n/n_s$ (top), for devices GH1-4, respectively.}
  \label{fig:Rn}
\end{figure}

The maximum voltage we can typically apply in our devices with the Si gate is $\pm \ 80$ V, enabling us to reach carrier densities of almost $\pm \ 6.0 \times 10^{- 12}$ cm\textsuperscript{-2}, which corresponds to a filling of about $n/n_s\sim\pm2.7$ for samples with low twist angle. In \autoref{fig:Rn} we show the normal resistance taken at 10 K for all samples, where we have applied a maximum value of $\pm \ 70$ V, approximately corresponding to $\pm \ 5.0 \times 10^{- 12}$ cm\textsuperscript{-2}, and $n/n_s\sim\pm$2.0 depending on the twist angle of the sample (see Table S1).

\begin{figure}[H]
  \centering
  \includegraphics[width=\linewidth]{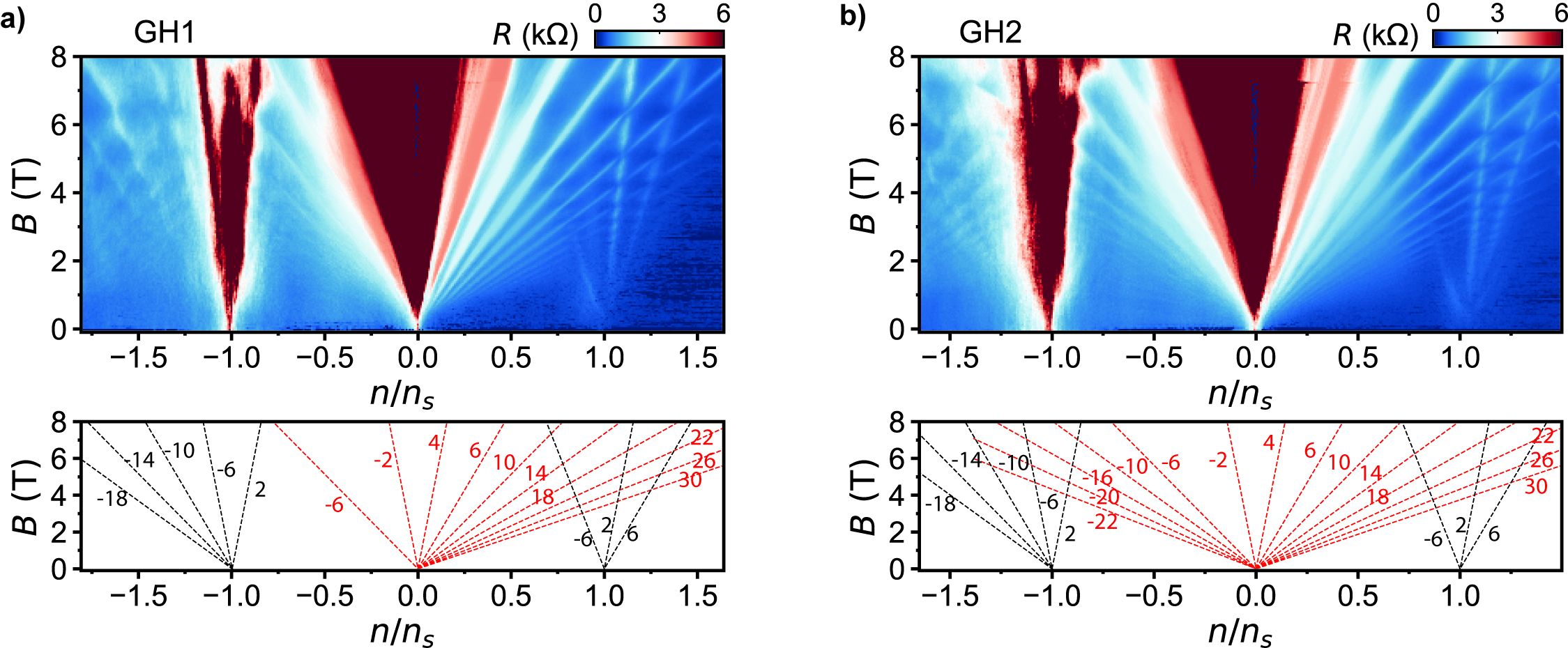}
  \caption{
  \textbf{High field phase diagram.}
  \textbf{a-b)} Resistance $R$ measured at base  100 mK, vs. magnetic field $B$ and vs. filling of the moiré bands $n/n_s$, for devices GH1-2, respectively. Bottom panels show the fitted Landau levels with their respective filling factor, which follow a 4-fold degeneracy over the typical graphene sequence $\pm2$, $\pm6$, $\pm10$, etc.}
  \label{fig:Landaufans_fits}
\end{figure}

 \autoref{fig:Landaufans_fits} shows the Landau fan measurements of the two main samples GH1 and GH2, where the fractal Hofstadter spectra of the LLs crossing, due to the moiré potential of the graphene/hBN aligned lattices, is observed. The semi 4-probe resistance that we measure, essentially a 2-terminal measurement, becomes the Hall resistance at high magnetic fields. From this measured quantized resistance of the Landau levels, apart from their slope in the $n$-$B$ diagram, we can extract their filling factor, as shown in the bottom panels of \autoref{fig:Landaufans_fits}. 

\begin{figure}[H]
  \centering
  \includegraphics[width=\linewidth]{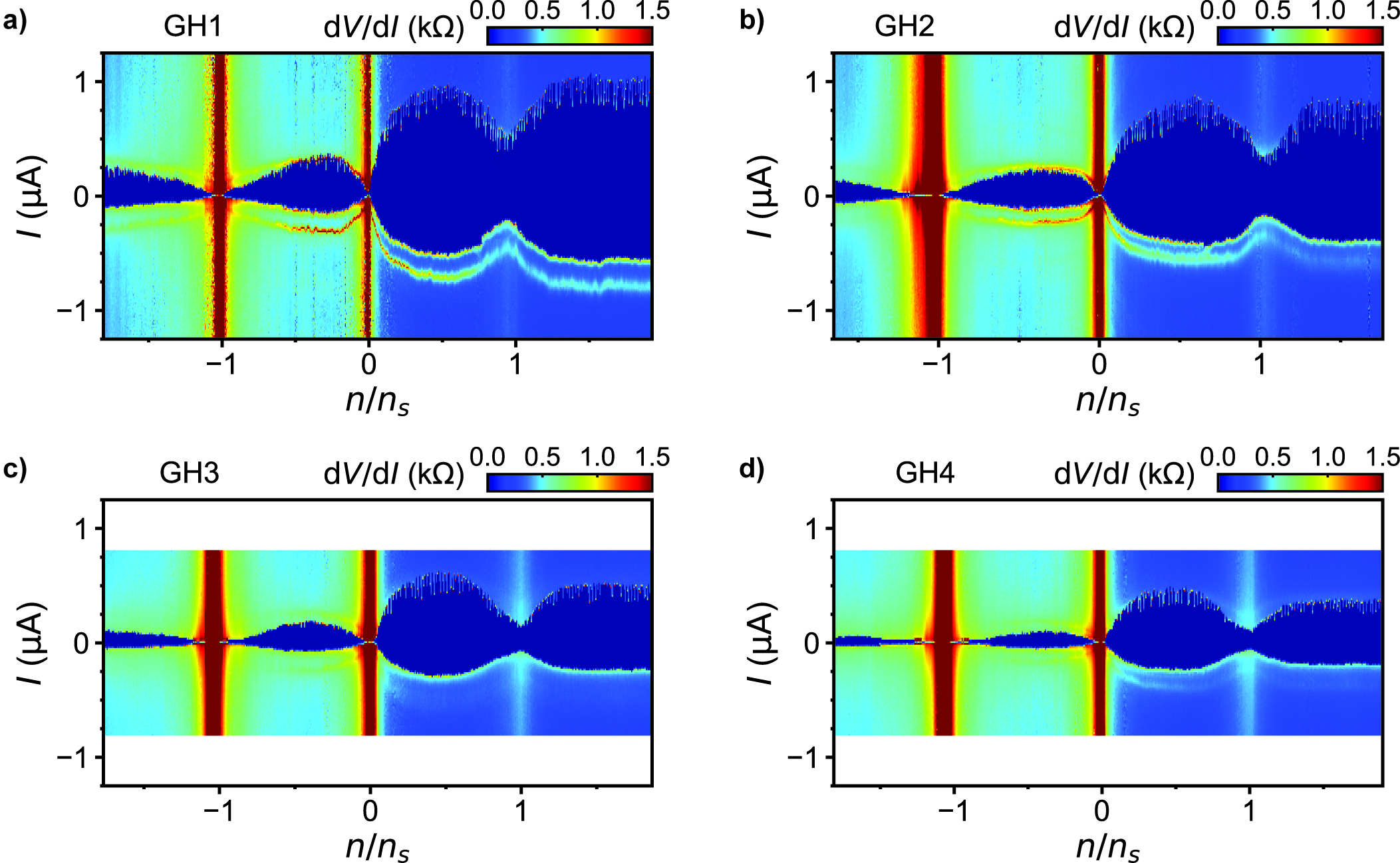}
  \caption{
  \textbf{Differential resistance maps at base temperature.}
  \textbf{a-d)} Differential resistance $\mathrm{d}V/\mathrm{d}I$ measured at base  100 mK, vs. d.c. current $I$ and vs. filling of the moiré bands $n/n_s$, for devices GH1-4, respectively.}
  \label{fig:dVdImaps}
\end{figure}

Finally, in \autoref{fig:dVdImaps} we show the characterization of the Josephson effect in all samples at base temperature 100 mK. As the length of the devices increases, the superconducting proximity effect becomes weaker, resulting in a lower critical current $I_c$. The apparent noise in these transitions with d.c. current when moving along the $n/n_s$ axis is due to the stochastic switching that is proper of underdamped JJs \cite{borzenets_ballistic_2016, tinkham_introduction_1996, lee_proximity_2018}. To study the Fabry-Pérot oscillations we do so at 2 K (see \autoref{fig:fabryperot_period} and Fig. 1c-d of the main text), so that the $I_c$ and reciprocal current $I_r$ match and the oscillations in $I_c$ are not hidden by a larger switching value.

\vspace{1.0\baselineskip}

\vspace{0.5\baselineskip}
\section{Fabry-Pérot oscillations: extraction, periodicity, and detection in the normal state}\label{fabry-puxe9rot-oscillations-in-the-normal-state}

In this section we provide additional information about the oscillations observed in the normal state resistance $R_N$ and the critical current $I_c$, which we presented in Fig. 1c-d of the main text. First we explain the extracting method we use from the raw data. We then show the analysis on the periodicity of the oscillations, and another set of measurements in the normal state above the superconducting gap of MoRe. These serve to further demonstrate that the origin of these oscillations lies in Fabry-Pérot interferometry; a key signature of electronic ballistic systems.

\begin{figure}[b!]
  \centering
  \includegraphics[width=1.0\linewidth]{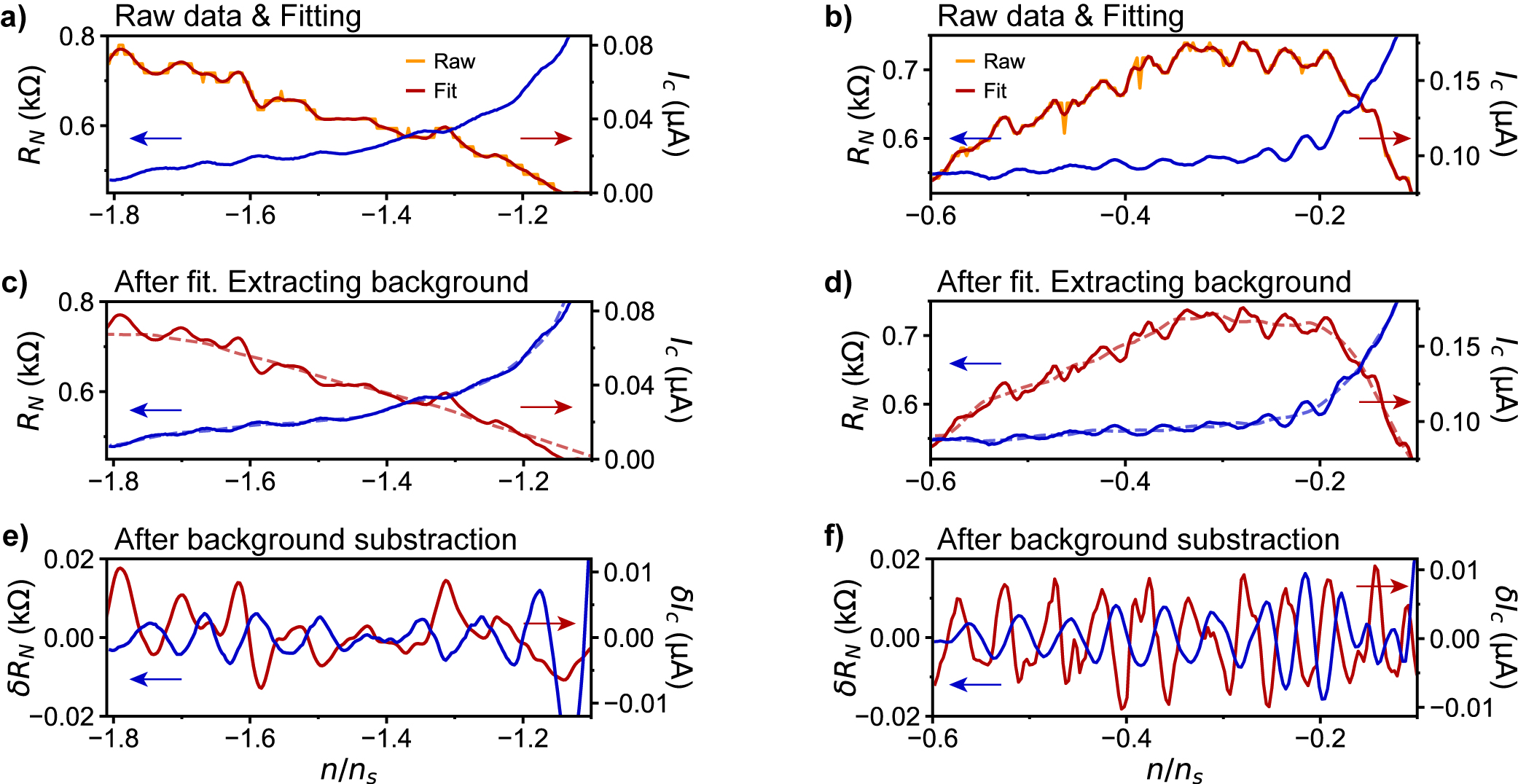}
  \caption{\textbf{Extraction of the oscillations in the critical current and normal state resistance.} 
  \textbf{a-b)} Normal-state resistance $R_N$ (blue, left axis) and critical current $I_c$ (red, right axis) both as a function of carrier density in units of the superlattice density $n/n_s$. For the $I_c$ we include two traces> the raw data in orange and a fitted curve to the raw data in red.
  \textbf{c-d)} In solid lines, $R_N$ (blue) and $I_c$ (red) from \textbf{a}-\textbf{b}. The dashed lines are the extracted underlying background curves for each. 
  \textbf{e-f)} Resulting $\delta R_N$ and $\delta I_c$ after subtracting the dashed curve from the solid curves in \textbf{c}-\textbf{d}.
  All data corresponds to device GH1. Left and right panels correspond to the oscillations in the moiré minibands and the main Dirac band, respectively.}
  \label{fig:fabryperot_extraction}
\end{figure}

\begin{figure}[t!]
  \centering
  \includegraphics[width=0.95\linewidth]{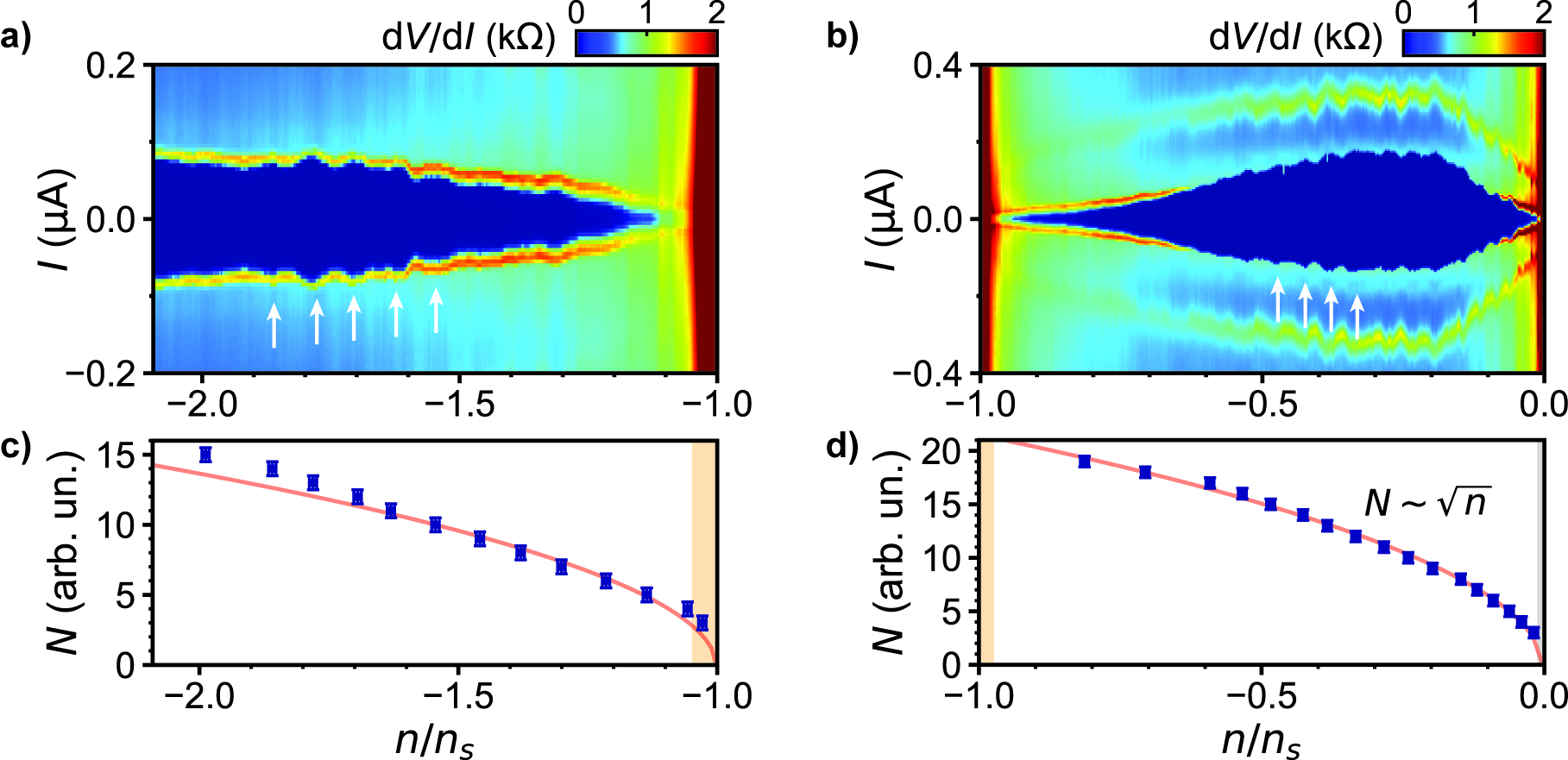}
  \caption{\textbf{Dependence of the Fabry-Pérot oscillations over the different moiré bands.} \textbf{a-b)} Map of differential resistance $\mathrm{d}V/\mathrm{d}I$ vs d.c. current bias $I$ for two different ranges of the normalized carrier density $n/n_s$, measured at 2 K. Some local maxima in the $I_c$ are indicated with white arrows. \textbf{c-d)} Extracted consecutive maxima in the $I_c$ (also minima in the $R_N$) from \textbf{a}-\textbf{b}. While for the Dirac cone band at $- 1 < n/n_s < 0$ in \textbf{d} we find $N\sim\sqrt{n}$ (red solid line), that is not the case for the bands at $n/n_s < - 1$ in \textbf{c}. The errorbars in the y-axis account for a potential missing extrema in the oscillating $I_c$ or $R_N$. All data corresponds to device GH1.}
  \label{fig:fabryperot_period}
\end{figure}

\autoref{fig:fabryperot_extraction} shows the extraction method we use to get the oscillations in $R_N$ and $I_c$ from the raw data, which are shown in Fig. 1c-d of the main text.
The $R_N$ presented in \autoref{fig:fabryperot_extraction} is the same as in \autoref{fig:Rn}a, which is measured at 10 K. The $I_c$ instead is extracted from the differential resistance maps \autoref{fig:fabryperot_period}a-b, which are measured at 2 K. These are zoom-ins of the same Fig. 1e of the main text. Our criteria is to define $I_c$ as the $I$ value at which $\mathrm{d}V/\mathrm{d}I$ reaches $50\%$ of the $R_N$. 

The left and right panels of \autoref{fig:fabryperot_extraction} correspond to the oscillations in the moiré minibands and the main Dirac band, respectively. \autoref{fig:fabryperot_extraction}a-b shows the raw data of $R_N$ and $I_c$. In the case of $I_c$, we show the raw data in orange, along with a smoother curve that is the result of fitting the data to a high order polynomials at each datapoint (Savitzky-Golay filter). From the perfect overlay of the two curves, one can see that the resulting smoother curve fit does not alter the results. Furthermore, the oscillations in the $I_c$ raw data can be very well visualized in \autoref{fig:fabryperot_period}a-b, with some of them being pointed out by white arrows.
The reason to apply such fit is that, when the $I_c$ is low, the difference between the values of two consecutive $n/n_s$ datapoints can be on the same scale as the step of 2 nA that we use when sweeping $I$. This creates an artificial step-like dependence that is detrimental to the data analysis when extracting the oscillations on top of the absolute $I_c$ values, i.e. artificial oscillations with sawtooth profiles appear. This is also done to avoid considering the sudden changes in the value of $I_c$ at certain datapoints, especially in \autoref{fig:fabryperot_extraction}b and \autoref{fig:fabryperot_period}b, which are due to a stochastic premature switching of the superconducting transition with current. This underdamped behavior of our JJs is much more present at base temperature (see \autoref{fig:dVdImaps}), which we thus try to avoid by measuring the $I_c$ at 2 K.
For $R_N$, in \autoref{fig:fabryperot_extraction}a-b we only show the raw data, as the fitted smooth curve to the former is almost identical due to the much higher precision of the lock-in technique in recording voltage (i.e. resistance) values.

\autoref{fig:fabryperot_extraction}c-d shows the resulting smooth solid curves after the fit. They also present dashed curves, which are the extracted overall tendency, or background values, of $I_c$ and $R_N$. These are extracted by fitting the data to a fifth degree polynomial. \autoref{fig:fabryperot_extraction}e-f shows the resulting $\delta R_N$ and $\delta I_c$ after subtracting the dashed curve (background) from the solid curves (data). These panels are exactly the same as the ones shown in Fig. 1c of the main text.
Here it can be very well observed that the beating of the oscillations have exactly opposite phases between $R_N$ and $I_c$, due to the ballistic regime in which our JJs lie.

We now discuss the periodicity of the oscillations shown in \autoref{fig:fabryperot_extraction} and \autoref{fig:fabryperot_period}a-b. Analogous to the interference of light in optical Fabry-Pérot interferometers, these oscillations arise when electrons reflect between two partially reflecting barriers, such as n-p–n junctions, forming an electronic cavity. Within this cavity of length $L_c$, standing wave patterns develop due to the constructive and destructive interference of all outgoing electron wavefunctions, leading to oscillations in the device's resistance as a function of carrier density or gate voltage \cite{young_quantum_2009, rickhaus_ballistic_2013}. The resonant condition for these standing waves is thus $N = 2L_c/\lambda_F$ where $N$ is the integer number of modes in the cavity and $\lambda_F=2\pi/k_F$ the Fermi wavelength. 
We are interested into evaluating the dependence of the FP oscillations with carrier density $n$, which we can tune in our experiment. Since in monolayer graphene $k_F = \sqrt{\pi n}$, for FP oscillations in this system .
\begin{equation}\label{eq:N_FP_graphene}
    N = L_c \sqrt{n/\pi}
\end{equation}
We note that in Eq.~\eqref{eq:N_FP_graphene}, the size of the Fermi velocity $v_F$ is unimportant. Thus, any linear dispersion will yield this same result.
By now extracting consecutive minima or maxima on the $I_c$ or $R_N$, $N_i$ and $N_{i+1}$, as shown in \autoref{fig:fabryperot_period}b,d, we can see that Eq.~\eqref{eq:N_FP_graphene} satisfies well the dependance of the FP oscillations in the main Dirac dcone band at $-1<n/n_s<0$, with the square root dependence fits extremely well. However, for densities $n/n_s<-1$ in \autoref{fig:fabryperot_period}c, we observe that the square root dependence of Eq.~\eqref{eq:N_FP_graphene} no longer predicts the positions of all the oscillations in $n/n_s$. The fit rather overestimates them at $n/n_s\lesssim-1.5$ and underestimates them very close to $n/n_s\sim-1$. This result indicates that the energy dispersion at these fillings of the moiré minibands cannot be fully linear, which is consistent with certain band structure models \cite{wallbank_generic_2013, moon_electronic_2014, lee_ballistic_2016}.

From the slope of the fits in \autoref{fig:fabryperot_period}c-d, one could in principle also estimate $L_c$. However, finding the exact number of modes $N$ is a difficult task, and requires to study the extra phase that these resonances acquire when applying low-magnetic fields \cite{calado_ballistic_2015}. To estimate $L_c$, we can nevertheless measure the distance between each FP oscillations, which in the case of Eq.~\eqref{eq:N_FP_graphene} gives:
\begin{equation}\label{eq:FP_lengthcavity_graphene}
    L_c = \frac{\pi}{\sqrt{n_{i+1}}-\sqrt{n_i}} .
\end{equation}
For the Dirac cone band at $-1<n/n_s<0$, in \autoref{fig:fabryperot_period}d we find values of $L_c=250\pm45\,\mathrm{nm}$, consistent with the length of our device $L\sim200\pm10\,\mathrm{nm}$. We also observe that the length of the cavity decreases down to $L_c\sim150\,\mathrm{nm}$ as the density decreases towards the CNP. This has been observed previously in these n-p-n junctions \cite{ben_shalom_quantum_2016}, and has been attributed to the reflectiveness and sharpness of the p-n interface, which decreases as the density gets closer to the n-type doping of the MoRe contacts, thereby decreasing $L_c$.
In the case of the other moiré minibands at $n/n_s<-1$ in \autoref{fig:fabryperot_period}c, the use of Eq.~\eqref{eq:N_FP_graphene} is less justified, since the band structure is not linear as we discussed before. Doing so gives $L_c=180\pm54\,\mathrm{nm}$, decreasing down to $L_c\sim100\,\mathrm{nm}$ at the hDP. When $n/n_s<-1.7$, (vHs is reached) $L_c\sim300\,\mathrm{nm}$, which indeed gives unphysical results \cite{handschin_fabry-perot_2017}.

The oscillations in \autoref{fig:fabryperot_extraction} and \autoref{fig:fabryperot_period} having their origin in a Fabry-Pérot interference can be unequivocally proven when measuring the differential resistance in the normal state as a function of a d.c. voltage bias $V_b$. 
As the FP resonances modify the transmission of carriers, the mesoscopic conductance $G_0$ is altered according to $G(E_F)=G_0 + \delta G \sin{(2\pi E/E_0)}$. This modulation of the electronic transport is a periodic function of $E_F/E_0$, where $E_0=h v_F/2L_c$ is the energy scale of the resonant standing waves, since $E_F=hv_F/\lambda_F$ for graphene \cite{young_quantum_2009, rickhaus_ballistic_2013, ben_shalom_quantum_2016}. The total current flowing through the cavity is then $I=\int_{E_F-eV_b/2}^{E_F+eV_b/2} dE \ G(E)=G_0V_b+\delta G \frac{E_0}{\pi e}\sin{(\frac{2\pi E_F}{E_0})}\sin{(\frac{\pi e V_b}{E_0})}$. From here we arrive at a general expression for the differential resistance $R=\mathrm{d}V/\mathrm{d}I$:
\begin{equation}\label{eq:FP_Vbias_general}
    R(E_F,V_b) = R_0 + \delta R \sin{\left(\frac{2\pi E_F}{E_0}\right)}\cos{\left(\frac{\pi e V_b}{E_0}\right)}.
\end{equation}

\begin{figure}[b!]
  \centering
  \includegraphics[width=0.9\linewidth]{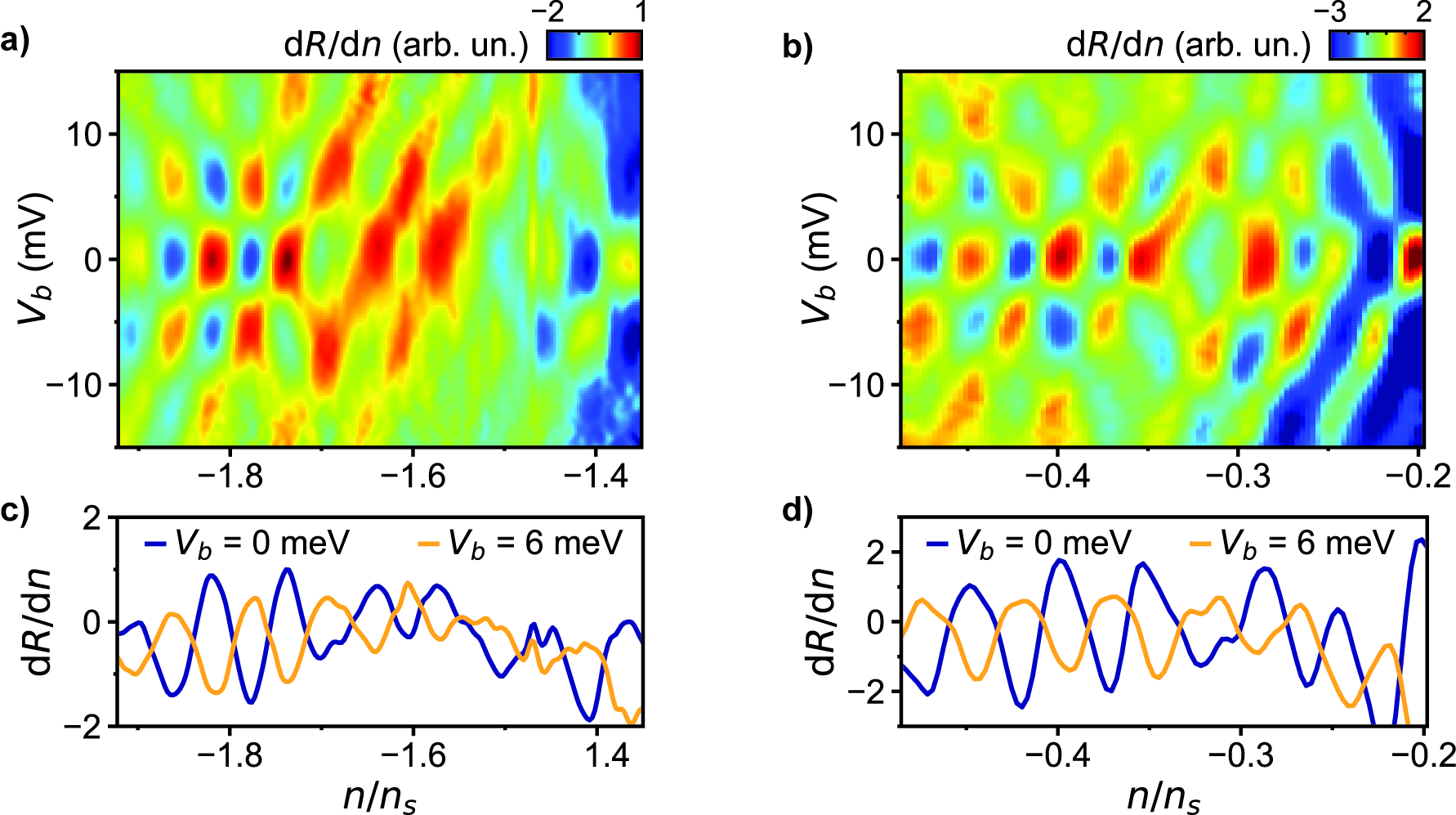}
  \caption{\textbf{Fabry-Pérot oscillations in the normal state.} \textbf{a-b)} Map of change in the differential resistance with carrier density, $\mathrm{d}R/\mathrm{d}n$, vs d.c. voltage bias $V_{b}$ and normalized carrier density $n/n_s$. Panel \textbf{a} shows the measured FP oscillations at the moiré minibands for $n/n_s < - 1$, while their behavior at the Dirac cone band around the CNP can be seen in \textbf{b}. The measurements are performed at 10 K. \textbf{c-d)} Linecuts of $\mathrm{d}R/\mathrm{d}n$ from \textbf{a} and \textbf{b} respectively, at two different values of $V_{b}$, where the oscillations show opposite phases. All data corresponds to device GH1.}
  \label{fig:FP_Vbias}
\end{figure}

\autoref{fig:FP_Vbias}a-b shows the measurements of the numerical derivative $\mathrm{d}R/\mathrm{d}n$ of the differential resistance $R$ in Eq.~\eqref{eq:FP_Vbias_general} with density $n$, and as a function of $V_b$, since $E_F$ is related to $n$. These are the same two different ranges of density indicated by the white arrows in \autoref{fig:fabryperot_period}a-b. In the colormaps, as well as in the linecuts in \autoref{fig:FP_Vbias}c-d, the periodicity both in $n$ and $V_b$ is very clear, demonstrating that the oscillations observed in \autoref{fig:fabryperot_period} are indeed FP resonances.

To analyze these results, we rewrite Eq.~\eqref{eq:FP_Vbias_general} in terms of $n$ for graphene, where $E_F=\hbar v_Fk_F=\hbar v_F\sqrt{\pi n}$, giving:
$$ R(n,V_b) = R_0 + \delta R \sin{\left(2L_c\sqrt{\pi n}\right)}\cos{\left(\frac{2\pi e L_c}{hv_F}V_b\right)}, $$
which is a function of two independent periodic $n$ and $V_b$ terms. 
The condition in the first periodic function is the same as Eq.~\eqref{eq:FP_lengthcavity_graphene}. For the second, another condition is given for the length of the cavity:
\begin{equation}\label{eq:FP_lengthcavity_Vbias_graphene}
    L_c = \frac{h v_F}{2e(V_{b_{i+1}}-V_{b_i})}
\end{equation}
This expression allows us to estimate the Fermi velocity of the bands, provided the dispersion is linear and if we know the length of the cavity, which we have estimated previously from Eq.~\eqref{eq:FP_lengthcavity_graphene}. We note that the decreasing value of $L_c$ as we approach the CNP and hDP can be visualized by the elongtaion of the checkerboard pattern in \autoref{fig:FP_Vbias}c-d along the $V_b$ axis. 
For the Dirac cone band at $-1<n/n_s$, the corresponding Fermi velocity gives $v_F=(0.7\pm0.1)\times 10^6 \,\mathrm{m/s}$, whereas for the moiré bands at $n/n_s<-1$, $v_F=(0.5\pm0.1)\times 10^6 \,\mathrm{m/s}$. These values are consistent with the renormalized Fermi velocity of the satellite Dirac points along one direction of the mBZ \cite{lee_ballistic_2016, handschin_fabry-perot_2017, kraft_anomalous_2020, mrenca-kolasinska_probing_2023, moon_electronic_2014}, although we note that the nonlinearities in the moiré bands can become important, as we have shown in \autoref{fig:FP_Vbias}e-f, so that the use of Eq.~\eqref{eq:FP_lengthcavity_Vbias_graphene} is less justified in this analysis. Nevertheless, by measuring the dependence of the critical current with temperature, we show an alternative method to estimate the $v_F$ of the bands, as we explain in the next section.

\section{Length dependence of the long-ballistic regime across four samples}\label{length-dependance-of-the-long-ballistic-regime-across-four-samples}

In this section we show that the temperature dependence of $I_c$ is also consistent with a ballistic, long regime in our JJs. For long Josephson junctions, $I_c$ is governed by an exponential scaling $I_c \propto e^{- k_{B}T/\delta E}$, where $\delta E \approx \hbar v_{F}/2\pi L$. This decaying rate $\delta E$ is related to the Thouless energy $E_{Th} = \hbar v_{F}/L$, which is the characteristic energy of a JJ in the long-ballistic regime \cite{borzenets_ballistic_2016, ben_shalom_quantum_2016, lee_proximity_2018}.

\begin{figure}[b!]
  \centering
  \includegraphics[width=0.95\linewidth]{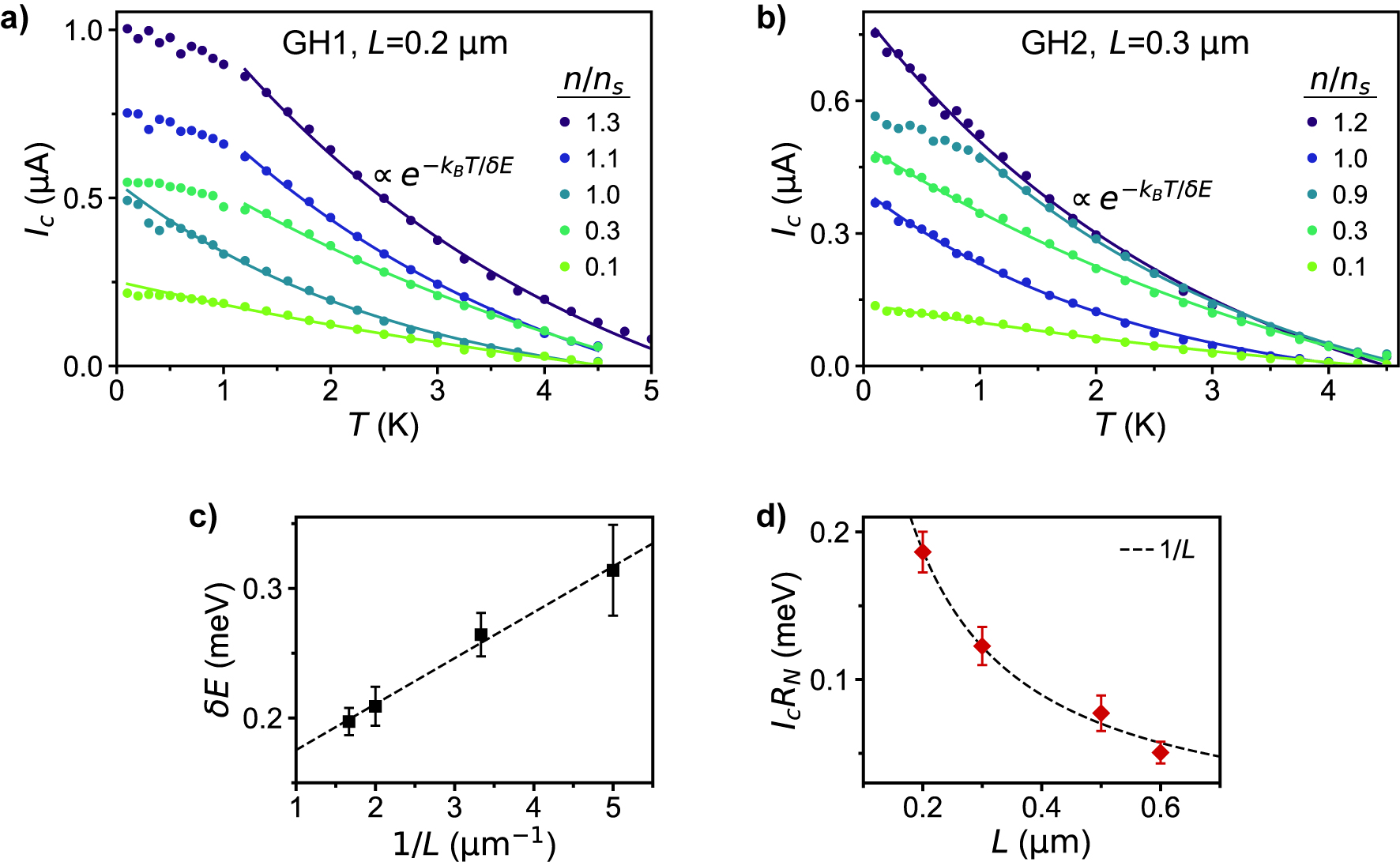}
  \caption{\textbf{Long-ballistic regime evidence in the length dependence of} $\mathbf{I}_{\mathbf{c}}\mathbf{(T)}$ \textbf{and} $\mathbf{I}_{\mathbf{c}}\mathbf{R}_{\mathbf{N}}$\textbf{.} \textbf{a-b)} Critical current $I_c$ vs temperature $T$ for different points in carrier density, and for devices GH1 and GH2, respectively. At high temperatures, $I_c$ follows an exponential trend. The solid lines represent a fit to the function $\propto e^{- k_{B}T/\delta E}$. \textbf{c)} Extracted $\delta E$ from the exponential fits in \textbf{a}-\textbf{b} at a fixed $n/n_s\sim1.3$ and for several devices with varying length $L$ (see devices GH1-4 in \autoref{tab:devices_GH1-4}). The dashed line represents a linear $1/L$ fit to the data. \textbf{d)} $I_cR_N$ product at a fixed $n/n_s\sim1.2$, for devices GH1-4 with varying $L$. The errorbars come from the average value of $I_c$, which is measured at base temperature over several measurements due to premature switching (see \autoref{fig:dVdImaps}).}
  \label{fig:Ic_vs_T}
\end{figure}

We thus measure the $I_c(T)$ dependence across the four different graphene/hBN aligned JJs with increasing length $L$ (see GH1-4 in Table S1). \autoref{fig:Ic_vs_T}a-b shows these measurements for the two shortest devices GH1-2, where we can see that the $I_c$ follows an exponential decay with temperature $T$. Most importantly, it does so for multiples values of density $n/n_s$, spanning over different bands; from the Dirac cone at the CNP to the moiré minibands.

By extracting the rate of the exponential decay $\delta E$ in all four GH1-4 devices, we find that it scales linearly with $1/L$, as shown in \autoref{fig:Ic_vs_T}c. Along with the observation of FP oscillations, this result provides another evidence that our graphene/hBN JJs are ballistic in their entire accessible band structure, and moreover in the long regime. From this fit we can also extract the Fermi velocity at the density $n/n_s\sim 1.3$ where $\delta E$ was extracted, giving $v_{F} \approx (0.34 \pm 0.05) \times 10^{6}$ m/s. This value is in good agreement with $v_{F} \approx (0.5 \pm 0.1) \times 10^{6}$ m/s reported in previous experiments \cite{yankowitz_emergence_2012, yu_hierarchy_2014, guarochico-moreira_thermopower_2023}.

Another evidence of our devices belonging to the long-ballistic regime follows from the $1/L$ dependence of the $I_cR_N$ product, as shown in \autoref{fig:Ic_vs_T}d. At zero temperature, $I_c \approx E_{Th}/R_N$ in the long junction regime \cite{ben_shalom_quantum_2016, lee_proximity_2018, dubos_josephson_2001}, and for a ballistic sample the $R_N$ does not depend on $L$ (as it is determined only by its cross-sectional width and the Fermi wavelength). Therefore, it is expected that $I_cR_N \approx E_{Th} \propto 1/L$ for long-ballistic junctions, in agreement with our measurements in \autoref{fig:Ic_vs_T}d.

\vspace{0.5\baselineskip}
\section{Interference patterns at low and high fields. Effects from flux creep} \label{additonal_interferene_patterns}

\begin{figure}[b!]
  \centering
  \includegraphics[width=0.85\linewidth]{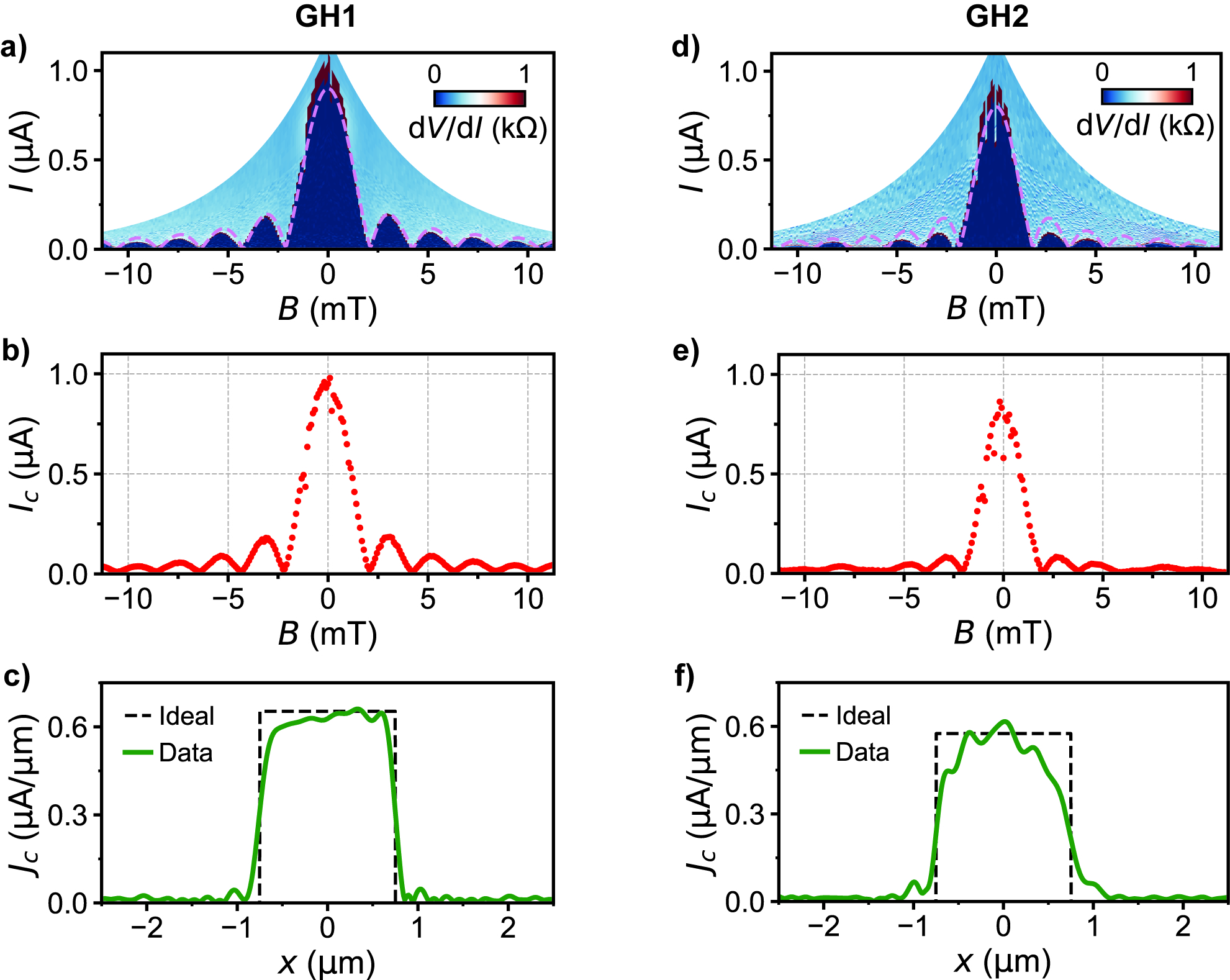}
  \caption{\textbf{Current homogeneity and interference patterns around zero field.} 
  \textbf{a,d)} Differential resistance vs d.c. current $I$ and field $B$. The violet dashed line is the expected dependence $I_c(B)$ for a homogeneous single slit with the same dimensions of our JJ.
  \textbf{b,e)} Extracted $I_c$ from \textbf{a}.
  \textbf{c,f)} Critical current density profile $J_c$ vs the lateral dimension of the junction $x$. The extracted values (green) from the data in \textbf{b} compare well with the expected dependence (dashed black) from the violet dashed line in \textbf{a}.
  Panels \textbf{a}-\textbf{c} were measured at $n/n_s\sim1.6$ in GH1, and \textbf{d-f)} in GH2 at $n/n_s\sim1.5$.
  }
  \label{fig:FH_zeroField}
\end{figure}

In this section we discuss the different regimes of the superconducting proximity effect with increasing magnetic field. While the JJs follow the typical Fraunhofer relation around zero field, in the semiclassical regime at intermediate and high fields, the $I_c$ oscillations become irregular in amplitude and period.

At zero field, the ABS follow straight trajectories. When a small perpendicular magnetic field $B$ is applied, in the order of mT, it imposes a phase gradient to these quasiparticles along the cross section of the junction, leading to an oscillating critical current $I_c(B)=I_c(0)\sin{(\pi B A/\phi_0)}/(\pi B A/\phi_0)$, which is a typical Fraunhofer interference diffraction pattern. This holds for our JJs as shown in \autoref{fig:FH_zeroField}a,d; where $A=W L_{eff}$ is the effective area of the junctions considering flux focusing effects, and $\phi_0=h/2e$. Such Fraunhofer pattern is proper of a homogeneous critical current profile $J_c$ along the lateral width of the junction $x$. Indeed, by extracting the experimental $I_c(B)$ (see \autoref{fig:FH_zeroField}b,e), and following the method from Dynes and Fulton \cite{dynes_supercurrent_1971} with the ansatz from \cite{hart_induced_2014}, we can compute $J_c(x)$ in our devices. \autoref{fig:FH_zeroField}c,f shows that it approaches very well the ideal limit of a perfectly uniform single slit junction with the same width $W\sim1.5$~µm as our devices (see \autoref{tab:devices_GH1-4}).

\begin{figure}[b!]
  \centering
  \includegraphics[width=0.8\linewidth]{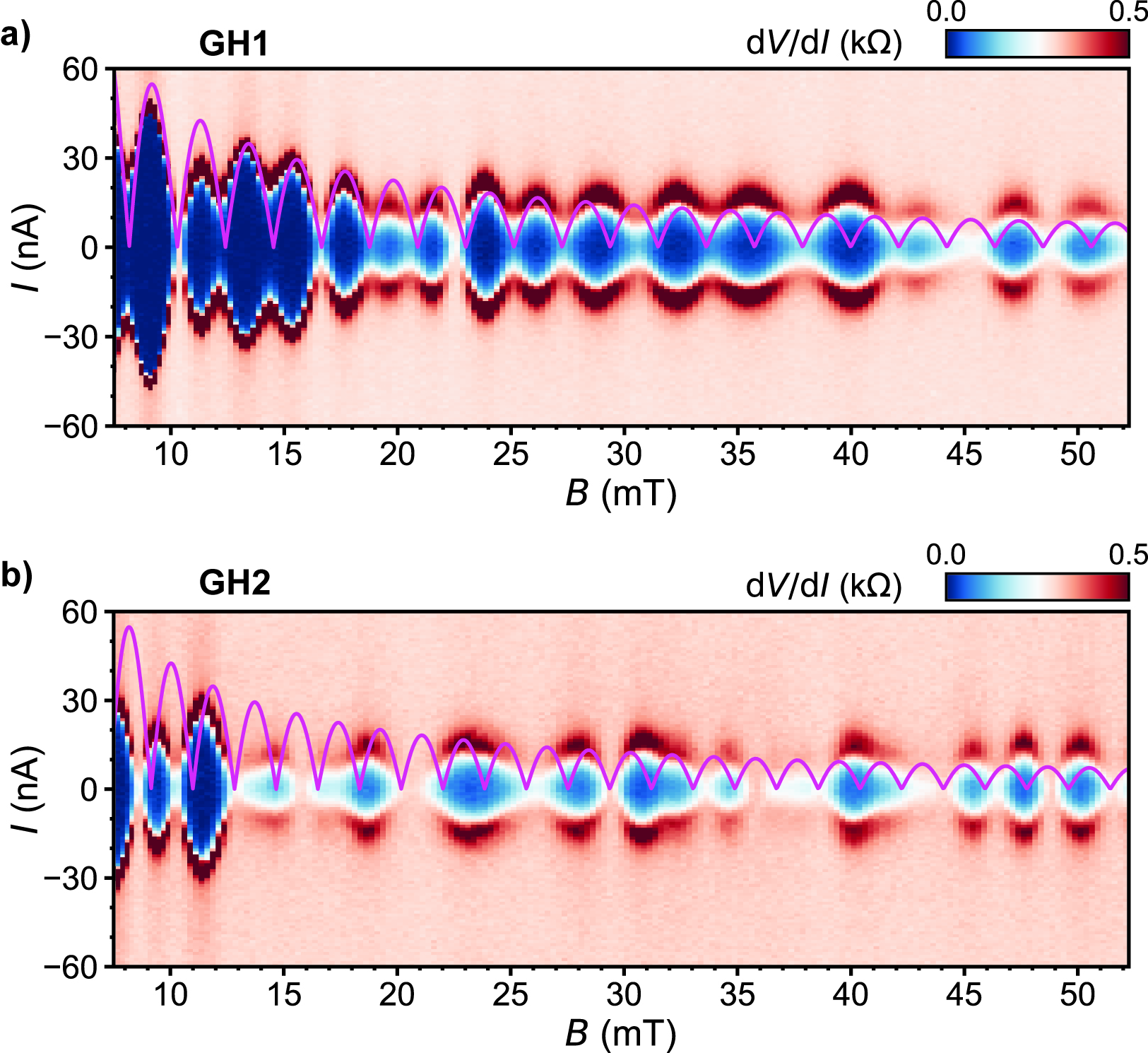}
  \caption{\textbf{Irregular oscillations at intermediate fields.} 
  \textbf{a)} Interference pattern measured at $n/n_s\sim1.8$ in sample GH1 and at base temperature. The violet solid line is the expected dependence $I_c(B)$ for a homogeneous single slit with the same dimensions of the JJ. From the misfit of the data to the solid line, the irregular amplitude and periodicity of the $I_c$ oscillations becomes evident. \textbf{b)} Equivalent for sample GH2, measured at $n/n_s\sim1.6$ and at base temperature. 
  }
  \label{fig:FH_lowFields}
\end{figure}

In a semiclassical picture, as the magnetic field $B$ is increased to moderate values of a few to tens of mT, the electron and holes forming the ABS acquire a finite cyclotron motion with opposite directions, which bends their trajectories. As long as $2r_c>L$, and the quasiparticles can reach the other end of the junction via Andreev reflections, they can retrace each other, such that ABS can still be formed. However, these mesoscopic trajectories are irregular, since also normal reflections instead of Andreev reflections will occur at the lateral edges of the junction. Thus, in this semiclassical regime, a chaotic billboard of mesoscopic trajectories dominates the ABS transport, which experimentally results in irregular oscillations of the $I_c(B)$ with amplitude and periodicity \cite{ben_shalom_quantum_2016, amet_supercurrent_2016}.
Also in Ref.~\cite{ben_shalom_quantum_2016}, the value of field where the ABS cease to conduct through the bulk and this semiclassical regime of a chaotic billboard of mesoscopic trajectories dominates, was estimated (and corroborated experimentally in graphene JJs) to be $B^*\sim\Delta/eLv_F$.

\begin{figure}[b!]
  \centering
  \includegraphics[width=1.0\linewidth]{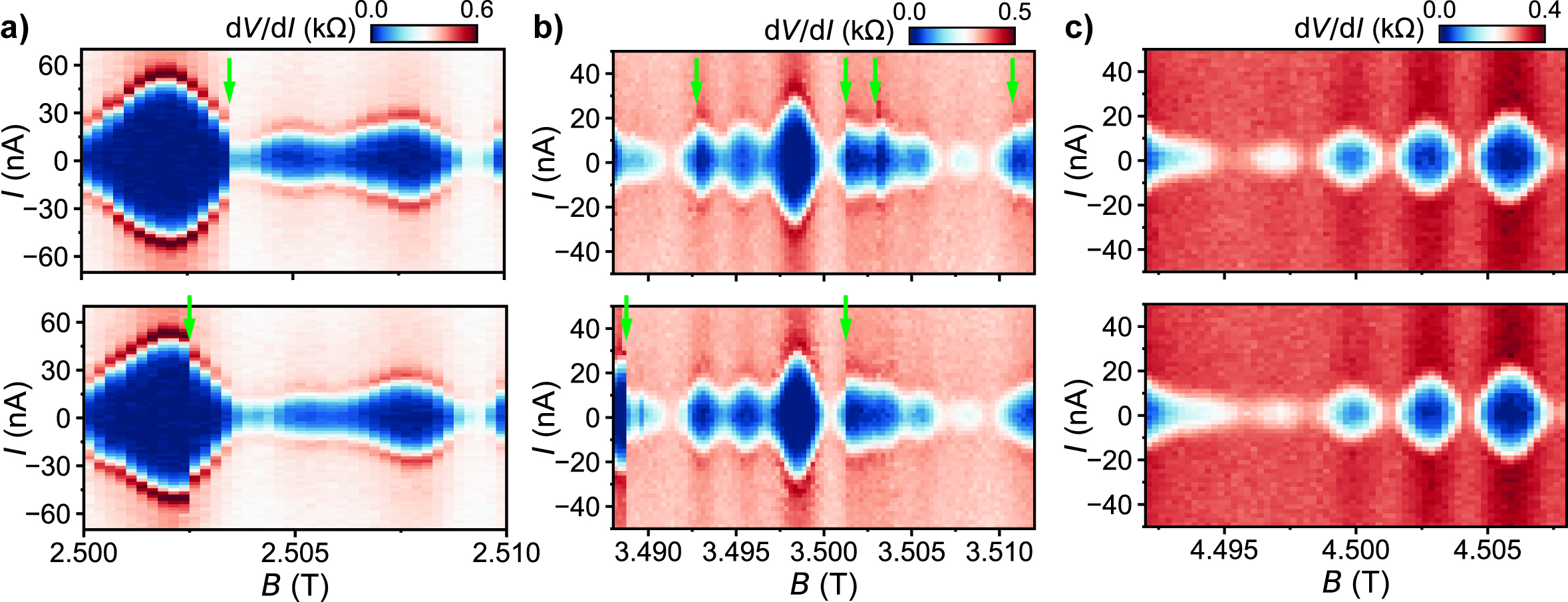}
  \caption{\textbf{Additional interference patterns at high field.} \textbf{a-c)} Interference patterns measured at around $2.5$~T, $3.5$~T and $4.5$~T, respectively; where resistance $R$ is taken vs perpendicular magnetic field $B$ and vs d.c. current bias $I$. The top and bottom panels show two consecutive measurements at the same fields. The green arrows mark the position in $B$ of a sudden discontinuity in the interference pattern, arising from flux creep into the junction. All data is measured at $n/n_s\sim1.8$ in sample GH1 and at base temperature. 
  }
  \label{fig:FH_HighField_Additional}
\end{figure}

In the case of samples GH1-2, setting $v_F\sim0.5\times10^6$~m/s, we find respectively $B^*\sim13$~mT and $B^*\sim8$~mT. This coincides with the values at which our recorded oscillations in \autoref{fig:FH_lowFields}a-b start to deviate from the Fraunhofer interference. Indeed, from those values onwards, the $I_c$ no longer homogeneously decays in amplitude. It rather increases, stays constant, or decreases randomly after each full oscillation. The periodicity of $I_c$ also becomes irregular, having an average value of $\Delta B=3.1\pm0.6$~mT in \autoref{fig:FH_lowFields}a. These periodicity values and their standard deviation have been extracted from the interference patterns by applying a FFT on the extracted $I_c$, which is defined as the $I$ value at which $\mathrm{d}V/\mathrm{d}I$ reaches $50\%$ of the $R_N$.
Upon further inspection, the minimum period found for these oscillations is the same as in \autoref{fig:FH_zeroField}a, $\Delta B=2.2\pm0.4$~mT, corresponds to one and two units of superconducting flux quanta threading the junction $\phi_0$. The maximum period recorded goes as high as $\Delta B=4.4\pm0.5$~m, which corresponds to $\sim2\phi_0$.

The sudden increase in these periods corresponds to these trajectories enclosing a smaller or bigger flux area threading the junction between consecutive oscillations, given their chaotic trajectories \cite{ben_shalom_quantum_2016, amet_supercurrent_2016}.

As discussed in the main text, these irregular mesoscopic orbits keep existing up to $\sim6$~T in the moiré minibands. \autoref{fig:FH_HighField_Additional} shows interference patterns at high fields, additionally to the one shown in Fig. 2c of the main text. In these, it can be very well visualized that the $I_c$ amplitude does go up and down randomly. We also note that the maximum $I_c$ values found are on the order of the quantum ballistic limit of a single ABS mode $I_Q=e\Delta/h\sim53\pm1$~nA for our MoRe JJs. As for the periodicities, the observed lobes also go from narrower to broader beatings. For example, in \autoref{fig:FH_HighField_Additional}c at 4.5~T, the oscillations take period values at around $\Delta B \sim 2.9\pm0.4$~mT, whereas in \autoref{fig:FH_HighField_Additional}b at 3.5~T we observe $\Delta B \sim 3.2\pm0.4$~mT.

In this high field regime, abrupt changes in the interference patterns can appear due to flux creeping from the adjacent MoRe contacts into the weak link, thus producing a magnetic flux instability. Nevertheless, we found that if $B$ is slowly swept at a rate of $\sim0.2$~mT/s, this flux creep is minimized and the measured interference patterns are rather stable over time, giving reproducible $I_c$ oscillations as shown in \autoref{fig:FH_HighField_Additional}.

\vspace{1.0\baselineskip}
\section{High-field superconductivity in a second device GH2}\label{high-field-superconductivity-in-a-second-device-gh2}

This section shows data at high magnetic fields from a second device GH2, with a different twist angle and length compared to those of the device GH1 which is presented in the main text.

\begin{figure}[t!]
  \centering
  \includegraphics[width=0.9\linewidth]{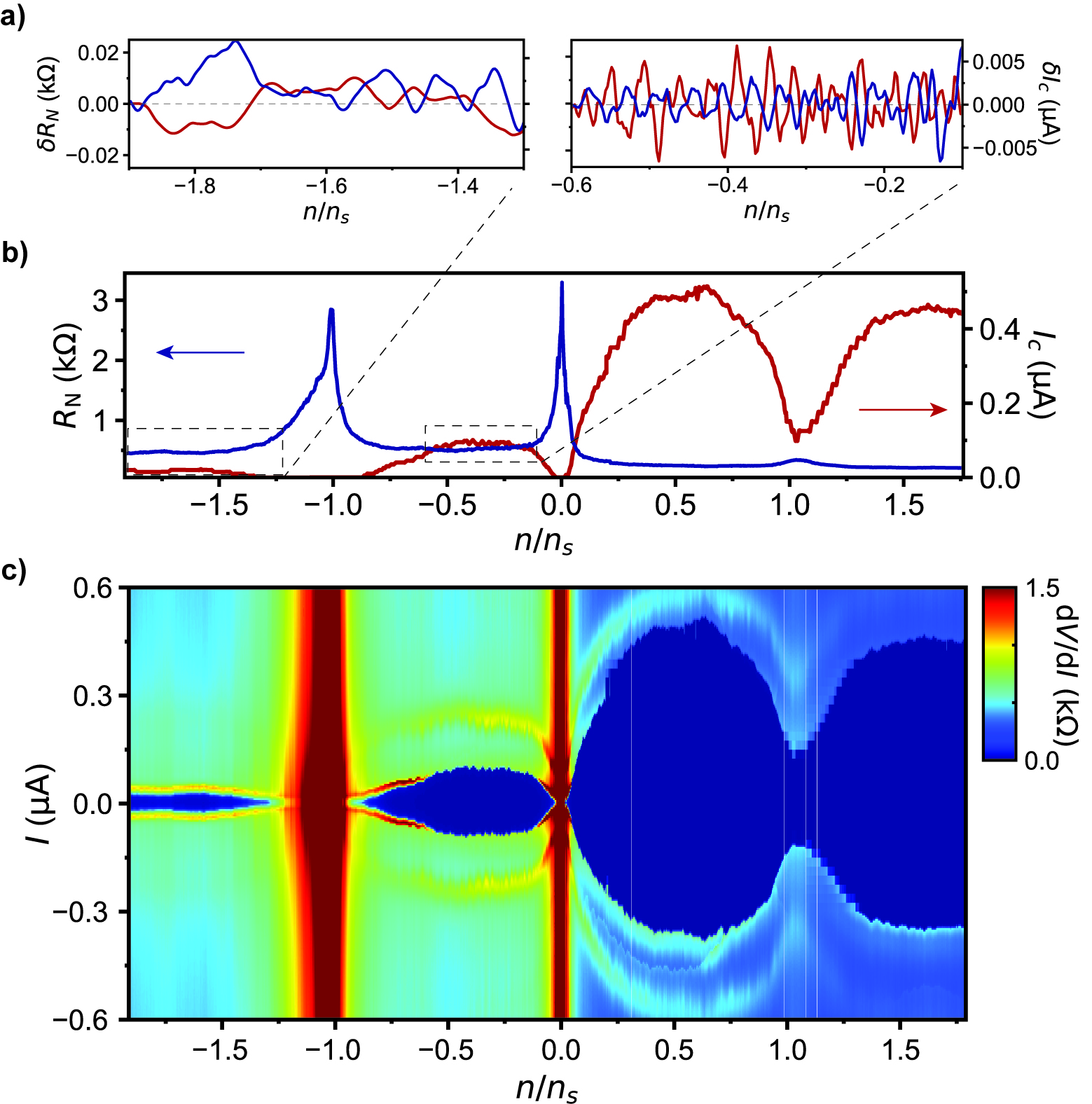}
  \caption{\textbf{Fabry-Pérots at the moiré minibands in a second device GH2.} 
    \textbf{a)} Fabry-Pérot oscillations in the moiré minibands (left) and the main Dirac cone (right) shown as a function of carrier density in units of the superlattice density $n/n_s$. These are obtained from \textbf{b} by subtracting a smooth fit of the normal-state resistance $R_N$ (blue) and critical current $I_c$ (red).
  \textbf{b)} $R_N$ at 10 K in blue (left axis), and $I_c$ at 2 K in red (right axis), both as a function of $n/n_s$.
  \textbf{c)} Map of differential resistance $\mathrm{d}V/\mathrm{d}I$ vs d.c. current bias $I$ and carrier density normalized to the superlattice density $n/n_s$. Dark blue regions are superconducting, and their contour along positive d.c. current bias $I$ yields the $I_c$ in \textbf{b}.
  All data corresponds to device GH2.}
  \label{fig:GH2_fabryperot}
\end{figure}

The appearance of Fabry-Pérot oscillations at the main Dirac band and at the moiré minibands is seen in \autoref{fig:GH2_fabryperot}. Their extraction follows the same method as the one shown in \autoref{fig:fabryperot_extraction}. Compared to GH1, the oscillations are less visible in the moiré minibands (left panel of \autoref{fig:GH2_fabryperot}a), given that the $I_c$ is much smaller. Nevertheless, after the background subtraction, it can be well seen that the $\delta I_c$ and $\delta R_N$ oscillations beat with opposite phases.

\begin{figure}[H]
  \centering
  \includegraphics[width=1.0\linewidth]{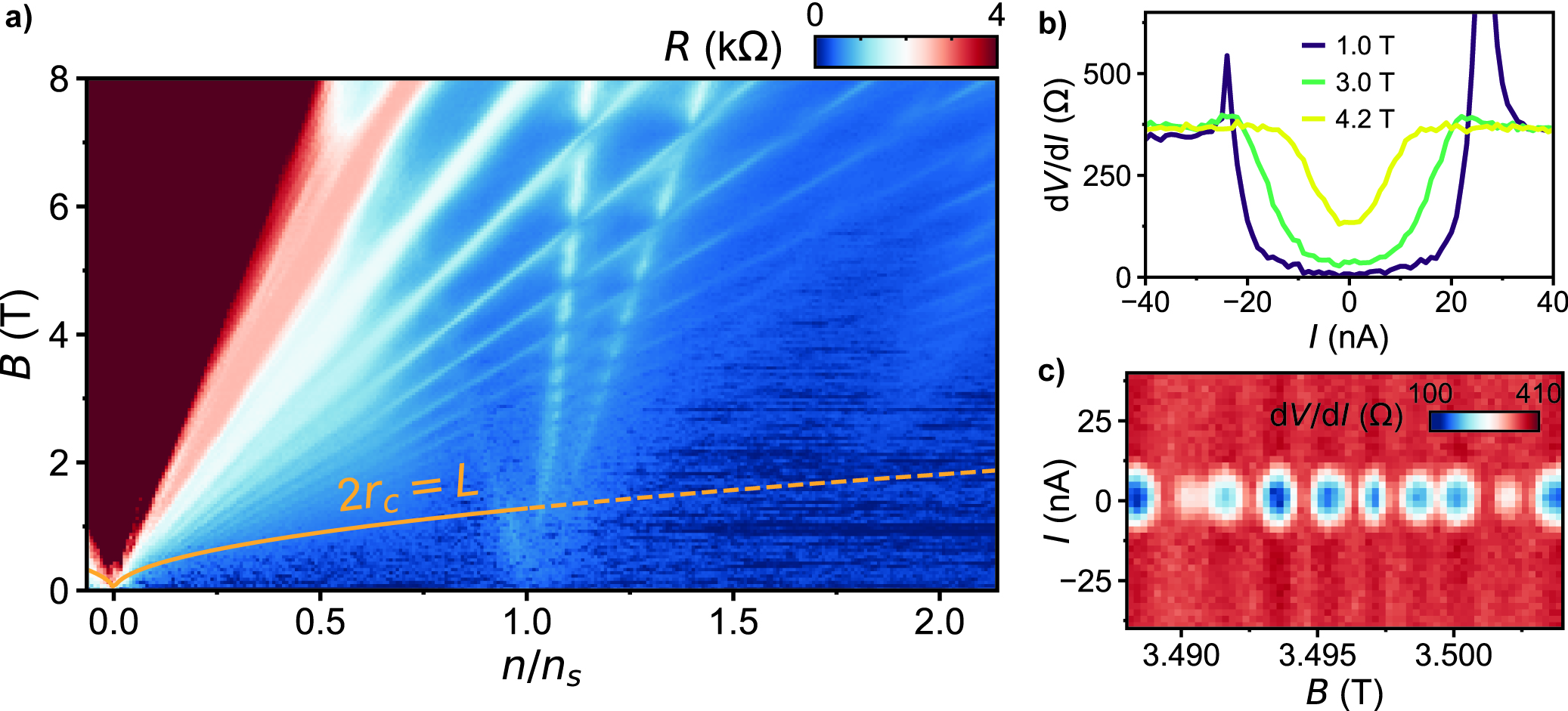}
  \caption{\textbf{High-field superconductivity at the moiré minibands in a second device GH2.} \textbf{a)} Resistance $R$ as a function of $B$ and $n/n_s$. The superconducting pocket regions are well delimited by the orange dashed line at the Dirac-cone band for $0 < n/n_s < 1$, while at the moiré bands for $n/n_s > 1$, superconductivity goes beyond this $2r_c=L$ limit, up to 5 T in the case of GH2.
  \textbf{b)} $\mathrm{d}V/\mathrm{d}I$ vs d.c. current $I$ for different values of magnetic field at a fixed $n/n_s = 1.6$.
  \textbf{c)} Interference pattern measured at $n/n_s = 1.6$ and around $B = 3.5$ T.
  All data corresponds to device GH2.
  }
  \label{fig:GH2_HighField}
\end{figure}

This device GH2 has a length $L\sim300$ nm, greater than the one from GH1 (200 nm, see \autoref{tab:devices_GH1-4}). This results in a $2r_c=L$ curve in \autoref{fig:GH2_HighField}a with lower values than that of GH1 (Fig. 2a of the main text). Nevertheless, such curve also limits well the boundary of the induced superconducting phase at the main Dirac cone band in the field--density phase diagram.
In the case of the moiré minibands, this second device GH2 shows equivalent trends as in Fig. 2a of the main text even though its twist angle is slightly different: superconductivity goes beyond this upper limit $2r_c=L$ (dashed orange line), reaching up to 5 T, gets cut around $n/n_s\sim 1.8$ and re-emerges again to high fields at $n/n_s\sim 2.0$. 
Non-linear $\mathrm{d}V/\mathrm{d}I$ characteristics and oscillations in the $I_c$ are further proven in \autoref{fig:GH2_HighField}b-c respectively.

The decrease of the maximum magnetic field at which we observe superconducting transport in GH2 with respect to GH1, as well as its $I_c$ values, can be explained by the greater length of GH2. Given that the normal state resistance $R_N$ increases with increasing length $L$, and the $I_cR_N$ product decreases with $1/L$ (see \autoref{fig:Ic_vs_T}d), the $I_c$ of GH2 is expected to be much smaller at zero field (\autoref{fig:Ic_vs_T}a-b), as well as in high magnetic fields. The increase in the area of the device GH2 is also noted in the much faster period of the $I_c$ oscillations compared to GH1, here ranging from $\Delta B\sim1.7\pm0.3$~mT to $\sim2.0\pm0.4$~mT. The fastest period also coincides with that of around zero field (see \autoref{fig:FH_zeroField}d-f and \autoref{fig:FH_lowFields}b).
We note that we cannot reach higher values of $n/n_s$ because in the case of GH2 the twist angle is greater than GH1, and in our devices we can only reach about $V_g\sim80$ V with our Si gate before it leaks.

No high field superconductivity nor clear Fabry-Pérot oscillations were observed for the other two devices GH3 and GH4. This can be expected, as their much longer lengths lead to very low critical currents (see \autoref{tab:devices_GH1-4}, \autoref{fig:dVdImaps}c-d and \autoref{fig:Ic_vs_T}d).

\vspace{1.0\baselineskip}
\section{Absence of Landau levels at the moiré bands in the normal state of the JJs}\label{absence-of-landau-levels-at-the-moiruxe9-bands-in-the-normal-state-of-the-jj}

\begin{figure}[H]
  \centering
  \includegraphics[width=1.0\linewidth]{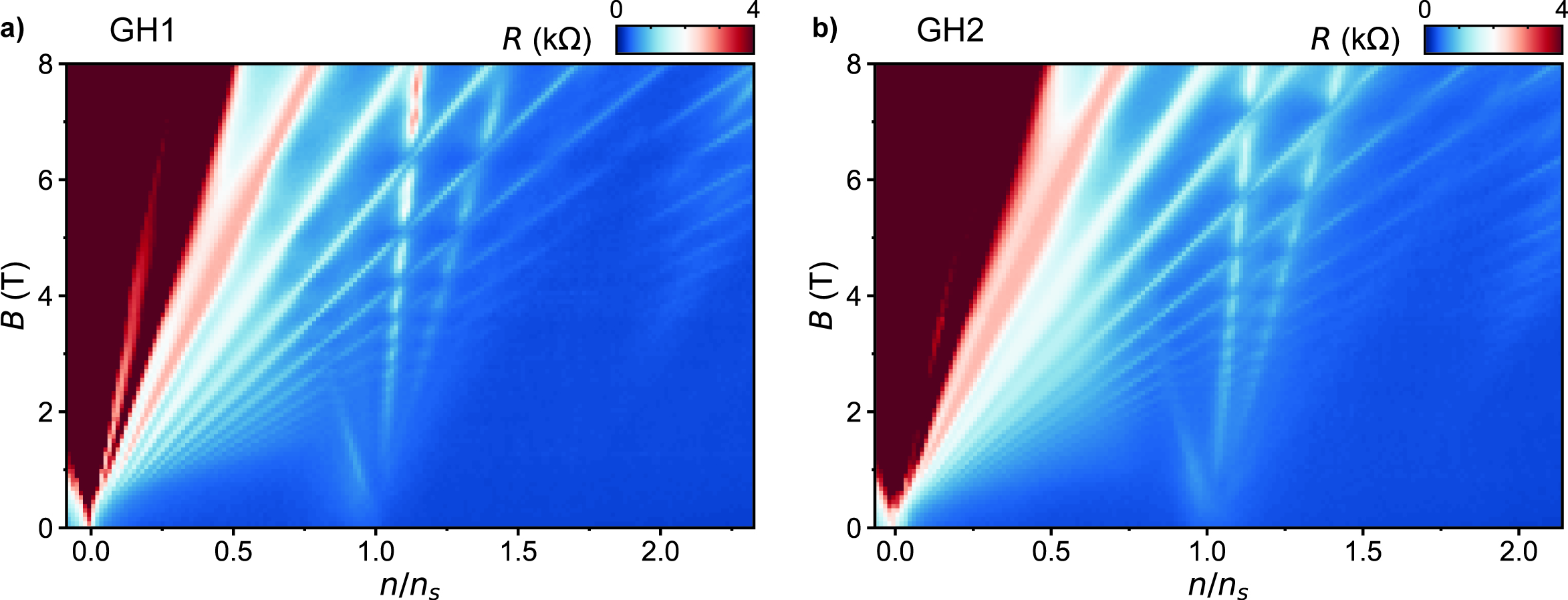}
  \caption{\textbf{Absence of Landau levels in the normal state.} \textbf{a-b)} Resistance $R$ as a function of magnetic field $B$ and $n/n_s$, for devices GH1 and GH2, respectively.  During both measurements we applied a constant d.c. current bias $I=200$ nA; much greater than the $I_c$ of the superconducting pockets at high fields.
  }
  \label{fig:HighField_noLLs}
\end{figure}

In the main text we have discussed that the high field superconducting states that we observe at the moiré minibands correspond to the semiclassical regime of chaotic mesoscopic ABS trajectories, instead of chiral Andreev edge states which would carry the ABS along the edges in the quantum Hall regime \cite{amet_supercurrent_2016, vignaud_evidence_2023}. \autoref{fig:HighField_noLLs} shows Landau Fan measurements from both devices GH1 and GH2, recorded while applying a fixed d.c. current $I = 200$ nA greater than any value of $I_c$ found at high magnetic fields, such that the JJ is in the normal state. By comparing \autoref{fig:HighField_noLLs} with Fig. 2a of the main text (GH1) and \autoref{fig:GH2_HighField}a (GH2), it can be seen that Landau levels are absent in the former at the same regions in the $B$--$n$ phase map where superconductivity is observed in the latter. The absence of these LLs proves that the superconducting proximity effect in our JJs does not occur in the quantum Hal regime.

\vspace{1.0\baselineskip}
\section{High field SC in the semiclassical regime of a non-moiré ballistic graphene JJ}\label{high-field-sc-in-the-semiclassical-regime-of-a-non-moiruxe9-ballistic-graphene-jj}

Here we briefly show the study of a monolayer graphene JJ not aligned with hBN, which also falls in the long-ballistic regime, so that we can compare the results of our graphene/hBN moiré JJs. This device has a length $L\sim 300$ nm and width $W\sim 1.5$ µm.

\begin{figure}[H]
  \centering
  \includegraphics[width=0.9\linewidth]{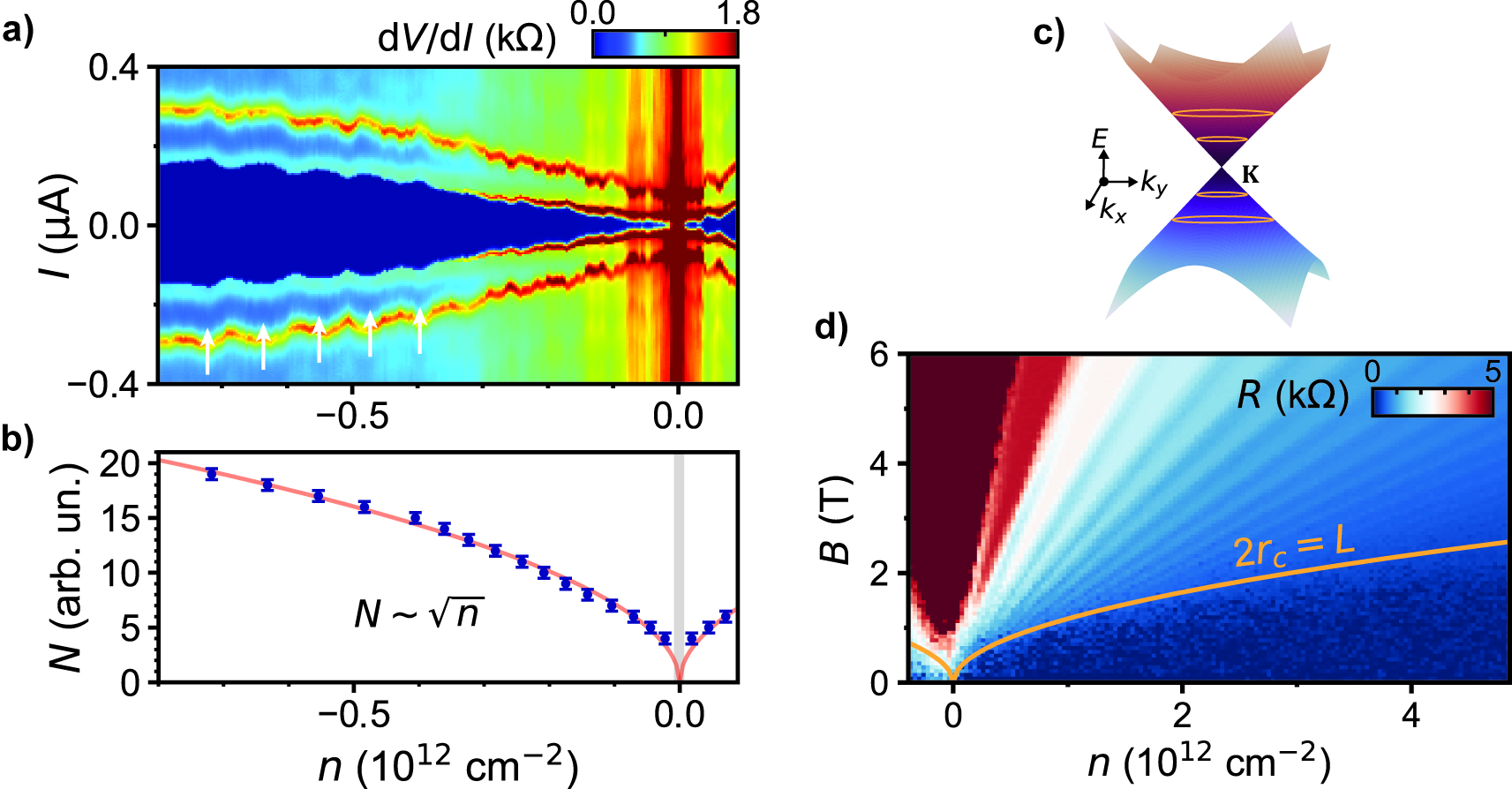}
  \caption{\textbf{Fabry-Pérots and high-field superconductivity in a non-moiré ballistic graphene JJ.} \textbf{a)} Map of differential resistance $\mathrm{d}V/\mathrm{d}I$ vs d.c. current bias $I$ and carrier density $n$, for a non-aligned monolayer graphene JJ, measured at 2 K. The dark blue regions indicate superconducting proximitized states, where the transition at positive $I$ corresponds to the critical current $I_c$. \textbf{b)} By extracting the consecutive maxima in the $I_c$ or the minima in the $R_N$ from \textbf{a}, we find that $N\sim\sqrt{n}$. Some of them at high carrier density are pointed in \textbf{a} with white arrows. The errorbars account for a potential missable extrema in the oscillating $I_c$ or $R_N$. \textbf{c)} Low energy band structure of graphene at the K point, where the circular Fermi surfaces are shown in the gray contours. \textbf{d)} Landau Fan diagram of the JJ measured at base temperature, where the superconducting states are visible below the $2r_c = L$ curve (orange). Here $r_c \propto \sqrt{n}$, due to the linear Dirac dispersion proper of graphene and its circular Fermi surfaces.}
  \label{fig:grapheneJJ}
\end{figure}

\autoref{fig:grapheneJJ}a shows a differential resistance map, where close to the CNP both the $I_c$ and $R_N$ clearly oscillates with a FP interference. Not only for hole doping (n-p-n junction), but as well in the electron side (n-n\textquotesingle-n junction), although for a lower density range. This is expected since the lower normal state resistance for electron doping reduces the number of reflections and the FP interferometry becomes weaker. The linear band structure of graphene (\autoref{fig:grapheneJJ}c) gives $N\sim\sqrt{n}$, corroborated in the measurements of \autoref{fig:grapheneJJ}b, where the square root dependence fits perfectly the observed oscillations.

As for the proximity effect at high magnetic fields, the superconducting region in \autoref{fig:grapheneJJ}d falls well below the $2r_c = L$ orange curve delimiting the semiclassical regime, where $r_c = \hbar\sqrt{\pi n}/eB$. We note that the densities that we reach here ($< 5 \times 10^{12}$ cm\textsuperscript{-2}) are the same that we reach in our graphene/hBN aligned JJs, so that a one-to-one comparison can be made with Fig. 2a of the main text and \autoref{fig:GH2_HighField}a. From these comparisons it can be seen that the moiré minibands in graphene/hBN moiré JJs enable a superconducting proximity effect at much higher fields than in bare monolayer graphene JJs.

%
%
%
%
%
\vspace{1.0\baselineskip}
\section{Theoretical Calculations}\label{theory}

\subsection{Non-magnetic model}
\begin{figure}[htb!]
 \centering
   \includegraphics[width=1\columnwidth]{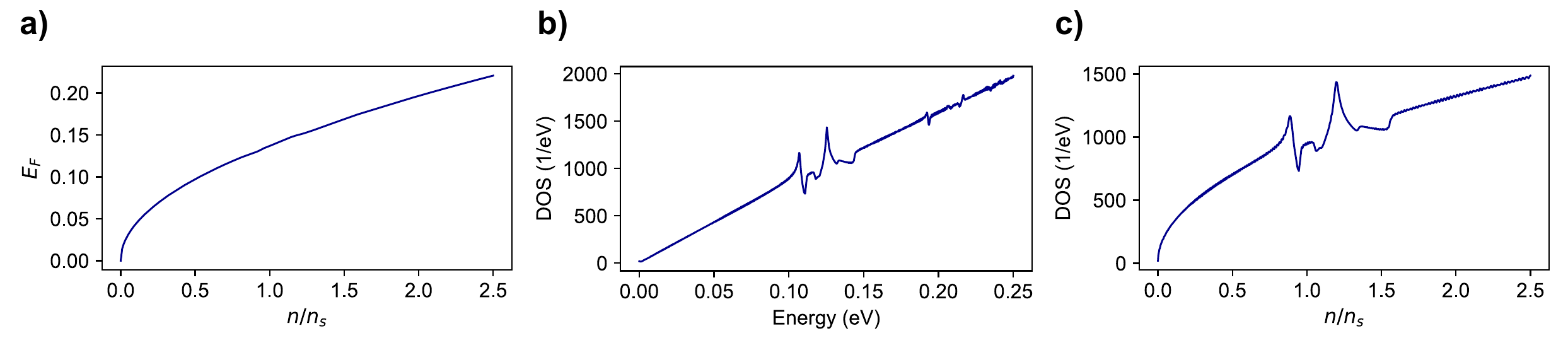}
 \caption{\textbf{a)} Fermi energy $E_F$ as a function of filling $n/n_s$. \textbf{b-c}) DOS as a function of energy $E$ and filling $n/n_s$ for a graphene/hBN moiré superlattice with twist angle $\theta=0.2$°.}
 \label{fig:figSI1}
\end{figure}

For the non-magnetic calculations, we use the model introduced by Moon and Koshino \cite{moon_electronic_2014}. In the absence of a magnetic field, the model is time-reversal symmetric, which allows us to perform calculations in a single valley. We couple $N = 5$ different mini Brillouin zones (mBZ) along each reciprocal moiré lattice vector $\mathbf{G^{1,2}_M}$. The resulting non-magnetic Hamiltonian has dimension $2(2N+1)^2$. The data in Fig.~2d of the main text are sampled along a grid of $(2N_k + 1)^2$ momentum points with $N_k = 200$, using a twist angle of $\theta = 0.2^{\circ}$, corresponding to device GH1. The corresponding Fermi energy is shown in \autoref{fig:figSI1}(a). We use a broadening $\eta = 0.0005\,\mathrm{eV}$ to calculate the density of states (DOS) as a function of energy $E$ (\autoref{fig:figSI1}b) and filling $n/n_s$ (\autoref{fig:figSI1}b-c). The van Hove singularities arise from saddle points of the moiré minibands, corresponding to energies at which neighboring equal energy contours touch and reconnect.

\paragraph{Momentum space gaps}
In the reconstructed miniband structure of graphene/hBN, two distinct types of momentum-space gaps $\Delta k$ can be identified from equal-energy contours, and they characterize different aspects of semiclassical transport. 
The first type, denoted $\Delta k_{\mathrm{inter}}$, is the separation in momentum space between distinct equal-energy pockets, for example pockets associated with Dirac cones in different minibands. For an approximately circular pocket centered at a Dirac point $\bm{k}_D$, we associate momenta satisfying $\abs{\bm{k}-\bm{k}_D}<k_F$ with that cone. $\Delta k_{\mathrm{inter}}$ therefore measures how far apart different semiclassical orbits are in $k$-space and controls the possibility of magnetic-breakdown tunneling \emph{between} different orbits. The second type, denoted as $\Delta k_{\mathrm{intra}}$, arises when a constant-energy contour is fragmented into disconnected segments, for example, between pockets originating from different bands along what would otherwise be the same Fermi contour. This quantity, therefore, measures the continuity of propagation \emph{along} a given equal-energy contour and determines whether a quasiparticle can remain on a single orbit or whether its motion is interrupted by forbidden momentum space regions. The characteristic magnetic field scale can be estimated as
\begin{equation}\label{eq:magnetic_length}
    B_l \sim \frac{\hbar}{e}\,(\Delta k)^2 \,,
\end{equation}
expressing the condition that magnetic breakdown becomes relevant when the orbit separation in momentum space is comparable to the inverse magnetic length \cite{alexandradinata_geometric_2017}. For $\Delta k_{\mathrm{inter}}$, this field describes the onset or suppression of tunneling between different Dirac cones or distinct Fermi pockets, and therefore controls whether semiclassical trajectories remain pocket-selective or instead form larger composite orbits involving multiple minibands. Additionally, one has to consider the size of the Fermi pockets.

For $\Delta k_{\mathrm{intra}}$, the corresponding field describes whether disconnected segments of a Fermi contour are effectively reconnected by breakdown, thereby restoring continuous propagation around the orbit. In the electrons in a graphene/hBN Josephson junction, $\Delta k_{\mathrm{inter}}$ determines whether electron- and hole-like trajectories can switch between different pockets in momentum space, while $\Delta k_{\mathrm{intra}}$ determines whether a given  trajectory is geometrically continuous in angle or fragmented into disconnected sectors. If two different values of $\Delta k_{\mathrm{intra}}$ are observed, they define two different characteristic fields, $B_{l,1} \sim \Delta k_1^2$ and $B_{l,2} \sim \Delta k_2^2$. The smaller field corresponds to the easiest bottleneck in momentum space, i.e., the first place where partial breakdown-assisted connectivity can occur or disappear, whereas the larger field corresponds to the hardest bottleneck, i.e., the last link required to maintain a fully connected trajectory. 

To extract the distances $\Delta k$ between different contours, we calculate $i \in [1,N(N-1)/2]$ pairwise distances $d_i$ between $N$ different loops, group unique distances $d^{\rm unique}_i$ with a relative error of 1\%, and retain the relevant unique distances. The minimum distance is determined as $\Delta k = \min_i(d^{\rm unique}_i)$. We calculate ${\Delta k}^j$ for all bands $j=1,2,3$ and between different bands ${\Delta k}^{j,l}$ such that $j \neq l$. For $M$ bands, we thus have $M + M(M-1)/2$ different values for ${\Delta k}$. We verified that the relative error in grouping unique distances between different loops does not lead to double counting or missed distances.

\begin{figure}[htb!]
 \centering
   \includegraphics[width=1\columnwidth]{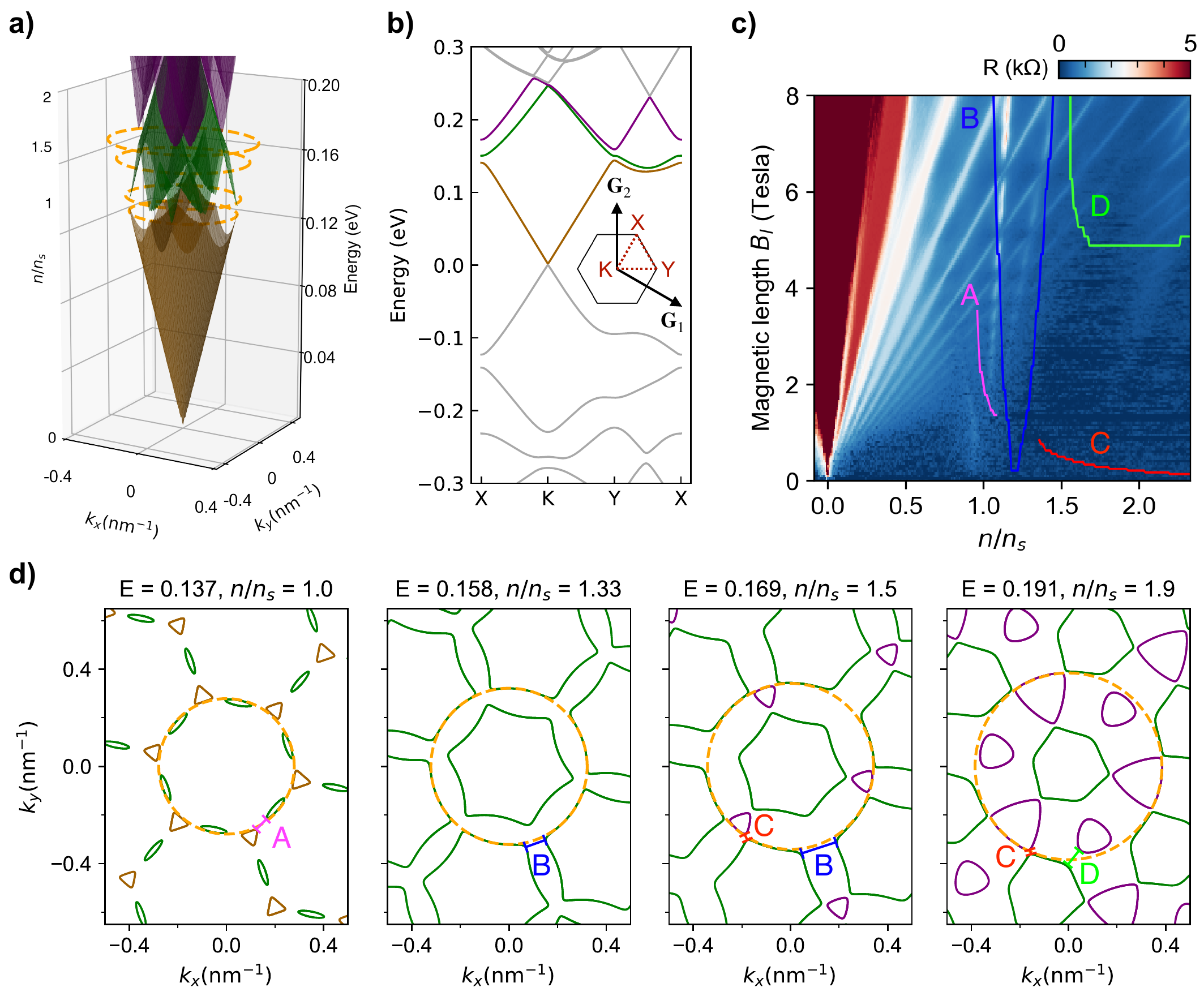}   
 \caption{Band structure of graphene/hBN at zero field for a twist angle of $\theta = 0.2^{\circ}$. \textbf{a)} Bands from Fig.~1b of the main text with respective Fermi momenta $k_F$ (circles) corresponding to the line cuts in panel \textbf{d}. \textbf{b)} First (orange), second (green), and third (purple) bands along a path in the first mini Brillouin zone (inset) spanned by $\mathbf{G}_1$ and $\mathbf{G}_2$. \textbf{c)} Minimum distance $\Delta k$ between contours, converted to the magnetic field $B_l$, allowing for an estimate of the low-resistance region. \textbf{d)} Equal energy contours at selected fillings $n/n_s = 1, 1.33, 1.5, 1.9$. Momentum space gaps are denoted by A-.D.}
 \label{fig:figSI0}
\end{figure}

\autoref{fig:figSI0}(a) and (b) show the first three bands, with circles denoting the Fermi momentum $k_F$ at the fillings $n/n_s = 1,1.33,1.5,1.9$, corresponding to the equal energy contours in panel (d). 
Panel (c) of \autoref{fig:figSI0} shows the field scale $B_l$ obtained from \autoref{eq:magnetic_length} for different momentum-space gaps. \autoref{fig:figSI0}(d) present the momentum space structure for different fillings with corresponding gaps: A (pink), between the first and second band at filling $n/n_s \lesssim 1$; B (blue), between the second band for $1 \lesssim n/n_s \lesssim 1.4$; C (red) and D (green), between the second and third band. Note that A,C, and D are momentum space gaps of type $\Delta k_{\mathrm{inter}}$, while B is $\Delta k_{\mathrm{intra}}$. C and D are two different gaps along $k_F$ between the same bands. For fillings $1.5 \lesssim n/n_s \lesssim 2$, we observe that the larger gap D allows for a qualitative estimate of the upper boundary of the low resistance region.

\paragraph{High-symmetry points in momentum space}
The moiré reciprocal lattice is generated by
\begin{equation}
    \mathbf{G}_{1} = \left( \mathrm{1} - M^{-1} R_\theta \right)\mathbf{b}_1,
    \qquad
    \mathbf{G}_{2} = \left( \mathrm{1} - M^{-1} R_\theta \right)\mathbf{b}_2,
\end{equation}
where $\mathbf{b}_{1,2}$ are the reciprocal lattice vectors of graphene,
$R_\theta$ is the rotation matrix with a twist angle $\theta$,
and $M = (a_{\rm hBN}/a)\,\mathrm{1}$ accounts for lattice mismatch.
The twist angle enters only through $R_\theta$ and therefore
rotates and rescales the moiré reciprocal basis vectors $\mathbf{G}_{i}$.
The first mBZ is the Wigner-Seitz cell of the
triangular lattice spanned by $\mathbf{G}_{1}$ and $\mathbf{G}_{M}$.
The high-symmetry points are defined directly in this basis:
\begin{align}
    \Gamma &= \mathbf{0}, \\
    K_{+} &= \frac{2\mathbf{G}_{1} + \mathbf{G}_{2}}{3}, 
    \qquad
    K_{-} = \frac{\mathbf{G}_{1} + 2\mathbf{G}_{2}}{3}, \\
    M_1 &= \frac{\mathbf{G}_{1}}{2}, \quad
    M_2 = \frac{\mathbf{G}_{2}}{2}, \quad
    M_3 = \frac{\mathbf{G}_{1} + \mathbf{G}_{2}}{2}.
\end{align}
In the band-structure calculations, we use $X \equiv M_1$ and $Y \equiv M_2$.
Band structures in a given valley $\xi=\pm1$ are calculated in momenta
measured w.r.t to the corresponding corner,
\begin{equation}
    \mathbf{q} = \mathbf{k} - K_\xi,
\end{equation}
such that the valley point is located at $\mathbf{q}=\mathbf{0}$.

\clearpage
\subsection{Magnetic Model}
\subsubsection{Landau levels and magnetic translation operators}

For electrons in a magnetic field $\bm B =\nabla\times \bm A = -B\hat{z}$, the kinetic momentum $\bm\pi = \bm p + e\bm A$ and the guiding center coordinates $\bm R= \bm r - \frac{l_B^2}{\hbar}\hat{z}\times \bm \pi$ with the canonical momentum $\bm p=-i \hbar \bm\nabla$ and magnetic length 
\begin{equation}\label{eq:lB}
    l_B = \sqrt{\hbar/eB}  
\end{equation}
obey the following commutation relations,
\begin{align}
    &[\pi_i, \pi_j] = \frac{i\hbar^2}{l_B^2} \epsilon_{ij}, \label{eq:commutator_pipi}\\  
    &[\pi_i, R_j] = -i\hbar \delta_{ij} + i\hbar \epsilon_{jk}\epsilon_{ik} = 0,\\
    &[R_{i}, R_{j}] = -l_B^2\epsilon_{ij},
\end{align}
where $\epsilon_{ij} = \begin{cases}
    +1 \, & \text{if} \quad i,j = (x,y) \,,\\
    -1 \, & \text{if} \quad i,j = (y,x) \,,\\
    0 \, & \text{else} \\
\end{cases}$.
Using Eq.~\eqref{eq:commutator_pipi}, we can define the annihilation and creation operators
\begin{equation}
    a = \frac{l_B}{\sqrt{2\hbar^2}}(\pi_x + i\pi_y), \ \ \ a^{\dagger}=\frac{l_B}{\sqrt{2\hbar^2}}(\pi_{x} - i\pi_y),
\end{equation}
with the commutation relation $[a,a^{\dagger}]=1$.
The magnetic translation operators
\begin{equation}
    T_{\bm a} =e^{i\bm a\cdot(\hat{z}\times \bm R)/l_B^2}
\end{equation}
commute with kinetic momentum $\bm \pi$ and satisfy
\begin{equation}\label{eq:commutation_T}
   T_{\bm a}T_{\bm a'} = e^{i\bm a\wedge\bm a'/l_B^2} T_{\bm a'}T_{\bm a},
\end{equation}
with $\bm a\wedge \bm a'\equiv a_{x}a_{y}' - a_{y}a_{x}'$. This relation can be proven using the Baker–Campbell–Hausdorff (BCH) formula $e^{A}e^{B} = e^{A+B+[A,B]/2}$ for operators $A,B$ whose commutator is constant.

For a superlattice with primitive lattice vectors $\bm a_{1,2}$, its unit cell area is $A_{\rm u.c.} = |\bm a_1\wedge \bm a_2|$, and the corresponding primitive reciprocal lattice vectors are denoted as $\bm G_{1,2}$.
At a rational magnetic flux quanta $B A_{\rm u.c.} = \phi_0p/q$ per unit cell, the magnetic length obeys
\begin{align}\label{eq:lB^2}
    l_B^2=\frac{|\bm a_1\wedge \bm a_2|}{2\pi} \frac{q}{p} = \frac{2\pi}{|\bm G_1\wedge \bm G_2|}\frac{q}{p},
\end{align}
and the magnetic translation operators $T_{q\bm a_1}$ and $T_{\bm a_2/p}$ commute, $[T_{q\bm a_1}, T_{\bm a_2/p}] = 0$, according to qs.~\eqref{eq:commutation_T} and~\eqref{eq:lB^2} \cite{kolar_hofstadter_2024}.

We choose a Landau-level basis of simultaneous eigenstates of $\bm\pi^2, T_{q\bm a_1}, T_{\bm a_2/p}$. The basis states $|n,\bm k\rangle$, labeled by a Landau level index $n$ (not to be confused with the filling $n/n_s$) and momentum $\bm k$, are defined as follows,
\begin{gather}
a^{\dagger}a|n, \bm k\rangle = n |n, \bm k\rangle,\\
T_{q \bm{a}_1}|n, \bm k\rangle=\exp \left(i q \bm{a}_1 \cdot \bm k\right)|n, \bm k\rangle, \\
T_{\bm{a}_2 / p}|n, \bm k\rangle=\exp \left(i \frac{1}{p} \bm{a}_2 \cdot \bm k\right)|n, \bm k\rangle ,
\end{gather}
where $\hat{a}^{\dagger}$ ($\hat{a}$) raises (lowers) the Landau level $\hat{a}^{\dagger} \ket{n} = \sqrt{n+1} \ket{n+1}$ ($\hat{a} \ket{n} = \sqrt{n} \ket{n-1}$). 
Here, $T_{q\bm a_1}|n,0\rangle = T_{\bm a_2/p}|n,0\rangle = |n,0\rangle$ defines the state in the $n$th Landau level that is invariant under the two commuting magnetic translations. Other eigenstates at momentum $\bm k = \frac{k_1}{q}\bm G_1 + k_2 p \bm G_2 $ are related by
\begin{equation}
    |n,\bm k\rangle = e^{ik_2 p\bm G_2\cdot \bm R} e^{i\frac{k_1}{q}\bm G_1\cdot \bm R} |n,\bm k=0\rangle,\ \ k_{1,2}\in [0,1).
\end{equation}
\subsubsection{Graphene/hBN superlattice}

The continuum model of a graphene/hBN moir\'e superlattice is diagonal in the valley space, and can be formally written as follows:
\begin{equation}\label{eq:H}
    \hat{H}^{\xi} = \hat{H}_{\text{G}}^{\xi} + \sum_{\bm g}\hat{V}_{\bm g}^{\xi} e^{i\bm g\cdot \bm r}
\end{equation}
where the kinetic term in the $\xi=\pm$ valley of a monolayer graphene reads
\begin{align}
    \hat{H}_\text{G}^{\xi}=-v_F\left(\begin{array}{cc}
0 & \xi \pi_x-i \pi_y \\
\xi \pi_x+i \pi_y & 0
\end{array}\right).
\end{align}
This can be rewritten as
\begin{align}
        \hat{H}_\text{G}^{+}= -\frac{\sqrt{2}\hbar v_F}{l_B}\left(\begin{array}{cc}
0 & a^{\dagger} \\
a & 0
\end{array}\right),\qquad     \hat{H}_\text{G}^{-}=\frac{\sqrt{2}\hbar v_F}{l_B}\left(\begin{array}{cc}
0 & a \\
a^{\dagger} & 0
\end{array}\right).
\end{align}
Note that we will use a convention different from Ref.~\cite{moon_electronic_2014}, where the magnetic field is applied in $+z$-direction.
In the basis of the $\xi=+$ valley, the Landau level spectrum $\epsilon_{\eta n}$ and wave functions $|\eta n;\xi\rangle$ of $\hat{H}_{\text{G}}^+$ read
\begin{align}
    &\epsilon_{0}=0, &&|0;\xi=+\rangle=\binom{|0\rangle}{0}, \\
    &\epsilon_{\eta, n}= - \eta \frac{\hbar v_F}{l_B} \sqrt{2 n}, &&|\pm n;\xi=+\rangle=\frac{1}{\sqrt{2}}\binom{|n\rangle}{\pm|n-1\rangle}, \quad n>0.
\end{align}
In this basis, $\hat{H}_{\text{G}}^{\xi=+}$ is diagonal
\begin{equation}
    \hat{H}_\text{G}^+ = -\Omega \, \text{diag}(-\sqrt{N_{LL}}, ..., -\sqrt{1}, 0, \sqrt{1}, ..., \sqrt{N_{LL}}),
\end{equation}
where $N_{LL}$ is the highest Landau level, and $\Omega = \sqrt{2} \hbar v_F / l_B$. The moir\'e potential has a $2\times2$ matrix form in the sublattice space
\begin{equation}
    \hat{V}_{\bm g}^{\xi} \equiv \begin{pmatrix}
        V_{00}^{\xi}({\bm g})& V_{01}^{\xi}({\bm g})\\
        V_{10}^{\xi}({\bm g})& V_{11}^{\xi}({\bm g})\\
    \end{pmatrix}.
\end{equation}

We notice that the eigenstates of $H_{\text{G}}^{\pm}$ are related by swapping the two sublattice components, $|\eta n;\xi=-\rangle = \sigma^y |\eta n;\xi=+\rangle$. In this notation,
\begin{equation}
    U_{ij}^{\xi=+}(\bm g) = V_{ij}^{+}(\bm g),\quad U_{ij}^{\xi=-}({\bm g}) \equiv (\sigma^y\hat{V}_{\bm g}^{-} \sigma^{y})_{ij},
\end{equation}
and the matrix elements of the moir\'e potential in the Landau level basis read:
\begin{align}
\left\langle 0, \bm {k}^{\prime};
\xi\right| \hat{V}_{\bm g}^{\xi} e^{i\bm g\cdot \bm r}|0, \bm {k};\xi\rangle= & U_{00}^{\xi} \Lambda_{0 \bm {k}', 0 \bm {k}}(\vec{g}), \label{eq:matrixlement1}\\
\left\langle \eta^{\prime} n^{\prime}, \bm {k}^{\prime};\xi\right| \hat{V}_{\bm g}^{\xi} e^{i\bm g\cdot \bm r}|0, \bm {k};\xi\rangle= & \frac{1}{\sqrt{2}} U_{00}^{\xi} \Lambda_{n' \bm {k}', 0 \bm {k}}(\vec{g})+\frac{\eta'}{\sqrt{2}} U_{10}^{\xi} \Lambda_{n'-1\bm {k}^{\prime}, 0 \bm {k}}(\vec{g}), \\
\left\langle 0, \bm {k}^{\prime};\xi\right| \hat{V}_{\bm g}^{\xi} e^{i\bm g\cdot \bm r}|\eta n, \bm {k};\xi\rangle= & \frac{1}{\sqrt{2}} U_{00}^{\xi} \Lambda_{0 \bm {k}',n \bm {k}} 
(\vec{g})+\frac{\eta}{\sqrt{2}} U_{01}^{\xi} \Lambda_{0\bm {k}^{\prime}, n-1 \bm {k}}(\vec{g}), \\
\left\langle \eta^{\prime}n^{\prime}, \bm {k}^{\prime};\xi\right| \hat{V}_{\bm g}^{\xi} e^{i\bm g\cdot \bm r}|\eta n, \bm {k};\xi\rangle= & \frac{1}{2} U_{00}^{\xi} \Lambda_{n' \bm {k}^{\prime}, n \bm {k}}(\vec{g})+\frac{\eta}{2} U_{01}^{\xi} \Lambda_{n' \bm {k}^{\prime}, n-1 \bm {k}}(\vec{g})  \notag\\
&+\frac{\eta'}{2} U_{10}^{\xi} \Lambda_{n^{\prime}-1 \bm {k}^{\prime}, n \bm {k}}(\vec{g})+\frac{\eta\eta'}{2} U_{11}^{\xi} \Lambda_{n^{\prime}-1 \bm {k}^{\prime}, n-1 \bm {k}}(\vec{g}).\label{eq:matrixlement4}
\end{align}
Here, we introduce the form factors
\begin{align}
    \Lambda_{n'\bm {k}', n\bm {k}}(\bm Q) &= \langle n', \bm {k}' | e^{i\bm Q\cdot \bm r} |n,\bm {k}\rangle \notag\\
    &= F_{n'n}(\bm Q) \langle \bm k'|e^{i\bm Q\cdot \bm R}|\bm k\rangle, \label{eq:Lambda}
\end{align}
with $F_{n'n}(\bm Q) = \langle n' | e^{i\frac{l_B^2}{\hbar}\bm Q\cdot (\hat{z}\times \bm\pi)}|n\rangle$. A simplified expression for $F_{n'n}(\bm Q)$ can be found in the literature, see Appendix of Ref.~\cite{goerbig_electronic_2011} for example,
\begin{equation}\label{eq:F}
     F_{n'n}(\bm Q) = \begin{cases}
         \sqrt{\frac{n!}{n'!}}\left[\frac{(Q_x+iQ_y)l_B}{\sqrt{2}}\right]^{n'-n} L_{n}^{n'-n}\Big(\frac{\bm Q^2 l_B^2}{2}\Big)e^{-\frac{\bm Q^2l_B^2}{4}},\quad &n'\geq n,\\
         F_{nn'}(-\bm Q)^*,&n'<n,
     \end{cases}
\end{equation}
where $L_{m}^{n}(x)$ represents the associated Laguerre polynomial. 

Plugging $\bm Q = \frac{Q_1}{q} \bm G_1 + Q_2 p\bm G_2$ into the second term on the right-hand side of Eq.~\eqref{eq:Lambda} and repeatedly using the BCH formula, we find
\begin{align}
    \langle \bm k'|e^{i\bm Q\cdot \bm R}|\bm k\rangle &= \langle 0| e^{-i \frac{k_1^{\prime}}{q} \bm G_1 \cdot \vec{R}} e^{-i k_2^{\prime} p \vec{G}_2 \cdot \vec{R}} e^{i \vec{Q} \cdot \vec{R}} e^{i k_2  p \bm G_2 \cdot \vec{R}} e^{i \frac{k_1}{q} \bm G_1 \cdot \vec{R}}|0\rangle \notag\\
    & =\langle 0| e^{-i \frac{k_1^{\prime}}{q} \bm G_1 \cdot \vec{R}} e^{-i k_2^{\prime} p \vec{G}_2 \cdot \vec{R}} e^{i \frac{Q_1}{q}\vec{G}_1 \cdot \vec{R}} e^{i Q_2p\vec{G}_2 \cdot \vec{R}} e^{i k_2  p \bm G_2 \cdot \vec{R}} e^{i \frac{k_1}{q} \bm G_1 \cdot \vec{R}}|0\rangle  e^{-i \pi Q_1 Q_2}\notag\\
    &= \langle 0| e^{-i \frac{k_1^{\prime}}{q} \vec{G}_1 \cdot \vec{R}} e^{i\left(Q_2+k_2 -k_2^{\prime} \right)p \vec{G}_2^ \cdot \vec{R}} e^{i\frac{Q_1+k_1}{q} \bm{G}_1 \cdot \vec{R}}|0\rangle e^{-i \pi Q_1 Q_2} e^{i m_1\left(Q_2+k_2\right)p G_1 \wedge G_2 l_B^2} \notag\\
    & = \langle 0 | e^{i\left(Q_2+k_2 -k_2^{\prime} \right) \bm{G}_2 \cdot \vec{R}} e^{i(Q_1+k_1-k_1^{\prime}) \bm{G}_1 \cdot \vec{R}}|0\rangle e^{-i \pi Q_1 Q_2} e^{2i \pi [Q_1\left(Q_2+k_2 \right)-k_1^{\prime}\left(Q_2+k_2  - k_2^{\prime}\right)]}\notag\\
    & = \delta_{\{Q_2+k_2\}, k_2'}\delta_{\{Q_1+k_1\}, k_1'}\ e^{-i \pi Q_1 Q_2} e^{2i \pi [Q_1\left(Q_2+k_2 \right)-\{Q_1+k_1\}\lfloor Q_2+k_2\rfloor]},
\end{align}
where $\{x\} = x - \lfloor x \rfloor$.
In particular, for reciprocal lattice vectors $\bm g = m_1 \bm G_1 + m_2\bm G_2\ (m_{1,2}\in\{0,\pm 1\})$, the momentum-dependent matrix element reads
\begin{equation}\label{eq:e^iqR}
    \langle \bm k'|e^{i\bm g\cdot \bm R}|\bm k\rangle = \delta_{k_1, k_1'}\delta_{\{\frac{m_2}{p}+k_2\}, k_2'} e^{-i \pi m_1 m_2 \frac{q}{p}} e^{2i \pi m_1\left(m_2+k_2 p\right)\frac{q}{p}}e^{-2i\pi k_1\lfloor \frac{m_2}{p}+k_2 \rfloor}.
\end{equation}
In this equation, $k_{1,2},k_{1,2}' \in [0,1)$.
The Kronecker deltas in Eq.~\eqref{eq:e^iqR} indicate that the superlattice potential can couple momentum eigenstates $\bm k, \bm k+ \bm G_{2},...,\bm k+(p-1)\bm G_{2}$. Therefore, the primitive reciprocal lattice vectors of the mini-Brillouin zone of the superlattice at $p/q$ magnetic flux quanta per unit cell become $\bm G_1/q$ and $\bm G_2$. Note that $k_1,k_2$ are the components of 
\begin{equation}\label{eq:momentum_vector_def}
  \bm k = k_1\frac{1}{q}\bm G_1 + k_2 p \bm G_2
\end{equation}
where $k_1\in [0,1), k_2\in [0,\frac{1}{p})$.  
In summary, at a given momentum $\bm k$
keeping $(2N_{LL}+1)$ Landau levels, the full Hamiltonian eq.~\eqref{eq:H} becomes a matrix of dimension $(2N_{LL}+1) p$ in the Landau level basis. 

The system has a magnetic translation symmetry $[T_{\bm a_1}, H]=0$.
Because $T_{\bm a}$ commutes with $\bm\pi$, we have 
\begin{equation}
    [T_{\bm a_1},H_G^\xi]=0 \,.
\end{equation}
Moreover, using
$\bm r=\bm R+\frac{l_B^2}{\hbar}\hat{\bm z}\times\bm\pi$ and $T_{\bm a}^\dagger \bm R T_{\bm a}=\bm R+\bm a$,
one finds $T_{\bm a}^\dagger e^{i\bm g\cdot\bm r}T_{\bm a}=e^{i\bm g\cdot\bm a}e^{i\bm g\cdot\bm r}$.
For $\bm a=\bm a_1$ and $\bm g=m_1\bm G_1+m_2\bm G_2$, this phase is $e^{i\bm g\cdot\bm a_1}=e^{i2\pi m_1}=1$,
hence
\begin{equation}
    [T_{\bm a_1},H]=0 \,.
\end{equation}
We can apply 
\begin{equation}
   T_{-\bm a_1} |n,\bm k\rangle = e^{i\frac{p}{q}\bm G_2\cdot\bm R}|n,\bm k\rangle = |n, \bm k+\frac{1}{q}\bm G_2\rangle, 
\end{equation}
the energy spectra at $\bm k$ and $\bm k + \frac{1}{q}\bm G_2$ are identical. Therefore, we can perform all calculations in a reduced mBZ of $k_1 \times k_2 \in [0,1/q) \times [0,1/q)$.

\subsubsection{Quantities of Interest}
\paragraph{Fermi energy}
The bands are calculated on a grid of $N_k \times N_k$ momentum points, each with $N_{\text{bands}} = \text{dim}(H)$ energy levels. The Fermi energy $E_F$ is calculated by sorting all $N_{\text{bands}} N_k^2$ states by energy, filling half of them, and determining the lowest energy of all unoccupied states. For a given flux $p/q$ and a single valley, we can calculate $E_F(n/n_s)$ as a function of $n/n_s$ by filling $\left(\frac{N_{\text{bands}}}{2} + q\,n/n_s\right)N_k^2$ states, where $q \rightarrow 2q$ if we combine both valleys. 

\paragraph{Band velocity}
The group velocity of a band with dispersion $E(k)$, $k = (k_x,k_y)$, is defined as
\begin{equation}
    v_{\alpha}(k) = \frac{1}{\hbar} \pdv{E(k)}{k_{\alpha}} \,,
\end{equation}
where ${\alpha} = x,y$. We calculate
$\expval{\mathbf{v}^2}$
\begin{equation}\label{eq:velocity}
    \expval{\mathbf{v}^2} = \frac{1}{N_k^2} \sum_{k} \left(v_x(k)^2 + v_y(k)^2 \right)
\end{equation}
for $N_k^2$ momentum points.
In our calculations, we use momentum components $k_1,k_2$ such that
\begin{equation}
    \begin{pmatrix}
        k_x \\ k_y
    \end{pmatrix}
     = k_1 \mathbf{G}_1 + k_2 \mathbf{G}_2 \quad \Leftrightarrow \quad \mathbf{k} = B \mathbf{u}
\end{equation}
with $\mathbf{u} = (k_1,k_2)^T$ and $B = (\mathbf{G}_1, \mathbf{G}_2)$. Hence, we have to calculate
\begin{equation}
    v_{\alpha}(k) = \frac{1}{\hbar} \pdv{E(k)}{k_{\alpha}}  = \frac{1}{\hbar} \left(\pdv{E(k)}{k_1} \pdv{k_1}{k_{\alpha}} + \pdv{E(k)}{k_2} \pdv{k_2}{k_{\alpha}} \right)
\end{equation}
or
\begin{equation}
    \mathbf{v} = \frac{1}{\hbar} \nabla_{\mathbf{k}} E = \frac{1}{\hbar} (B^{-1})^T \nabla_{\mathbf{u}} E
\end{equation}
with $\nabla_{\mathbf{k}} = (\partial_{k_x}, \partial_{k_y})^T$, $\nabla_{\mathbf{u}} = (\partial_{k_1}, \partial_{k_2})^T$, and  
\begin{equation}
    B^{-1} = \begin{pmatrix}
         \pdv{k_1}{k_x} & \pdv{k_1}{k_y} \\
         \pdv{k_2}{k_x} & \pdv{k_2}{k_y} \\         
    \end{pmatrix}
\end{equation}

\paragraph{Filling}
The density of states (DOS) is defined as a function of filling:
\begin{equation}
     \text{DOS}(n/n_s) = \frac{1}{q N_k^2  \vert \mathbf{G_1} \wedge \mathbf{G_2} \vert}\sum_{k_1=1}^{N_k} \sum_{k_2=1}^{N_k} \sum_{i=1}^{N_{\text{bands}}} \delta^L_{\eta}(E_i(k_1,k_2) - E_F(n/n_s))
 \end{equation}
where we use a Lorentzian function with small broadening $\eta$:
\begin{equation}
    \delta^L_{\eta}(x) = \frac{1}{\pi} \frac{\eta}{x^2 + \eta^2} \,.
\end{equation}

Similarly, we can calculate the band velocity $v^i_{\alpha}(k_1,k_2)$ for $\alpha = x,y$ for all bands $i$ and momenta $(k_1,k_2)$ and define $\expval{\mathbf{v}^2}$ as a function of filling:
\begin{equation}\label{eq:velocity_filling}
    \expval{\mathbf{v}^2}(n/n_s) = \frac{1}{\mathcal{N}_{v}} \sum_{k_1=1}^{N_k} \sum_{k_2=1}^{N_k} \sum_{i=1}^{N_{\text{bands}}} w_{\eta}(E_i(k_1,k_2) - E_F(n/n_s)) \left(v^i_x(k_1,k_2)^2 + v^i_y(k_1,k_2)^2 \right)
\end{equation}
where $\mathcal{N}_{v}$ is the number of energy levels in the energy range $[E_F(n/n_s)-\eta,E_F(n/n_s)+\eta]$. 
The delta function $w_{\eta}(...)$ for the band velocity is implemented as a window function:
\begin{equation}
    w_{\eta}(x) = \begin{cases}
        1 \,,& \abs{x} < \eta \\
        0 \,,& \abs{x} > \eta
    \end{cases} \,.
\end{equation}

\paragraph{Band gap}
We can define the direct and indirect band gaps $\Delta_{\text{direct},\xi}$ and $\Delta_{\text{indirect},\xi}$ between two neighboring bands $i$ and $i+1$ in the same valley $\xi = \pm$ as:
\begin{equation}
    \Delta_{\text{direct},\xi} = \text{min}_{k_1,k_2}\left(E_{\xi,i+1}(k_1,k_2) - E_{\xi,i}(k_1,k_2)\right)
\end{equation}
and
\begin{equation}
    \Delta_{\text{indirect},\xi} = \text{min}_{k_1,k_2}\left(E_{\xi,i+1}(k_1,k_2)\right) - \text{max}_{k_1,k_2}\left(E_{\xi,i}(k_1,k_2)\right)
\end{equation}
respectively.

\subsection{Comparison of Small and Large JJ Device}

\begin{figure}[b!]
 \centering
   \includegraphics[width=1\columnwidth]{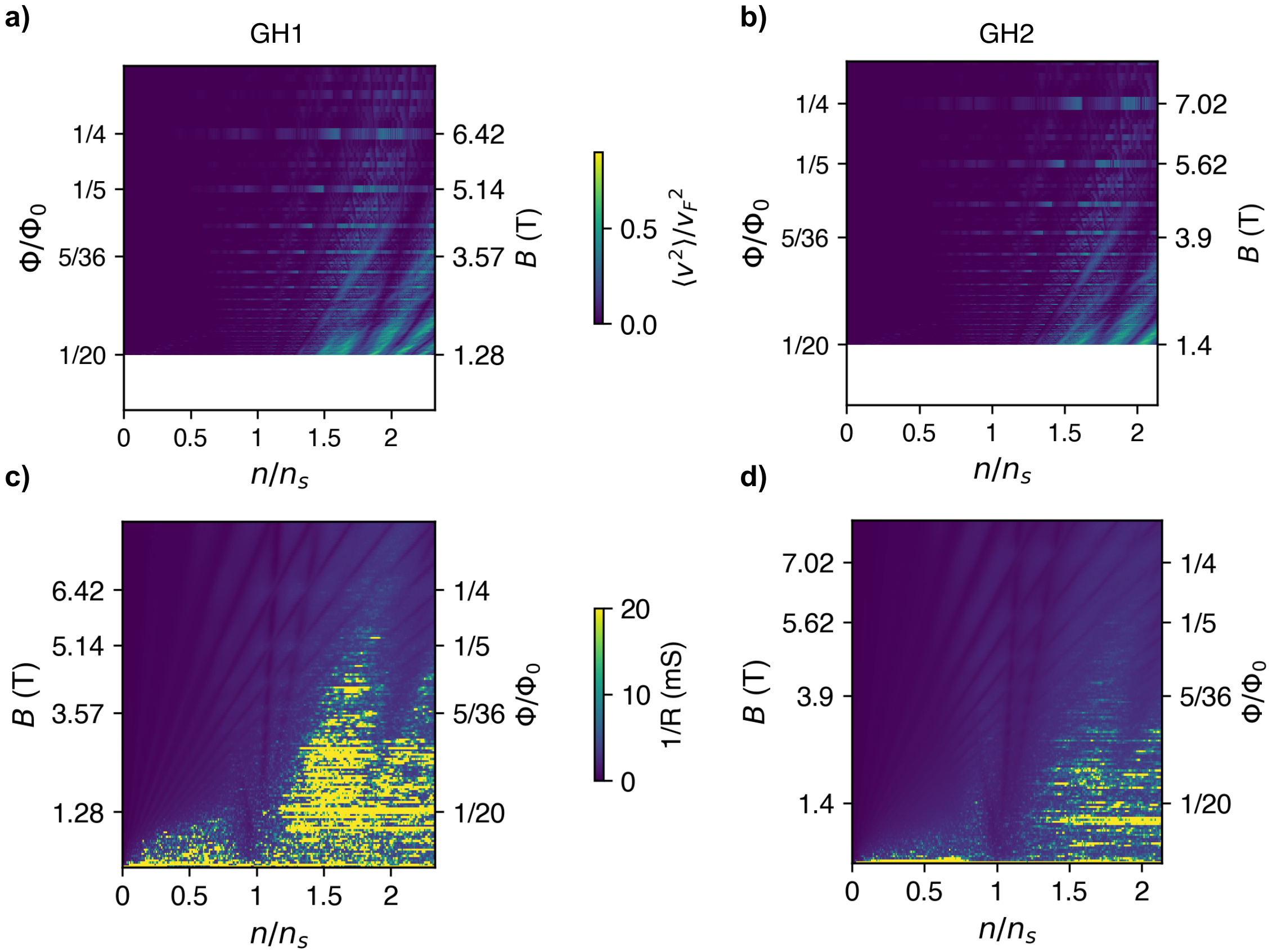}
 \caption{\textbf{a,b)} Average squared band velocity calculated for flux quanta between $p/q = 1/3$ and $p/q = 1/20$ for a twist angle of $\theta = 0.2^{\circ}$ and $\theta = 0.38^{\circ}$, corresponding to GH1 and GH2, respectively.  \textbf{c,d)} Map of inverse resistance $1/R$ with cutoff of $20 \, \text{mS}$ for GH1 and GH2, respectively.}
 \label{fig:figSI4}
\end{figure}

For our calculations, we consider the two devices GH1 and GH2 as described in section \ref{transport-characterization-of-all-samples}.
Panels Fig. 3d-e of the main text establish a direct connection between the emergence of low resistance along a broad range of fillings $n/n_s$ persisting to high magnetic fields and large band velocity associated with overlapping bands. Increasing magnetic flux leads to a reduced average squared band velocity \cite{krishna_kumar_high-order_2018}, reflecting the flattening of magnetic bands. We calculate $\langle v^2 \rangle$ relative to the Fermi velocity $v_F^2$, indicating how the band velocity is suppressed relative to pristine graphene.

In \autoref{fig:figSI4}, we present the band velocity as a function of magnetic field and filling for both devices GH1 and GH2 in panels (a) and (c), respectively. Note that all data are presented for magnetic fields up to a cutoff of 8 Tesla. Due to the different twist angles, the conversion of flux quanta $p/q$ to field $B$ is different for both devices. Panels (b) and (d) show the corresponding experimental inverse resistance data. 
The largest filling value measured is slightly lower for GH2 compared to GH1. In both devices, we observe a large region of low resistance around $n/n_s \approx 1.5$, which persists up to large flux $1/5$. We observe that this region matches well with the calculated average band velocity $\langle v^2 \rangle / v_F^2$ for both devices.

\begin{figure}[H]
 \centering
   \includegraphics[width=1\columnwidth]{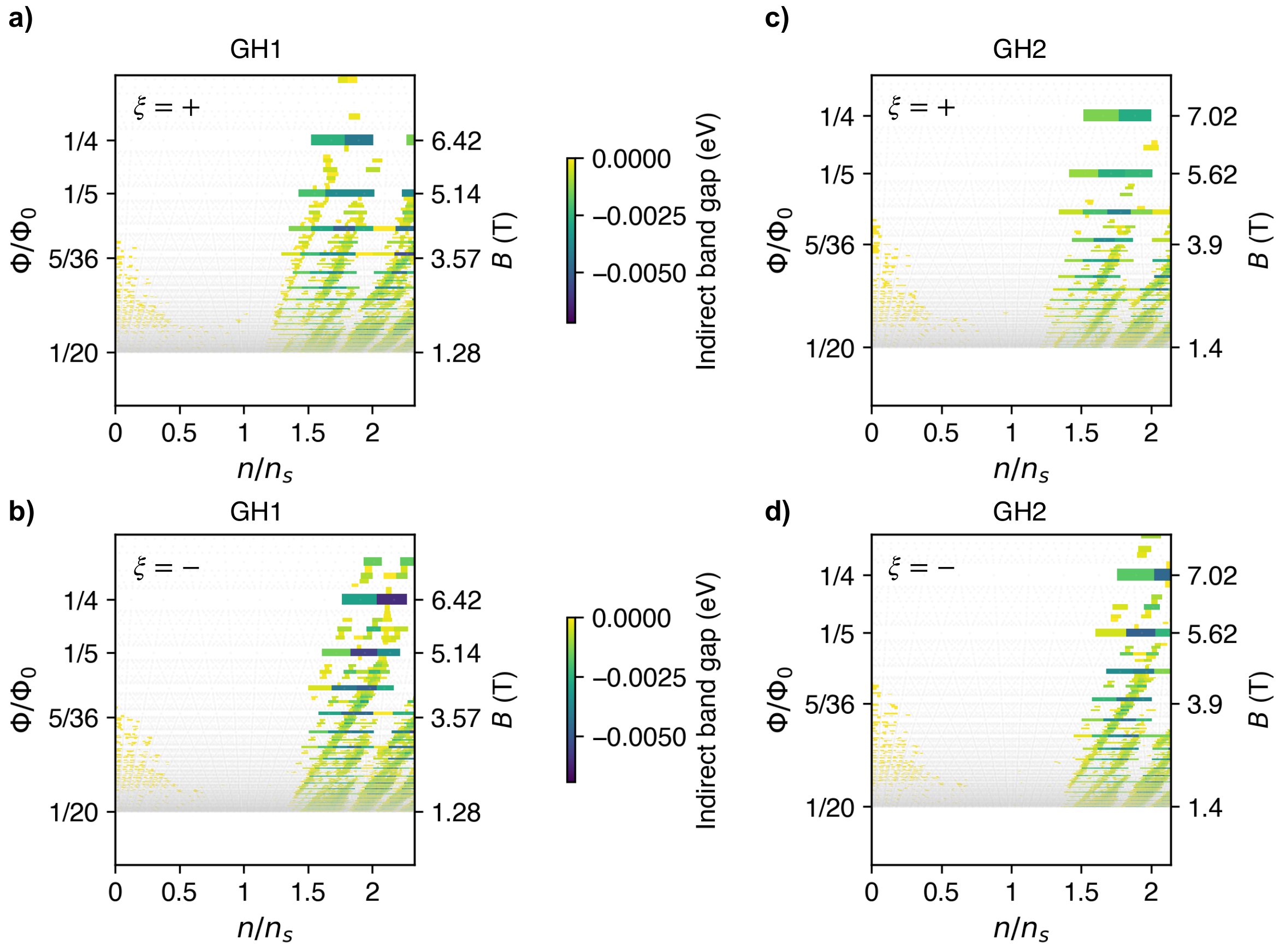}
 \caption{\textbf{a,b)} Negative indirect band gap of $\xi = +$ and $\xi = -$ valleys versus filling for a range of magnetic flux values p/q for device GH1. \textbf{c,d)} Same data for GH2.}
 \label{fig:figSI6}
\end{figure}

\begin{figure}[H]
 \centering
   \includegraphics[width=1\columnwidth]{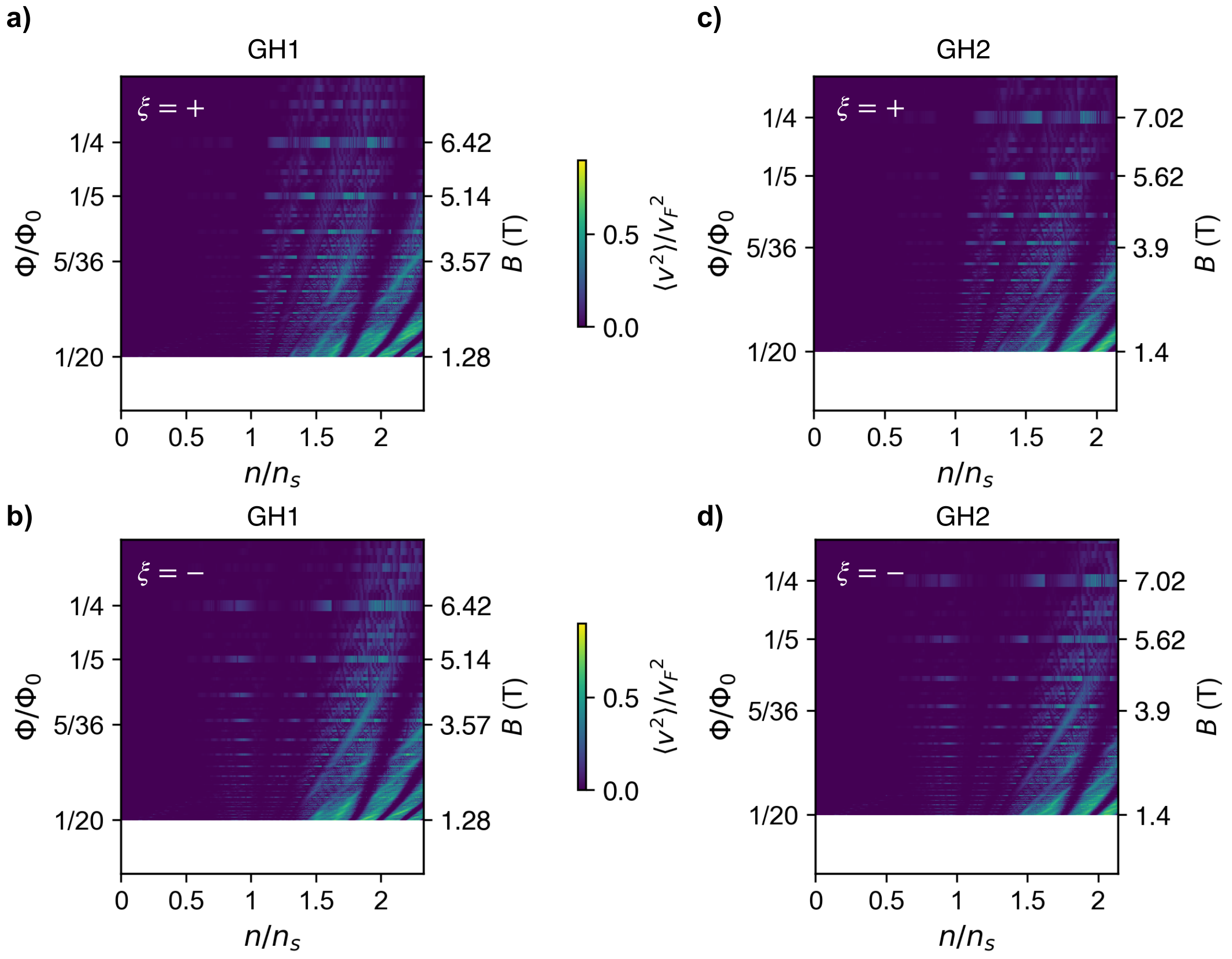}
 \caption{\textbf{a,b)} Average squared band velocity of $\xi = +$ and $\xi = -$ valleys versus filling for GH1. \textbf{c,d}) Same data for larger device GH2.}
 \label{fig:figSI7}
\end{figure}

Both devices have overlapping bands in the region of low resistance, as seen in \autoref{fig:figSI6} for both valleys $\xi = \pm$. Note that this plot only shows data for the negative indirect band gap. Each data point corresponds to the band gap between two bands $i$ and $i+1$, where the  filling value is determined via the Fermi energy of the fully filled band $i$. In both cases, we find large band velocities in these regions, which are presented in \autoref{fig:figSI7}.

\vspace{1.0\baselineskip}

\subsection{Magnetic Model: Fermi Energy and Indirect Band Gap}
\begin{figure}[h!]
 \centering
   \includegraphics[width=1\columnwidth]{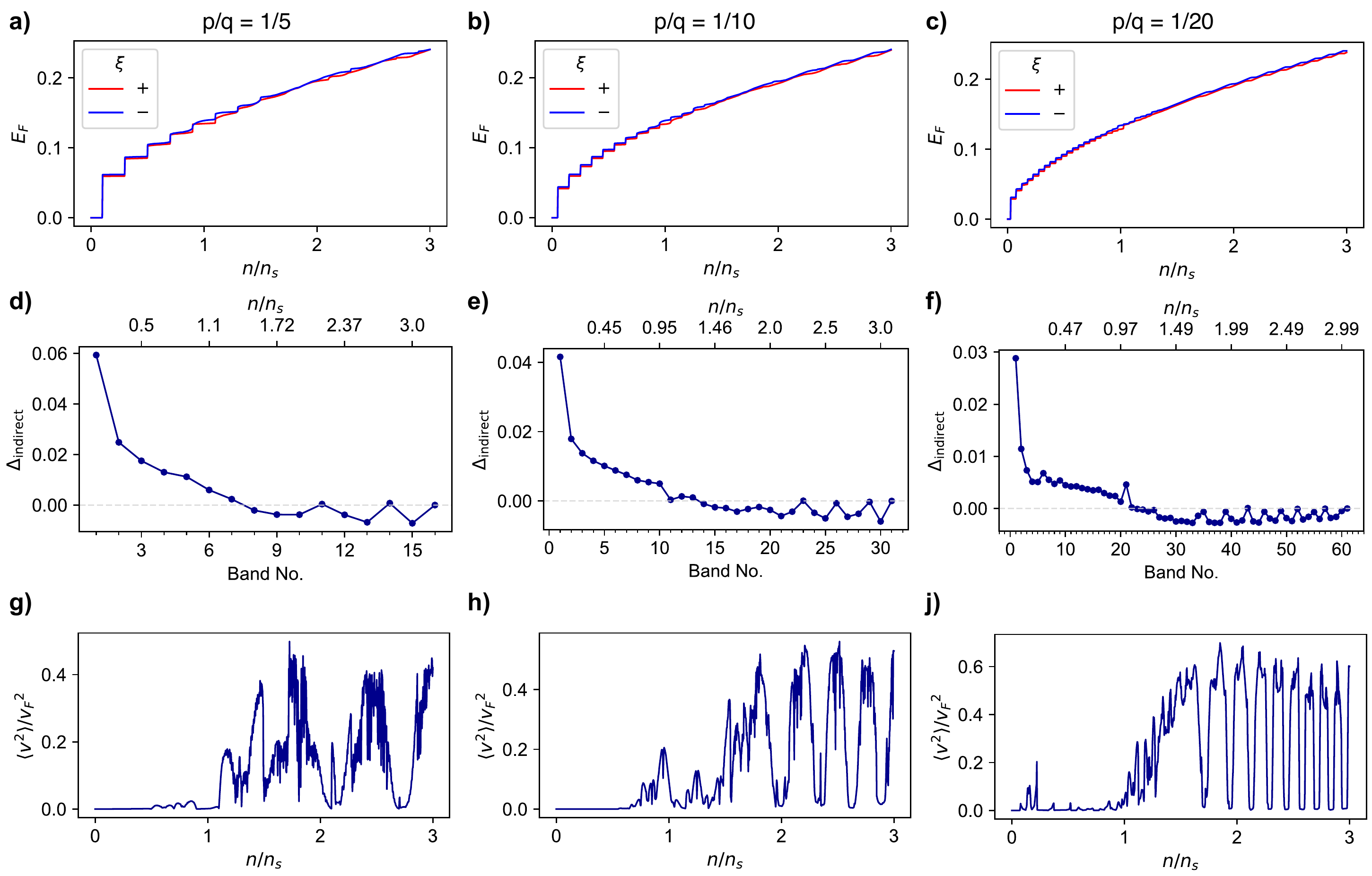}
 \caption{Top row: Fermi energy $E_F$ of the $\xi = +$ ($\xi = - $) valley in red (blue) as a function of filling $n/n_s$ for different magnetic flux quanta  $p/q = 1/5$, $1/10$, and $1/20$ in panels \textbf{a)}-\textbf{c)}, respectively. \textbf{d-f)} Indirect band gap of the $\xi = +$ valley as a function of band number for the same flux quanta. The top axis indicates corresponding filling values. \textbf{g-h)} Band velocity $\langle v^2\rangle/v_F^2$ vs filling for both valleys.}
 \label{fig:figSI2}
\end{figure}

In \autoref{fig:figSI2}a,c, we present an overview of the Fermi energy $E_F(n/n_s)$ as a function of filling for different flux quanta $p/q = 1/5$, $1/10$, and $1/20$, respectively. Noticeable, below filling $n/n_s = 1$, we observe that the bands are well separated and flat. For larger fillings $n/n_s > 1$, the bands become more dispersive and eventually overlap, and the indirect band gap becomes negative; see panels \autoref{fig:figSI2}d-f. Panels \autoref{fig:figSI2}g-j corresponds to horizontal linecuts of Fig. 3d of the main text.

\begin{figure}[H]
 \centering
   \includegraphics[width=0.33\columnwidth]{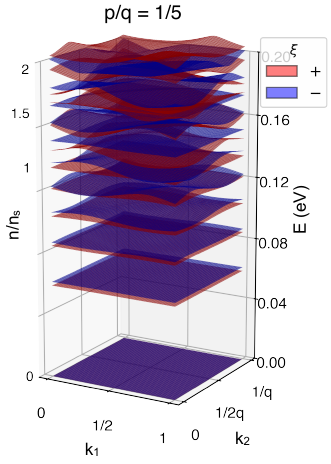}
 \caption{Individual bands of the $\xi = +$ ($\xi = -$) valley in red (blue) for $p/q = 1/5$.}
 \label{fig:figSI3}
\end{figure}

\autoref{fig:figSI3} shows the bands from both valleys $\xi = \pm$ for all $p/q = 1/5$ over the full mBZ. Note that this data is presented as a function of $(k_1,k_2)$, see eq. \ref{eq:momentum_vector_def}. In agreement with the literature, the bands from the two valleys have slightly shifted Fermi energies for a given filling, except at $n/n_s = 0$, but exhibit similar dispersion. 

\vspace{1.0\baselineskip}

\subsection{Numerical calculations}
\subsubsection{Magnetic model}
All calculations are performed with $N_{LL} = 100$ Landau levels on a grid of $N_k^2$ many momentum points for both valleys $\xi = \pm$, where we use $N_k = 50$. Note that the Hamiltonian in eq. \ref{eq:H} is block-diagonal in valley-space, allowing us to perform separate calculations for each valley. 
We calculate all possible rational flux quanta $p/q$ between $p/q = 1/3$ (corresponding to 8.54 T in GH1) and $p/q = 1/20$ (equivalent to 1.28 T for GH1) for $p=1,2,3,4,5,6$ with all $q$, corresponding to 205 different magnetic field values.

We obtain the coefficients of $\hat{V}_{\bm g}^{\xi}$ from eq. 16 in \cite{moon_electronic_2014} with the same values of $V_0 = 0.0238 \,\text{eV}, V_1 = 0.0210 \,\text{eV}, \psi = -0.29 \,\text{rad}$, and $\omega = e^{2\pi \mathrm{i}/3}$. Note that the moir\'e potential parameters depend on the valley index. 
We use the following lattice parameters for graphene $a =  2.46 \times 10^{-10} \,\text{m}$ and hBN $a_{\rm hBN} = 1.018\, a$, and a Fermi velocity of $v_F = 0.8 \times 10^6 \,\text{m/s}$. Due to the weak scattering between the two valleys, we calculate the band velocity $\expval{\mathbf{v}^2}$ (see eq. \ref{eq:velocity}) separately and present the average:
\begin{equation}
    \expval{\mathbf{v}^2} = \frac{\expval{\mathbf{v}^2}_{\xi=+} + \expval{\mathbf{v}^2}_{\xi=-}}{2}
\end{equation}
We use a broadening of $\eta = 0.0001$ eV to calculate the band velocity as a function of filling $n/n_s$. 

\vspace{1.0\baselineskip}

\pagebreak
\section*{References}
\printbibliography[heading=none]

\end{refsection}

\end{document}